\documentclass[iop]{emulateapj-rtx4}
\usepackage{epsfig}
\usepackage{graphicx}
\usepackage{longtable}
\usepackage{natbib}
\usepackage{subfigure}

\newcommand{\Msolar}{\textit{M}$_{\odot}$}
\newcommand{\Rsolar}{\textit{R}$_{\odot}$}
\newcommand{\kms}{km s$^{-1}$}

\newcommand{\PRV}{\textit{P}$_\mathrm{RV}$}
\newcommand{\PPM}{\textit{P}$_{\mu}$}

\newcommand{\ClusAmp}{112.99}
\newcommand{\ClusAvgRV}{2.45}
\newcommand{\ClusSig}{1.02}
\newcommand{\FldAmp}{8.81}
\newcommand{\FldAvgRV}{$-$14.59}
\newcommand{\FldSig}{25.79}

\newcommand{\numcatSM}{566}
\newcommand{\numcatSN}{1381}
\newcommand{\numcatBM}{93}
\newcommand{\numcatBN}{79}
\newcommand{\numcatBLM}{20}
\newcommand{\numcatBLN}{172}
\newcommand{\numcatBU}{22}
\newcommand{\numcatU}{1562}

\newcommand{\numMSbin}{70}
\newcommand{\Pcirc}{$6.2^{+1.1}_{-1.1}$}

\begin{document}

\title{WIYN OPEN CLUSTER STUDY. LX. SPECTROSCOPIC BINARY ORBITS IN NGC 6819}

\author{Katelyn E. Milliman\altaffilmark{1,}\altaffilmark{6}, Robert D. Mathieu\altaffilmark{1,}\altaffilmark{6}, Aaron M. Geller\altaffilmark{2,}\altaffilmark{3}, Natalie M. Gosnell\altaffilmark{1}, S{\o}ren Meibom\altaffilmark{4}, and Imants Platais\altaffilmark{5}}
\email{email: milliman@astro.wisc.edu}

\altaffiltext{1}{Department of Astronomy, University of Wisconsin-Madison, 475 North Charter St, Madison, WI 53706, USA}
\altaffiltext{2}{Center for Interdisciplinary Exploration and Research in Astrophysics (CIERA) and Department of Physics and Astronomy, Northwestern University, 2145 Sheridan Rd, Evanston, IL 60208, USA}
\altaffiltext{3}{Department of Astronomy and Astrophysics, University of Chicago, 5640 S. Ellis Avenue, Chicago, IL 60637, USA}
\altaffiltext{4}{Harvard-Smithsonian Center for Astrophysics, 60 Garden Street, Cambridge, MA 02138, USA}
\altaffiltext{5}{Department of Physics and Astronomy, Johns Hopkins University, 3400 North Charles Street, Baltimore, MD 21218, USA}
\altaffiltext{6}{Visiting Astronomer, Kitt Peak National Observatory, National Optical Astronomy Observatory, which is operated by the Association of Universities for Research in Astronomy (AURA) under cooperative agreement with the National Science Foundation.}

\begin{abstract}
We present the current state of the WOCS radial-velocity (RV) survey for the rich open cluster NGC 6819 (2.5 Gyr) including \numcatBM~spectroscopic binary orbits with periods ranging from 1.5 to 8,000 days. These results are the product of our ongoing RV survey of NGC 6819 using the Hydra Multi-Object Spectrograph on the WIYN 3.5 m telescope. We also include a detailed analysis of multiple prior sets of optical photometry for NGC 6819. Within a 1$^{\circ}$ field of view, our stellar sample includes the giant branch, the red clump, and blue straggler candidates, and extends to almost 2 mag below the main-sequence (MS) turnoff. For each star observed in our survey we present all RV measurements, the average RV and velocity variability information. Additionally, we discuss notable binaries from our sample, including eclipsing binaries (WOCS 23009, WOCS 24009, and WOCS 40007), stars noted in $Kepler$ asteroseismology studies (WOCS 4008, WOCS 7009, and WOCS 8007), and potential descendants of past blue stragglers (WOCS 1006 and WOCS 6002). We find the incompleteness-corrected binary fraction for all MS binaries with periods less than 10$^{4}$ days to be 22\% $\pm$ 3\% and a tidal circularization period of \Pcirc~days for NGC 6819.
\end{abstract}

\section{Introduction}
The rich intermediate-age open cluster NGC 6819 (2.5 Gyr; $\alpha$= 19$^\mathrm{h}$41$^\mathrm{m}$17$^\mathrm{s}.$5 (J2000), $\delta$=$+$40$^{\circ}$11$'$47$''$; $l$=74$^{\circ}.$0, $b$=+8$^{\circ}.$5) is one of four open clusters located within the field of view of the $Kepler$ Mission and as such has experienced an increase in interest in recent years. Research on this cluster includes deep photometric studies (\citealt{RV1998}, \citealt{K2001}, \citealt{Yang2013}), a time-series study (\citealt{Street2002}, \citealt{Street2005}), a moderately high-precision radial-velocity (RV) survey (\citealt{Hole2009}), studies of metal abundance (\citealt{Brag2001}, \citealt{AnthonyTwarog2013}), and proper-motion measurements (\citealt{Sanders1972}, \citealt{Platais2013}). Other research has studied the X-ray properties of NGC 6819 (\citealt{Gosnell2012}) resulting in the discovery of a dozen sources including a candidate quiescent low-mass X-ray binary. $Kepler$ data have enabled a number of asteroseismology studies of red giants in NGC 6819 that have investigated cluster membership (\citealt{Stello2010}, \citealt{Stello2011}), provided model-independent estimates of cluster parameters such as distance and age (\citealt{Basu2011}), and integrated red giant branch (RGB) mass-loss (\citealt{Miglio2011}). $Kepler$ data have also been used to detect eclipsing binaries in NGC 6819 that have yielded stellar mass, radius, and age measurements (\citealt{Sandquist2013}, \citealt{Jeffries2013}).

As part of the WIYN Open Cluster Study (WOCS; \citealt{Mathieu2000}) we have continued the moderately high-precision RV survey of NGC 6819 discussed in \cite{Hole2009}. Our ongoing RV survey covers a 1$^{\circ}$ diameter region on the sky and includes giant branch, red clump, upper main-sequence (MS), and blue straggler candidates. In this paper we present \numcatBM\ spectroscopic binary orbits, as well as variability information for all stars observed as part of our RV survey. In Section 2, we discuss the photometry used in this study. Section 3 details the proper-motion membership information presented in \cite{Platais2013} that we have incorporated into our RV survey and analysis of NGC 6819. We describe the details and results of our RV survey in Section 4 including our observations, completeness, and RV membership probabilities. In Section 5, we provide \numcatBM\ single-lined (SB1) and double-lined (SB2) orbital solutions for binary cluster members of NGC 6819. Binaries identified as interesting stars from $XMM$-$Newton$ or $Kepler$ observations and stars that are potential triple systems are discussed in Section 6. We analyze the circularization period (CP) of NGC 6819 in Section 7 and the binary frequency for stars with periods under 10$^{4}$ days in Section 8. Section 9 summarizes our current results for the WOCS RV survey of NGC 6819. 

\section{Cluster Photometry}
\label{sec:phot}

\begin{figure*}[htbp]
\begin{center} 
\includegraphics[scale=0.8]{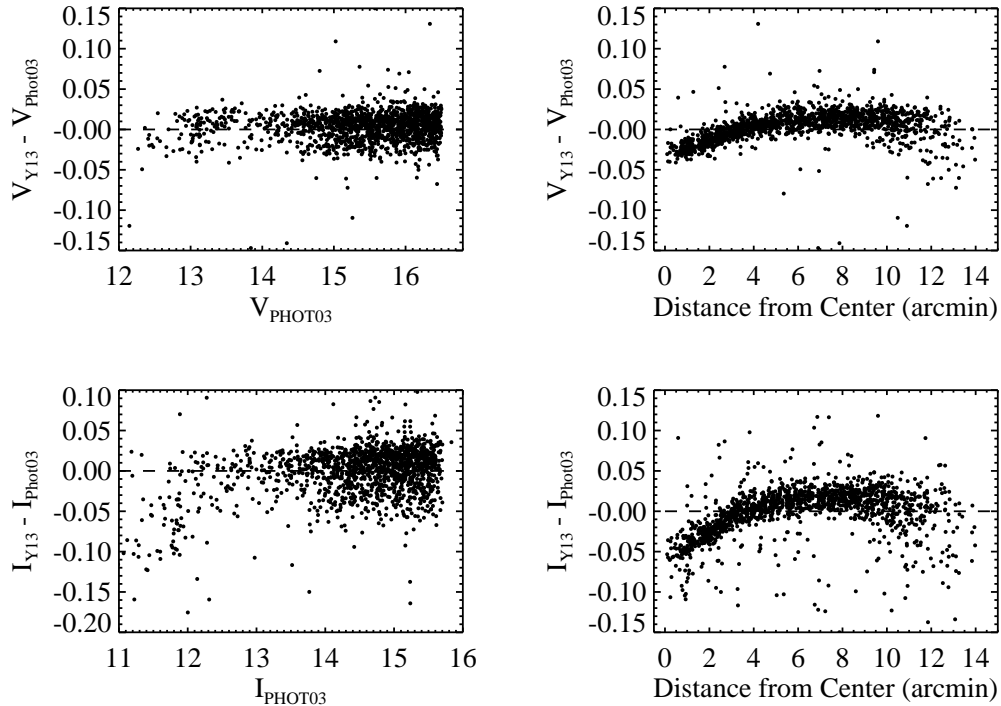}
\caption{Comparison of the photometry for overlapping sources between Y13 and Phot03. The top row of figures shows $V_\mathrm{Y13}$ minus $V_\mathrm{Phot03}$ as a function of $V_\mathrm{Phot03}$ (left) and as a function of radial distance from the center of the cluster (right). The bottom row shows the same comparisons for $I_\mathrm{Y13}$ and $I_\mathrm{Phot03}$. The dashed line in all the figures marks zero difference between the photometry sets.}
\label{fig:Y13}
\end{center}
\end{figure*}

The first WOCS NGC 6819 RV paper, \cite{Hole2009}, utilized four sources of CCD $BV$ photometry: \cite{RV1998}, \cite{K2001}, and two sets of WOCS photometry taken at the WIYN 0.9 m telescope in 1998 and 2003 (\citealt{Sarrazine2003}). We will use RV98, K01, Phot98, and Phot03 to refer to each photometry set, respectively. \cite{Hole2009} calibrated each set of photometry to the same zero point as Phot03 and derived the final photometry by averaging all $V$ and ($B$$-$$V$) information available for each star from RV98, Phot98, Phot03, and K01 (more details available in \citealt{Hole2009}, and references therein). This provided $BV$ photometry for 2220 stars in our sample, yet left a large number of stars, especially in the outer part of the field, with no $BV$ information. Recent photometry of \cite{Yang2013} and \cite{UBV2012}, hereafter Y13 and EHK12 respectively, include almost all the stars in our NGC 6819 sample. We have used these photometry sets to revise and expand the photometry for the stars in our RV survey.

\subsection{Yang et al. (2013)}
 Y13 provides CCD $VI$ photometry for 3831 stars in our RV sample and covers the entire radial extent of our survey. The data were taken in 2000 July on the WIYN 0.9 m telescope with the NOAO MOSAIC imager at the \textit{f}/7.5 secondary and provided a field of view of approximately 1$^{\circ}$$\times$1$^{\circ}$. \cite{Yang2013} used Phot03 to calibrate their instrumental magnitudes and therefore Y13 and Phot03 have no significant offsets. This is seen in Figure \ref{fig:Y13} and in the average differences, $\left \langle V_\mathrm{Y13} - V_\mathrm{Phot03} \right \rangle$ $=$ 0.005 $\pm$ 0.001 and $\left \langle I_\mathrm{Y13} - I_\mathrm{Phot03} \right \rangle$ $=$ 0.000 $\pm$ 0.001. The $I$ band shows a larger range in the scatter between Phot03 (root-mean-square error of 0.039) and Y13 than the $V$ band (rms error of 0.036), but neither band has a dependence on magnitude except in $I$ for the very brightest stars.

\subsection{\cite{UBV2012}}
EHK12 provides $UBV$ photometry of $4,414,002$ sources in the $Kepler$ field, including NGC 6819. They observed the $Kepler$ field using the NOAO Mosaic-1.1 Wide Field Imager and the WIYN 0.9 m telescope with a set of 209 pointings over five nights in 2011 June and achieved typical completeness limits of $U$ $\sim$ 18.7, $B$ $\sim$ 19.3, and $V$ $\sim$ 19.1, well below the magnitude limits of our target sample.

\begin{figure*}[htbp]
\begin{center}
\includegraphics[scale=0.8]{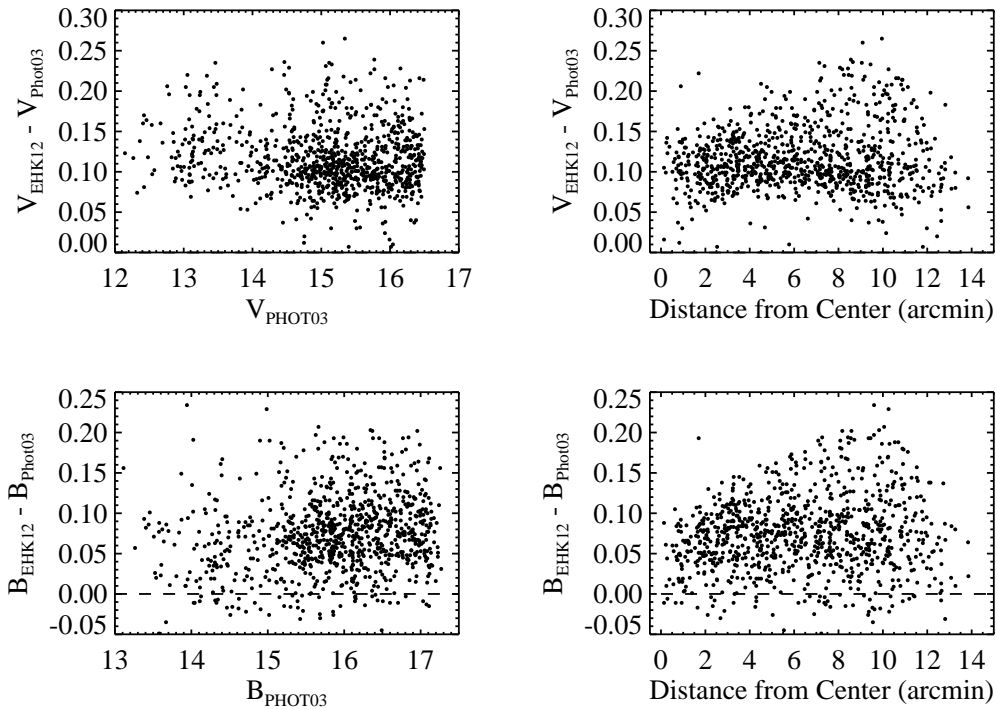}
\caption{Comparison of the photometry for overlapping sources between \cite{UBV2012} and Phot03. The top row of figures shows $V_\mathrm{EHK12}$ minus $V_\mathrm{Phot03}$ as a function of $V_\mathrm{Phot03}$ (left) and as a function of radial distance from the center of the cluster (right). The bottom row shows the same comparisons for $B_\mathrm{EHK12}$ and $B_\mathrm{Phot03}$. 
}
\label{fig:EHK12}
\end{center}
\end{figure*}

Figure \ref{fig:EHK12} shows a large offset between Phot03 and EHK12 in both $V$ and $B$ as well as a large amount of scatter at all magnitudes and stellar positions. $\left \langle V_\mathrm{EHK12} - V_\mathrm{Phot03}\right \rangle$ and $\left \langle B_\mathrm{EHK12} - B_\mathrm{Phot03}\right \rangle$ are 0.115 $\pm$ 0.001 and 0.073 $\pm$ 0.002, respectively. Also, $V_\mathrm{EHK12}$ $-$ $V_\mathrm{Phot03}$ has a rms error of 0.127 and $B_\mathrm{EHK12}$ $-$ $B_\mathrm{Phot03}$ has an rms error of 0.0935, almost triple the rms errors found with the Y13 photometry comparison.

\subsection{Photometry Summary}
We currently use only the $VI$ photometry from Y13 for our RV survey and in this paper. We made this decision due to the limited radial extent of the previous $BV$ photometry sets and the large offsets and lack of precision in EHK12. Y13 $VI$ photometry provides photometry for 98\% of our RV survey and covers the entire radial extent of our survey.

As seen clearly in Figure~\ref{fig:Y13} our photometry comparisons have revealed spatial trends in $V_\mathrm{Y13} - V_\mathrm{Phot03}$ and $I_\mathrm{Y13} - I_\mathrm{Phot03}$ with distance from the center of the cluster. \cite{Clem2013} note a similar trend in the difference between their calibrated magnitudes and those of Landolt standards which vary as a function of $x$ and $y$ position relative to the center of the two different detectors used in their study. They attribute this systematic trend to the failure of the flat fields to completely correct for large-scale illumination patterns in the detector systems. This detector-independent flat fielding effect may be the cause of the trends seen in the Y13 data as well. Regardless, the typical magnitude and color errors are less than 0.05 and will not affect any of the kinematic results discussed in the rest of this paper.

Finally, based on Y13 $VI$ photometry of stars within 10$'$ of the cluster center that are proper-motion members \cite{Platais2013} find differential reddening in NGC 6819. As \cite{Platais2013} only analyzed the inner part of our total sample region and we expect the specific $\Delta$$E(V-I)$ quantities may need to be revisited based on the radial trend in Y13 discussed above, we do not incorporate these findings in the photometry in Table 2 or the color$-$magnitude diagram (CMD) presented in this paper. However, the maximum differential reddening found by \cite{Platais2013} was $\Delta$$E(V-I)$ = 0.09 and we expect associated scatter in the CMD.
	
\section{Proper Motions}
\label{sec:ppm}
\cite{Platais2013} presents proper-motion measurements for $15,408$ stars in the NGC 6819 field with a precision that ranges from $\sim$0.2 mas yr$^{-1}$ within 10$'$ of the cluster center to 1.1 mas yr$^{-1}$ outside of 10$'$. The study uses CCD observations taken with the MegaCam imager at the CFHT as well as 23 scanned photographic plates from four different telescopes, all with different epochs, depth, scales, and fields of view.

The proper-motion membership probabilities, \PPM, are calculated with local-sample techniques (\citealt{Koz1995}) which utilize separate two-dimensional Gaussian frequency distributions of the field and the cluster for a limited sample of stars. The limited samples are chosen to contain stars with similar magnitudes and distance from the cluster center as the specific star being analyzed. 

\begin{figure*}[htbp]
\begin{center}
\includegraphics[width=\linewidth]{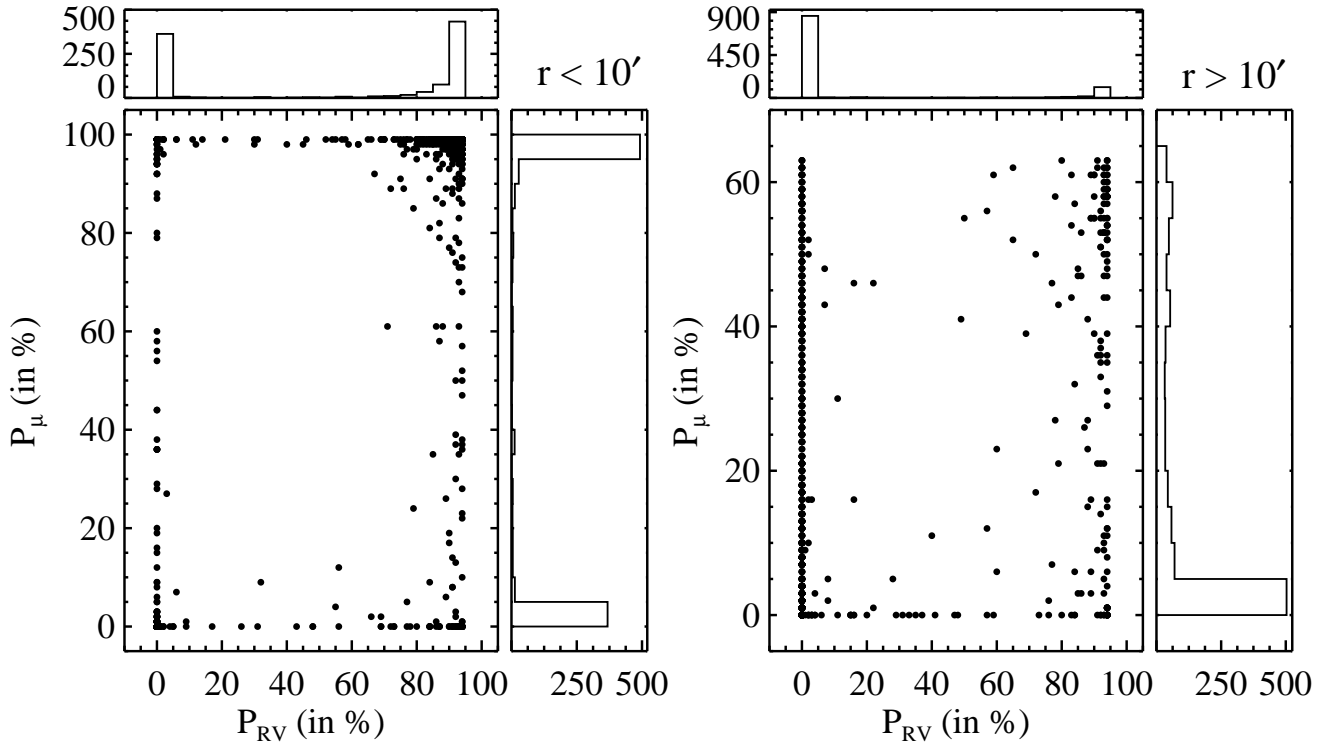}
\caption{Comparison of cluster membership probabilities from radial-velocity and proper-motion measurements. Proper-motion membership probabilities, \PPM, are from \cite{Platais2013}. Radial-velocity membership probabilities are determined from Equation (3). Sources within 10$'$ of the cluster center are presented on the left and sources outside of 10$'$ are on the right. Proper-motion membership probabilities outside of 10$'$ have lower precisions, peak at \PPM\ of~63\%, and have more inconsistencies with our radial-velocity membership determination. }
\label{fig:Prv_Ppm}
\end{center}
\end{figure*}

This survey provides proper-motion information for 78\% of the targets in our RV survey of NGC 6819. The large majority of stars missing proper-motion measurements are outside the area covered by the photographic plates. Proper-motion measurements are also missing for stars that fall near the MegaCam interchip gaps, suffer from image confusion on the photographic plates, or have proper-motions exceeding the upper limit imposed by \cite{Platais2013} of 20 mas yr$^{-1}$ in either the \textit{x} or \textit{y} coordinate. 

In Figure~\ref{fig:Prv_Ppm} we compare the membership probabilities from proper-motion and our RV (Section~\ref{sec:Prv}) measurements, splitting the sample at a radius of 10$'$ due to the large difference in the precisions of the proper-motion data at that radius. Overall we have good agreement between the two studies. A large number of stars have both \PPM~ and \PRV~$=$ 0\% and within 10$'$ we see a large concentration of targets with high RV and proper-motion probabilities. In both plots we note a large population of stars around the perimeter (e.g., the vertical lines at \PRV~$=$ 0\%). These over-densities are due to the vast majority of the stars in both studies falling at either end of the distribution, either 0\% or 99\%, providing a large population to fill up the edges and leaving very few stars to populate the centers of the plots of Figure~\ref{fig:Prv_Ppm}.

\cite{Platais2013} choose a membership threshold at a \PPM~of 4\%, but note a substantial overlap between the proper-motion distributions of the cluster and field stars in the direction of NGC 6819. They find that a CMD of stars with \PPM\ $\geq$ 4\% within 10$'$ shows a clean MS, sub-giant branch, and RGB. A CMD that includes stars with \PPM\ $<$ 4\% shows no distinguishable cluster features. 

We also adopt the proper-motion membership threshold of 4\%. We do caution that proper-motion information outside of 10$'$ from the cluster center have larger errors and disagree with our membership determinations from RVs more frequently (Figure~\ref{fig:Prv_Ppm}) than for stars within 10$'$. 

\section{Radial-Velocity Study}
The details of our RV survey of NGC 6819 including the observing procedure, data reduction, and membership classification are discussed in depth in \cite{Hole2009} and \cite{Geller2008}. The following sections summarize the details relevant to the results presented in this paper and provide updated information where appropriate. 

\begin{deluxetable*}{cccrcrrcrc}
\tablecolumns{10}
\centering
\tabletypesize{\footnotesize}
\tablecaption{Radial-velocity Measurements}
\tablehead{ \colhead{ID} & \colhead{HJD$-$2400000} & \colhead{Telescope\tablenotemark{a}} & \colhead{RV$_{1}$} & \colhead{Correlation Height$_{1}$} & \colhead{$O-C_{1}$} & \colhead{RV$_{2}$} & \colhead{Correlation Height$_{2}$} & \colhead{$O-C_{2}$} & \colhead{Phase}}
\startdata
  9030  &&&&&&&& \\
 & 55374.6633 &  W &     1.874 &    0.89 &       4.00 &      $-$19.272 &    0.63 &       3.44 &  0.234 \\
 & 55374.7375 &  W &   $-$13.202 &    0.89 &       0.01 &      \nodata & \nodata &    \nodata &  0.257 \\
 & 55375.6625 &  W &   $-$88.546 &    0.63 &      $-$3.10 &       60.357 &    0.57 &       5.29 &  0.545 \\
 & 55375.7354 &  W &   $-$80.345 &    0.72 &       2.46 &       54.662 &    0.65 &       2.06 &  0.568 \\
 & 55376.6627 &  W &    32.298 &    0.68 &       3.05 &      $-$50.376 &    0.56 &       1.62 &  0.856 \\
 & 55376.7367 &  W &    42.208 &    0.73 &       3.55 &      $-$59.876 &    0.54 &       0.90 &  0.879 \\
 & 55412.6959 &  W &    58.410 &    0.73 &      $-$2.79 &      $-$80.087 &    0.61 &       1.73 &  0.066 \\
 & 55413.6906 &  W &   $-$62.518 &    0.73 &       0.39 &       37.516 &    0.62 &       3.49 &  0.375 \\
 & 55414.6935 &  W &   $-$45.001 &    0.76 &       3.83 &       23.552 &    0.63 &       2.67 &  0.687 \\
 & 55440.8442 &  W &    11.868 &    0.81 &      $-$1.99 &      $-$34.090 &    0.64 &       3.54 &  0.822 
\enddata
\tablenotetext{a}{The observatory at which the observations were taken, using ``C'' for CfA facilities and ``W'' for the WIYN 3.5 m. \\ [3.5pt]
(This table is available in its entirety in machine-readable and Virtual Observatory (VO) forms in the online journal. A portion is shown here for guidance regarding its form and content.)}
\end{deluxetable*}

\subsection{Observations, Data Reduction, and Precision}
\label{sec:obs}
The WOCS RV target sample for NGC 6819 has 3895 stars that span 1$^{\circ}$ on the sky centered at $\alpha$ = 19$^\mathrm{h}$41$^\mathrm{m}$17$^\mathrm{s}.$5, $\delta$ = $+$40$^{\circ}$11$'$47$''$ (J2000). \cite{Hole2009} discuss the target selection and evolution of the RV target sample in great detail. In brief, the initial target sample was based on RV98 and PHOT98 photometry. Cutoffs based on $V$ and $(B-V)$ were chosen in order to select the upper MS, giant branch, red clump, and potential blue straggler candidates in NGC 6819. Later, the sample was updated and extended in radius with the addition of K01 and Phot03 photometric studies. Stars without optical photometry (primarily at large radii) were included based on $V$ magnitudes converted from 2MASS $J$ and $K$ information. In this paper, as discussed in Section~\ref{sec:phot}, we have incorporated Y13 $VI$ photometry that provides photometry for 98\% of our RV survey.
In 2010, we incorporated the proper-motion information from \cite{Platais2013} into our RV survey. Stars identified by \cite{Platais2013} as proper-motion non-members, \PPM~$<$ 4\%, were moved to the lowest priority for observations. 

Observations of NGC 6819 with the Hydra Multi-Object Spectrograph (MOS; \citealt{Barden1994}) on the WIYN\footnote{The WIYN Observatory is a joint facility of the University of Wisconsin-Madison, Indiana University, Yale University, and the National Optical Astronomy Observatory.} 3.5 m telescope began in 1998 June and are still ongoing. We have almost 14,000 spectra for over 2600 stars. These observations are augmented with 733 RV measurements for 170 stars taken at the Harvard-Smithsonian Center for Astrophysics (CfA) facilities between 1988 May and 1995 by R. D. Mathieu \& D. W Latham (\citealt{Hole2009}). 

The WIYN data include for each configuration: three science exposures, one 100 s dome flat, and two 300 s thorium$-$argon emission lamp spectra, one taken before and one taken after the science exposures. The science observations are split into three integrations of equal time for cosmic ray rejection. All the image processing is done within IRAF where the data are bias subtracted and dispersion corrected, and the extracted spectra are flat-fielded, throughput corrected, and sky subtracted. 

RVs for single-lined stars are derived from the centroid of a one-dimensional cross-correlation function (CCF) with an observed solar template, converted to a heliocentric velocity, and corrected for the unique fiber offsets of the Hydra MOS. RVs for double-lined stars are derived using two$-$dimensional correlation (TODCOR) and we provide more details in Section~\ref{sec:SB2}. The $\chi^{2}$ analysis of \cite{Hole2009} finds the precision for NGC 6819 data to be $\sigma_\mathrm{WIYN}$ = 0.4 \kms\ and $\sigma_\mathrm{CfA}$ = 0.7 \kms, which we have confirmed remains true for the additional RVs presented here.

We present all of our RV measurements for each star in Table 1, along with the Heliocentric Julian Date (HJD) of the observation and the height of the CCF. For binary stars with completed orbital solutions we also include the RV residual and the orbital phase of the observation. 

\subsection{Velocity Variables}
\label{sec:VV}
We define velocity variables stars as having RV standard deviations (the external error, $e$) greater than four times the precision (the internal error, $i$) or $e/i$ $>$ 4 (\citealt{Geller2008}).
We take into account the different data precisions if a star has RV measurements from both observatories using the following:
\begin{equation} 
\left ( \frac{e}{i} \right )^2=\frac{N^2}{N-1}\frac{\sum_{i}^{N}\left (\textup{RV}_{i} - \overline{\textup{RV}} \right )^2/\sigma_{i}^2}{\sum_{i}^{N}\sigma_{i}^2\sum_{i}^{N}{1}/{\sigma_{i}^2}},
\label{eq:e.i}
\end{equation}
where $\overline{\textup{RV}}$ is the mean RV weighted by the precision values of each observatory:
\begin{equation}
\overline{\textup{RV}}=\frac{\sum_{i}^{N}\textup{RV}_i/\sigma_i^2}{\sum_{i}^{N}1/\sigma_i^2}.
\label{eq:avgRV}
\end{equation}

For a star with three or more observations we calculate the $\overline{\textup{RV}}$ and $e/i$ and classify it as single or velocity variable. Given our RV measurement precisions binaries that are long-period and/or low-amplitude may be classified as single stars. We quantify our detection completeness for binaries using the Monte Carlo analysis described in Section~\ref{sec:bin.detect}.

In Table 2, we present the coordinates, $VI$ photometry, number of observations at each observatory, $\overline\mathrm{RV}$, $\overline\mathrm{RV}$ standard error, $e/i$, membership probabilities, and membership classification for all of the stars we have observed in NGC 6819 as of 2014 February. For each binary star with a completed orbital solution the center-of-mass velocity, $\gamma$, its error, and whether it is a single- or double-lined binary are also listed in Table 2.
\begin{deluxetable*}{l c c c c c c c c c c c c c c c}
\tablecaption{Radial Velocity Summary Table \label{RVtab}}
\tabletypesize{\footnotesize}
\tablehead{\colhead{ID$_W$\tablenotemark{a}} & \colhead{$\alpha$ (J2000)} & \colhead{$\delta$ (J2000)} & \colhead{$V$} & \colhead{$V-I$} & \colhead{$N_{W}$} & \colhead{$N_{C}$} & \colhead{$\overline\mathrm{RV}$} & \colhead{Std. Err.} & \colhead{$e/i$} & \colhead{$P_\mathrm{RV}$} & \colhead{$P_\mathrm{PM}$} & \colhead{Class\tablenotemark{b}} & \colhead{$\gamma_\mathrm{RV}$} & \colhead{$\gamma_\mathrm{RVe}$} & \colhead{Comment}}
\startdata
  1001 & 19   41   18.71 & 40   11   42.9 &       12.69 &       1.41 &  3 & 20 &    6.045 &   3.40 &  23.88 &      90 &      19 &      BM &       1.42 &      0.17 &     SB1 \\
  1002 & 19   41   17.05 & 40   10   51.8 &       11.65 &       1.73 &  2 & 11 &    0.943 &   0.18 &   0.92 &      84 &      99 &      SM &    \nodata &   \nodata & \nodata \\
  1004 & 19   41   26.59 & 40   11   41.8 &       12.22 &       1.50 &  5 &  5 &    1.547 &   0.22 &   1.15 &      91 &      99 &      SM &    \nodata &   \nodata & \nodata \\
  1005 & 19   41   11.85 & 40   13   30.1 &     \nodata &    \nodata &  5 &  6 &  $-$38.297 &   0.19 &   1.02 &       0 &      87 &      SN &    \nodata &   \nodata & \nodata \\
  1006 & 19   41   17.76 & 40    9   15.9 &       12.83 &       1.23 &  5 & 13 &    1.473 &   0.96 &   6.30 &      94 &      99 &      BM &       2.72 &      0.13 &     SB1 \\
  1007 & 19   41   13.20 & 40   14   56.7 &     \nodata &    \nodata &  3 &  6 &    3.415 &   0.28 &   1.31 &      91 &       0 &      SN &    \nodata &   \nodata & \nodata \\
  1010 & 19   41   42.64 & 40   11   40.2 &       12.82 &       0.29 & 17 & 15 &    2.330 &   0.87 &   8.66 &      86 &      87 &      BM &       1.07 &      0.35 &     SB1 \\
  1011 & 19   41   44.98 & 40   12   50.3 &       11.80 &       0.59 &  2 &  6 &  $-$22.497 &   0.27 &   1.13 &       0 &       0 &      SN &    \nodata &   \nodata & \nodata \\
  1012 & 19   41   46.76 & 40   10    2.3 &       12.40 &       0.73 & 11 &  2 &   14.813 &   9.70 &  73.19 &       0 &       0 &     BLN &    \nodata &   \nodata & \nodata \\
  1013 & 19   41   44.81 & 40    8    1.9 &       11.97 &       0.62 &  6 &  9 &  $-$27.611 &   0.37 &   2.34 &       0 &       0 &      SN &    \nodata &   \nodata & \nodata 
\enddata
\tablenotetext{a}{WOCS ID number. This number is based on $V$ magnitude and radial distance from the cluster center. Full details can be found in \cite{Hole2009}.}
\tablenotetext{b}{See Section~\ref{sec:class} for explanation of class codes.\\[3.5pt]
(This table is available in its entirety in machine-readable and Virtual Observatory (VO) forms in the online journal.  A portion is shown here for guidance regarding its form and content.)}
\end{deluxetable*}
\subsection{Completeness}
\label{sec:comp}
The entire target sample is comprised of 3895 stars with RV measurements from the CfA dating back to 1988 and WIYN observations beginning in 1998. Over the entire course of our survey we constantly reevaluate the observing priority of every star. Based on Monte Carlo analyses and observing efficiency arguments (\citealt{Mathieu1983}, \citealt{Geller2012}) a star is classified as single or velocity variable after three observations. In general, stars classified as single are given a low priority for further observation. Stars that are classified as velocity variable are given a high observing priority and are observed until an orbital solution is determined. However, stars identified as proper-motion non-members, \PPM\ $<$ 4\%, are given the lowest observing priority. 

\begin{figure*}
\begin{center}
\includegraphics[width=0.95\linewidth]{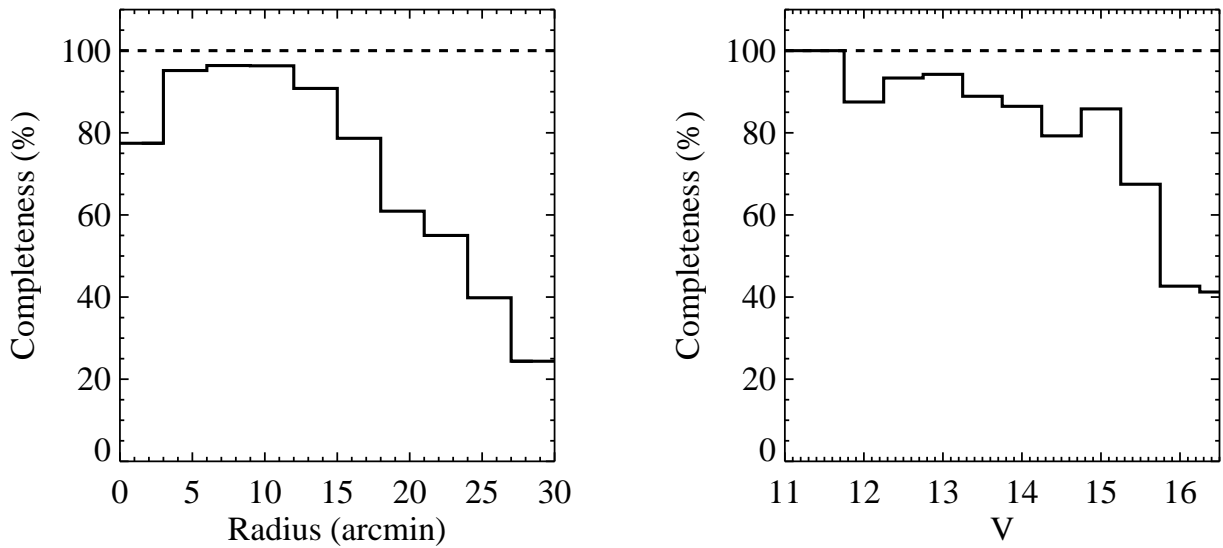}
\caption{Percentage of stars in our sample that have three or more RV observations and either no \PPM\ information or a \PPM\ $\geq$ 4\% with respect to distance from the cluster center (left) and $V$ magnitude (right).
}
\label{fig:complete}
\end{center}
\end{figure*}

The percentage of stars with either no \PPM\ or \PPM\ $\geq$ 4\% that we are able to classify as single or binary as a function of radius and $V$ magnitude are shown in Figure~\ref{fig:complete}. Lower classification percentages at the center of cluster are due to fiber separation requirements limiting the number of central stars we are able to observe in each pointing. The percentages decrease at larger radii in part because early in the survey we prioritized stars with color information and stars more than $\sim$18$'$ from the cluster center did not have color information until the inclusion of Y13 $VI$ information in 2013. We are able to classify more than 75\% of stars with $V$ magnitude less than $\sim$15.5 as single or binary and this number jumps to 96\% for stars within 18$'$ of the cluster center. Stars dimmer than this are observed less frequently because they require better observing conditions (clearer skies, dimmer moon, etc.) and longer exposures to meet our signal-to-noise requirements.

\subsection{RV Membership Probabilities}
\label{sec:Prv}
\begin{figure}
\begin{center}
\includegraphics[width=1.03\linewidth]{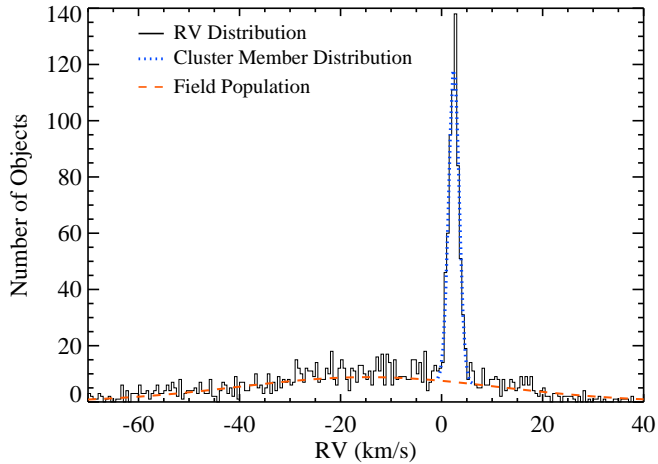}
\caption{Histogram of the RV distribution of single stars, $e/i$ $<$ 4, with three or more RV observations. Also plotted are the Gaussian distributions simultaneously fit to the cluster, the large peak at a mean velocity of 2.4 \kms~(blue dotted line), and the field (orange dashed line).}
\label{fig:clusgauss}
\end{center}
\end{figure}

The RV membership probability of a given star, \PRV, listed in Table 2 is calculated using the equation:
 \begin{equation}
 P_\mathrm{RV}(v)=\frac{F_\mathrm{cluster}(v)}{F_\mathrm{field}(v) + F_\mathrm{cluster}(v)},
 \end{equation}
where $F_\mathrm{cluster}$ and $F_\mathrm{field}$ are separate Gaussian functions fit to the cluster and field-star populations using our sample of single stars with three or more RV observations (Figure~\ref{fig:clusgauss}). The parameters for these Gaussian fits are shown in Table~\ref{tab:gauss}. With the observation and completion of more single stars these values have shifted only slightly from the numbers published in \cite{Hole2009} and we continue to use the membership threshold of \PRV\ $\geq$ 50\% adopted by \cite{Hole2009} to identify cluster members. We estimate from the cluster and field Gaussian functions a field star contamination of 13\% at this membership threshold (Figure~\ref{fig:clusgauss}). Interestingly, using \PPM\ $<$ 4\% as a criterion for non-membership, we find a field-star contamination of 12\% $\pm$ 2\% among stars with \PRV\ $\geq$ 50\%.

\begin{deluxetable}{ccc}
\tablewidth{0pt}
\tablecolumns{3}
\tablecaption{Gaussian Fit Parameters for Cluster and Field RV Distributions\label{t:dist}}
\tablehead{ \colhead{Parameter} & \colhead{Cluster} & \colhead{Field}}
\startdata
Ampl. (number) & \ClusAmp & \FldAmp \\[2pt]
$\overline\mathrm{RV}$ (\kms) & \ClusAvgRV & \FldAvgRV\\[3pt]
$\sigma$ (\kms) & \ClusSig& \FldSig
\enddata
\label{tab:gauss}
\end{deluxetable}

\subsection{Membership Classification of Stars}
\label{sec:class}
Stars in our RV survey of NGC 6819 can fall into one of eight different membership classifications determined by membership probability, number of observations, and variability. The number of stars in each membership classification is listed in Table~\ref{t:cat}. For single stars and velocity variable stars without completed orbital solutions we calculate \PRV\ from the mean velocity, $\overline\mathrm{RV}$, and for binary stars with completed orbital solutions we use the center-of-mass velocity, $\gamma$, from the orbital solution. For stars without \PPM\ information, membership is determined by \PRV\ alone.

\textit{Single member (SM)}. Stars that have $e/i$ $<$ 4, \PRV\ $\geq$ 50\%, and \PPM\ $\geq$ 4\%.

\textit{Single non-member (SN)}. Stars that have $e/i$ $<$ 4, and \PRV\ $<$ 50\% or \PPM\ $<$ 4\%.

\textit{Binary member (BM)}. Velocity-variable stars that have a completed orbital solution. \PPM\ is $\geq$ 4\% and \PRV\ $\geq$ 50\%. 

\textit{Binary non-member (BN)}. Velocity-variable stars that have a completed orbital solution. Either \PPM\ $<$ 4\% or \PRV\ $<$ 50\%. \

\textit{Binary likely member (BLM)}. Velocity-variable stars that do not have a completed orbital solution, but \PRV\ $\geq$ 50\% and \PPM\ $\geq$ 4\%. Without a completed orbital solution \PRV\ is calculated from $\overline{\textup{RV}}$ and this classification is subject to change because observations are ongoing.

\textit{Binary likely non-member (BLN)}. Velocity-variable stars that do not have a completed orbital solution. \PPM\ $<$ 4\% or the \PRV\ is $<$ 50\%. If \PPM\ $\geq$ 4\% but \PRV\ $<$ 50\%, the individual RVs do not cross the cluster mean, making it unlikely that the orbital solution will place the star within the cluster distribution. Without a completed orbital solution \PRV\ is calculated from $\overline{\textup{RV}}$ and this classification is subject to change because observations are ongoing.

\textit{Binary unknown (BU)}. Velocity-variable stars that do not have a completed orbital solution. \PPM\ $\geq$ 4\% and \PRV\ is $<$ 50\%, plus the range of individual RVs includes the cluster mean, making it more likely that the binary could be a member. Without a completed orbital solution \PRV\ is calculated from $\overline{\textup{RV}}$ and this classification is subject to change because observations are ongoing.

\textit{Unknown (U)}. Any star that does not have three or more RV measurements, along with stars such as rapid rotators for which we are unable to derive accurate RVs from our spectra. 
\begin{deluxetable}{cc}
\tablewidth{0pt}
\tablecolumns{2}
\tablecaption{Number of Stars Within Each Classification
\label{t:cat}}
\tablehead{ \colhead{Classification} & \colhead{N Stars}}
\startdata
SM & \numcatSM \\[2pt]
SN & \numcatSN \\[2pt]
BM & \numcatBM \\[2pt]
BN & \numcatBN \\[2pt]
BLM & \numcatBLM \\[2pt]
BLN & \numcatBLN \\[2pt]
BU & \numcatBU \\[2pt]
U & \numcatU 
\enddata
\end{deluxetable}

\subsection{Color$-$Magnitude Diagram}
\label{sec:cmd}
\begin{figure}[htbp]
\begin{center} 
\includegraphics[width=\linewidth]{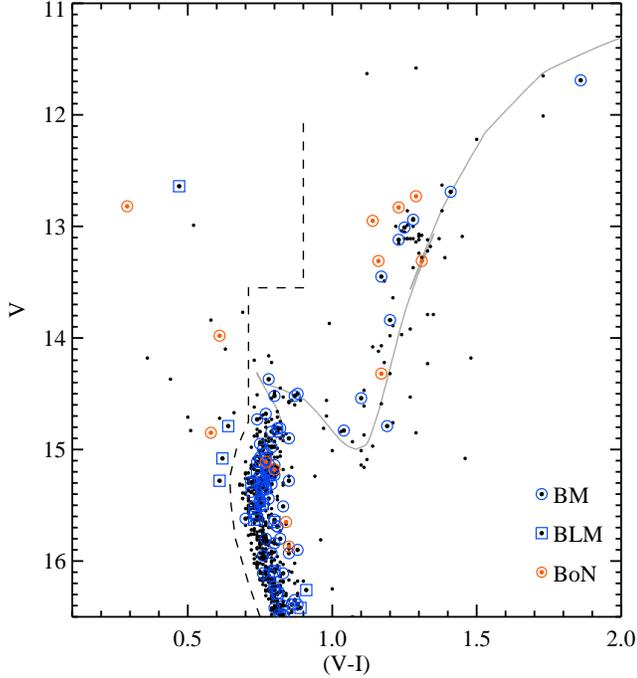}
\caption{CMD of all radial-velocity and proper-motion members of NGC 6819. Black dots represent single members. The binary members (BMs, with completed orbital solutions) are indicated by circles, the binary likely members (BLMs, without completed orbital solutions) are indicated by squares. The binaries of note (BoNs) discussed in Section~\ref{sec:BoN} are indicated in orange. A 2.5 Gyr PARSEC isochrone is shown in gray. The dashed line indicates the upper magnitude and color limit for binaries that could be compromised of two normal stars from the cluster turnoff region. Stars bluer than the dashed line are candidate blue stragglers.}
\label{fig:cmd.member}
\end{center}
\end{figure}

We incorporate all our membership classifications, the proper-motion information of~\cite{Platais2013} and the photometry of~\cite{Yang2013} to produce a cleaned CMD of NGC 6819 in Figure~\ref{fig:cmd.member} that includes BM, BLM, and SM stars. Binary stars of note (BoNs) are marked in orange on this CMD and will be discussed in more detail in Section~\ref{sec:BoN}. The CMD shows a clear MS turnoff, blue hook, and red clump. To guide the eye we plot a 2.5 Gyr PARSEC isochrone (\citealt{PARSEC2012}) using the $E(V-I)$ $=$ 0.20 and $(m - M)_{o}$ $=$ 11.93 from \cite{Yang2013}.

Candidate blue stragglers, defined as being to the blue of the dashed line in Figure~\ref{fig:cmd.member}, are also apparent. The dashed line indicates the upper magnitude and color limit for binaries that could be compromised of two normal stars from the cluster turnoff region or blue hook. A large fraction of these potential blue stragglers are RV variable stars (7/17, $\sim$40\%) and three have completed orbital solutions (WOCS 1010, WOCS 4004, and WOCS 14012; see Section~\ref{sec:BoN} for a more detailed discussion of these stars).

The RGB shows a large amount of scatter that is not caused by the blended light from binary stars and is not improved by incorporating differential reddening (\citealt{Platais2013}). This scatter on the RGB is very different from the very narrow RGB in M67 (\citealt{Mathieu2003}), but it is similar to the spread in the RGB of NGC 188. The cause of this scatter in NGC 6819 remains unknown.

\section{Spectroscopic Binary Orbits}
\label{sec:Specs}

\subsection{Single-lined Orbital Solutions}
\label{sec:SB1}
\begin{figure*}[htbp]
\begin{center}
\includegraphics[width=0.3\linewidth]{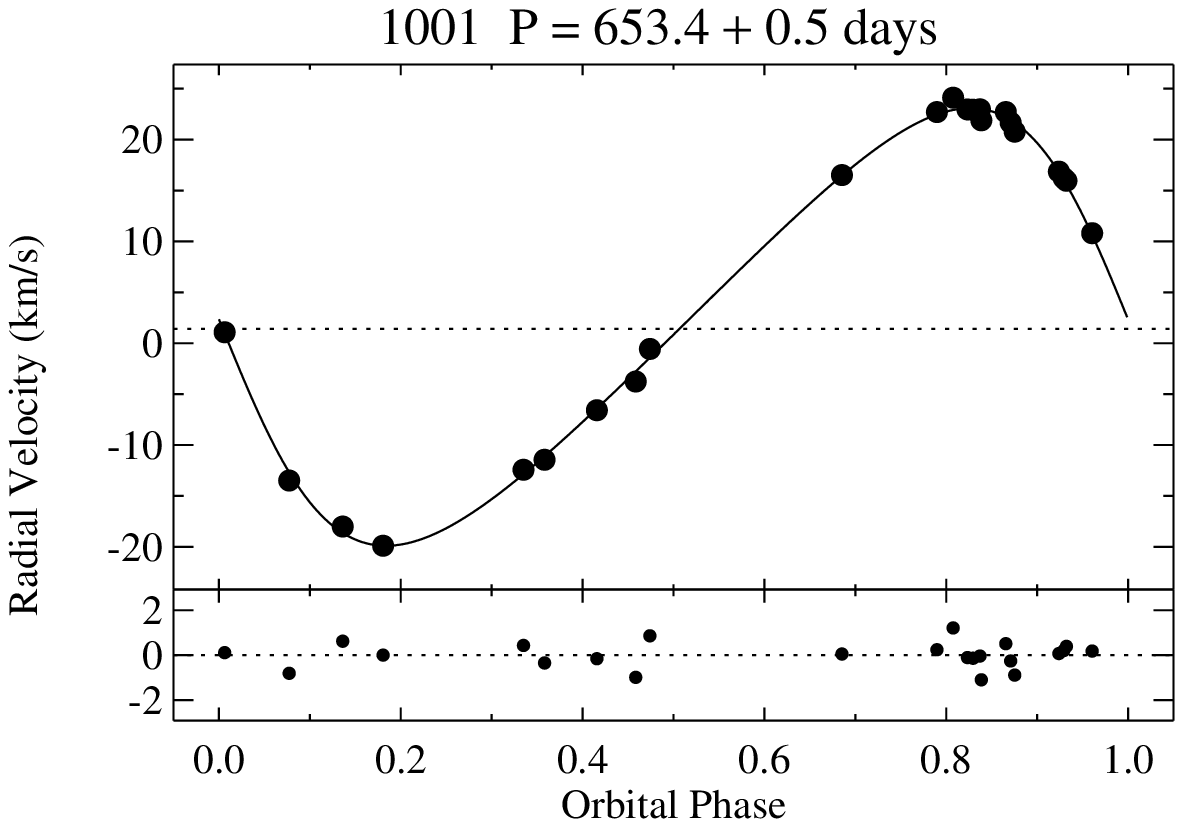} 
\includegraphics[width=0.3\linewidth]{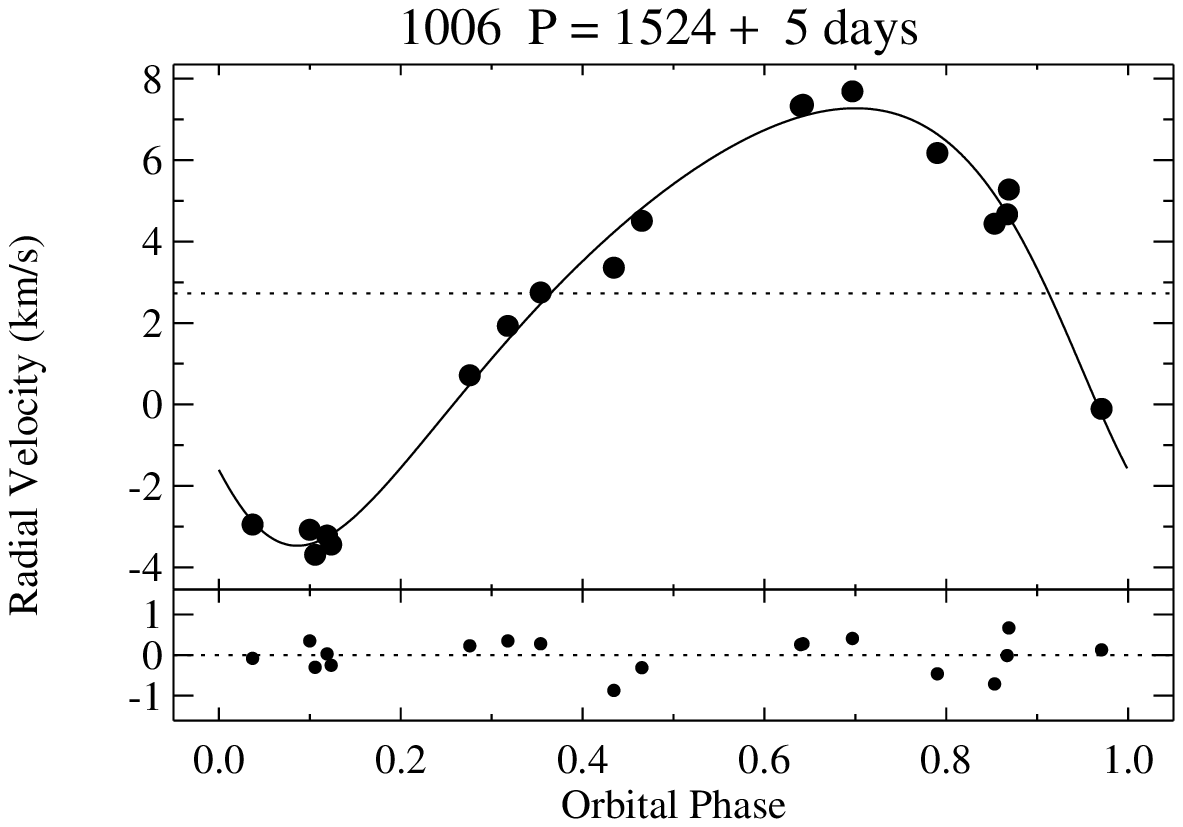}
\includegraphics[width=0.3\linewidth]{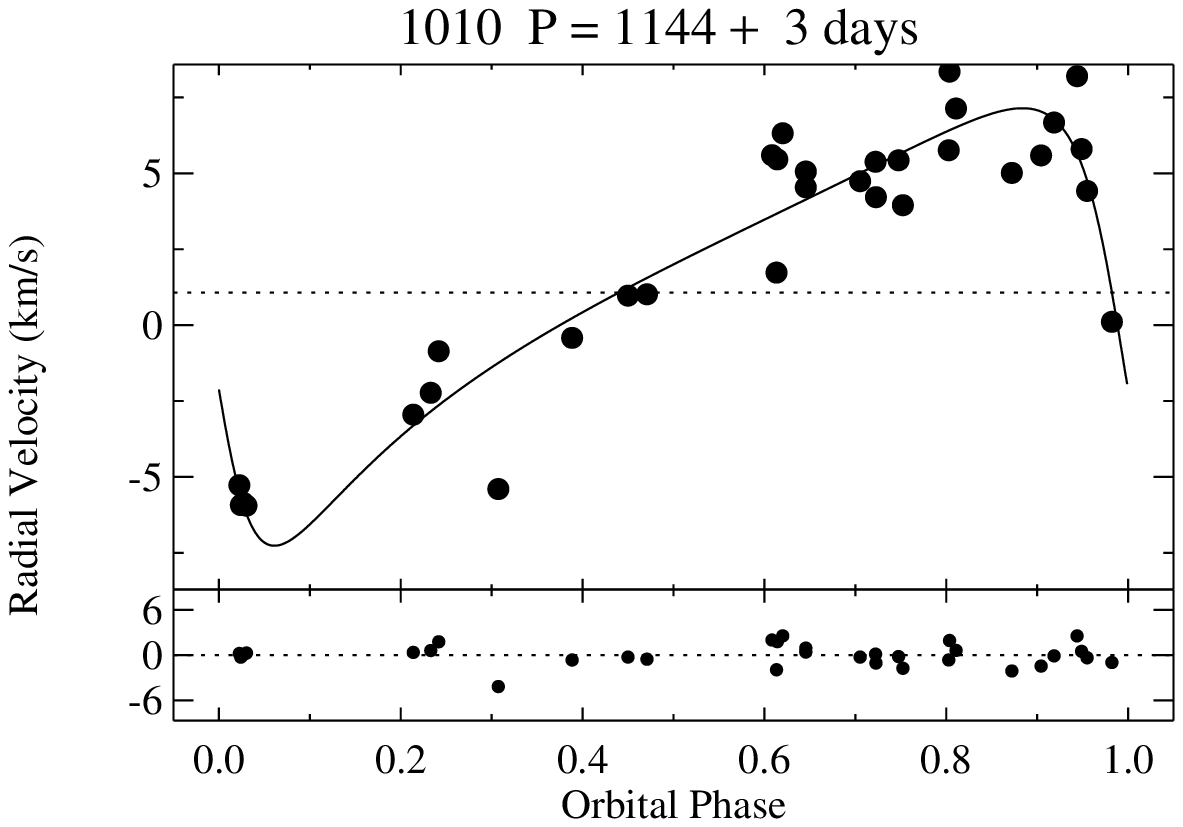}
\includegraphics[width=0.3\linewidth]{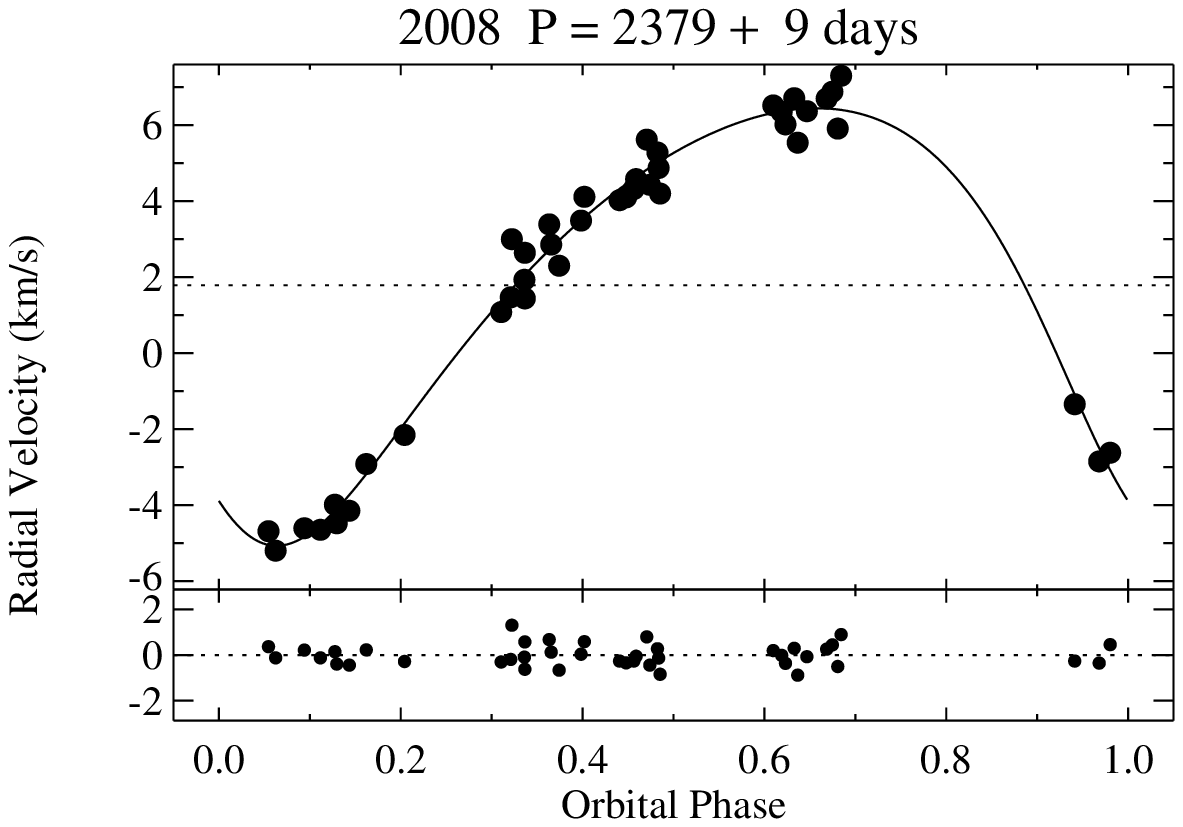}
\includegraphics[width=0.3\linewidth]{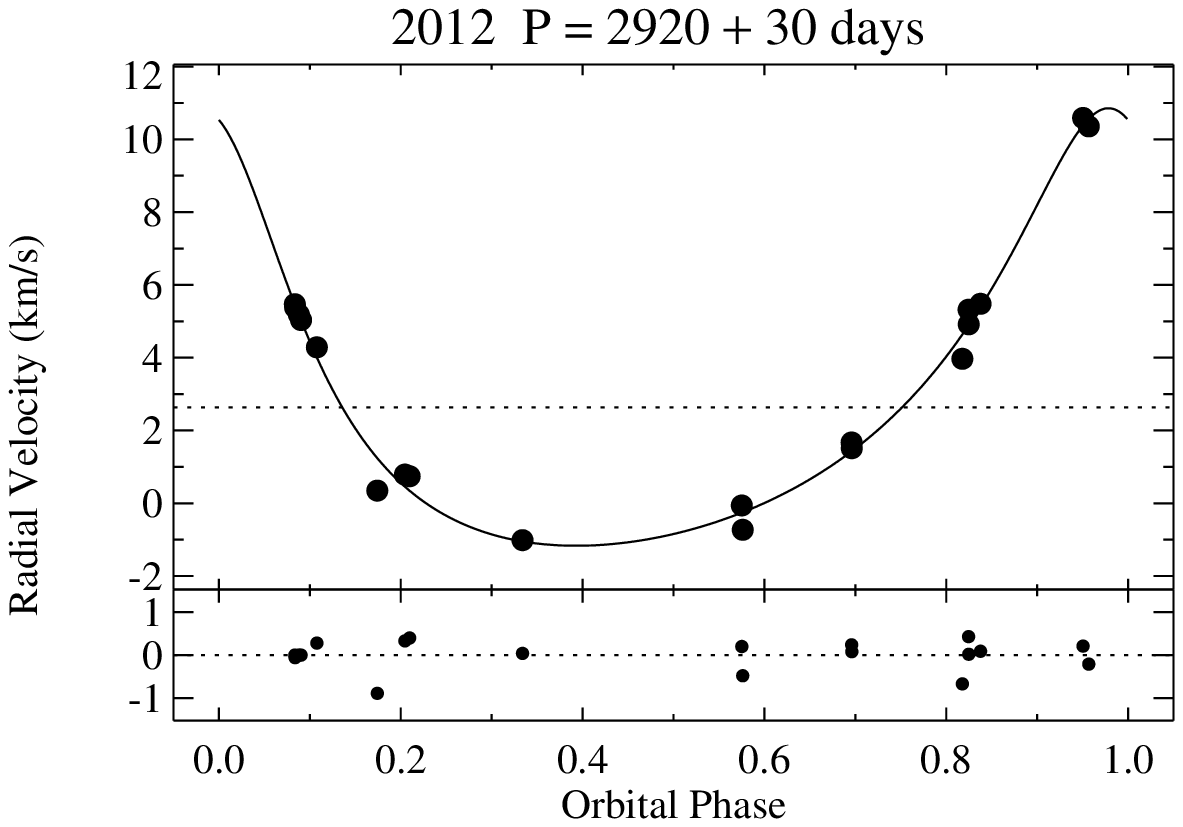}
\includegraphics[width=0.3\linewidth]{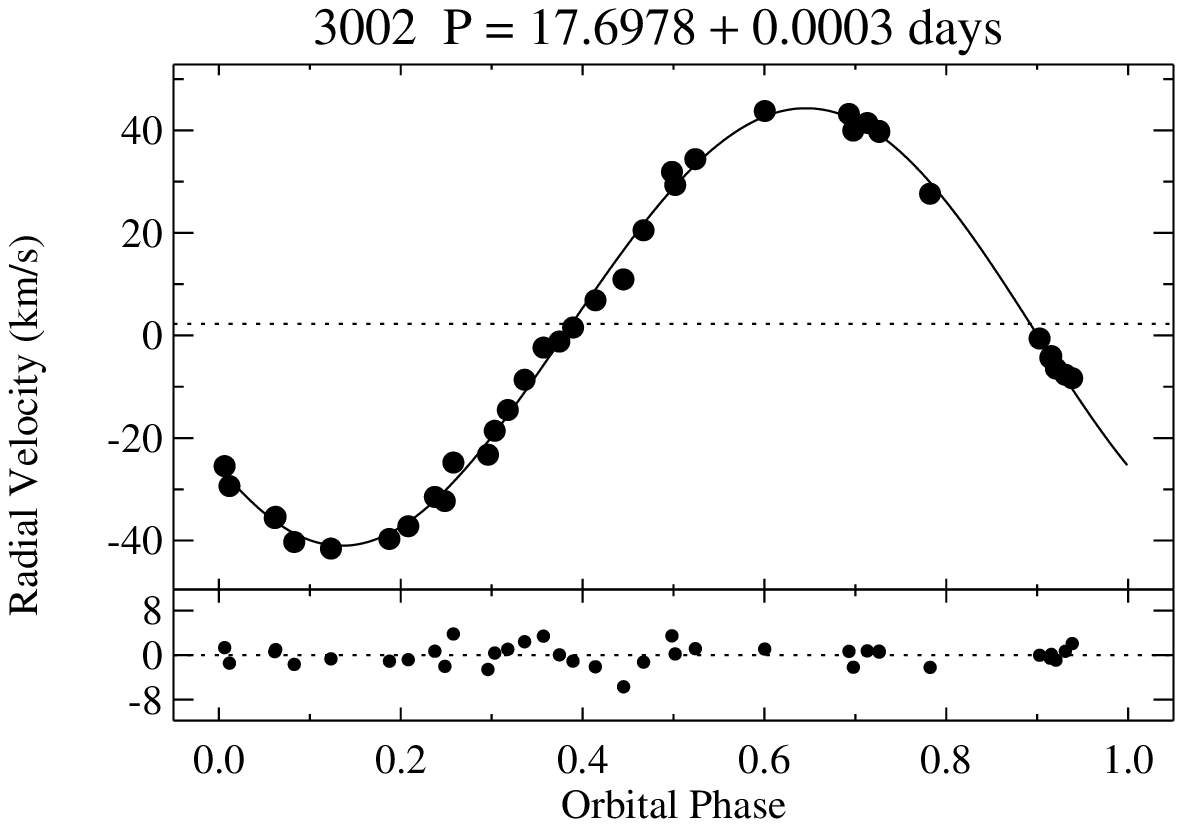}
\includegraphics[width=0.3\linewidth]{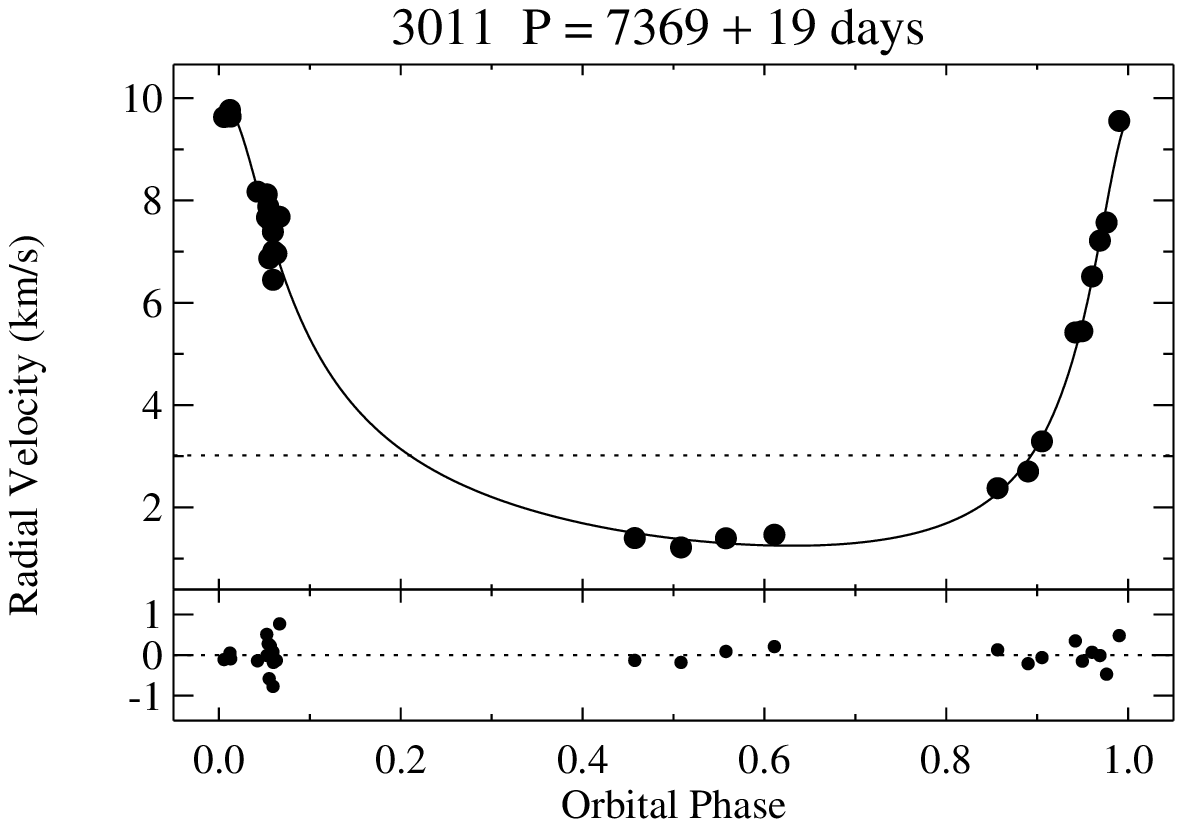}
\includegraphics[width=0.3\linewidth]{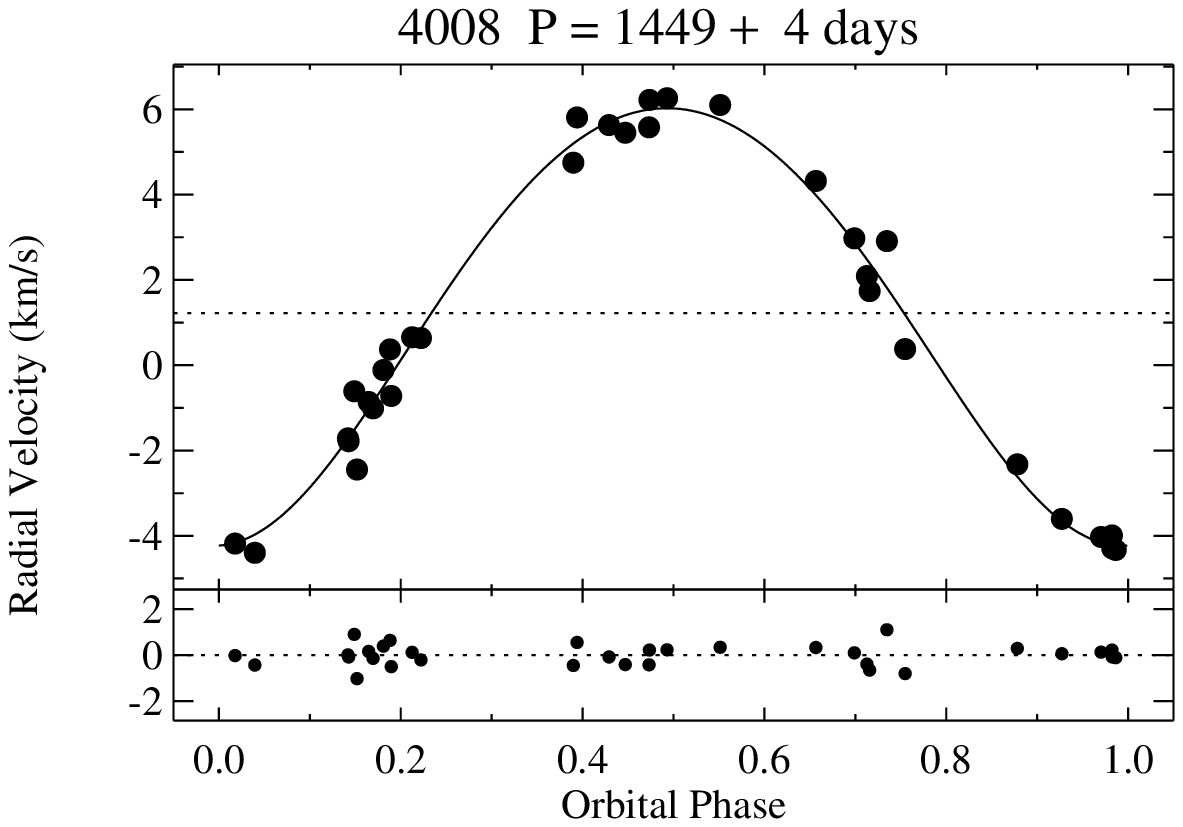}
\includegraphics[width=0.3\linewidth]{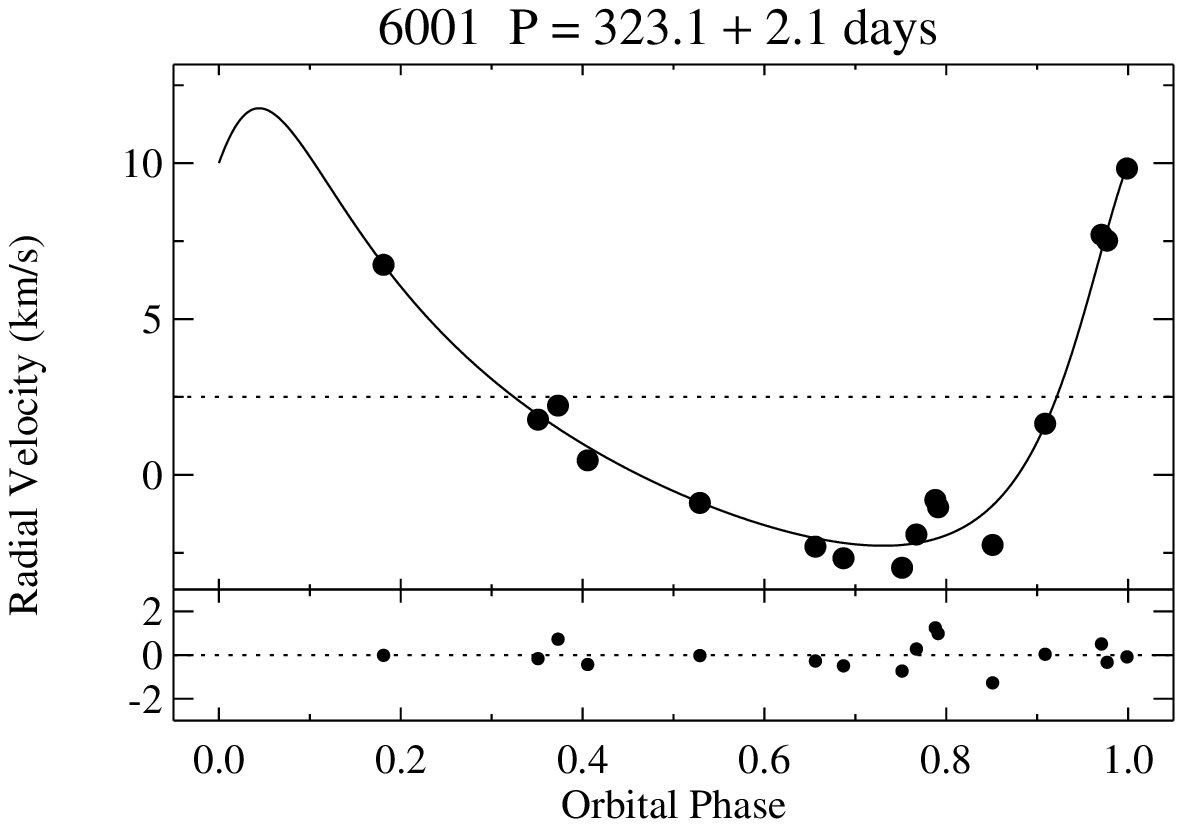}
\includegraphics[width=0.3\linewidth]{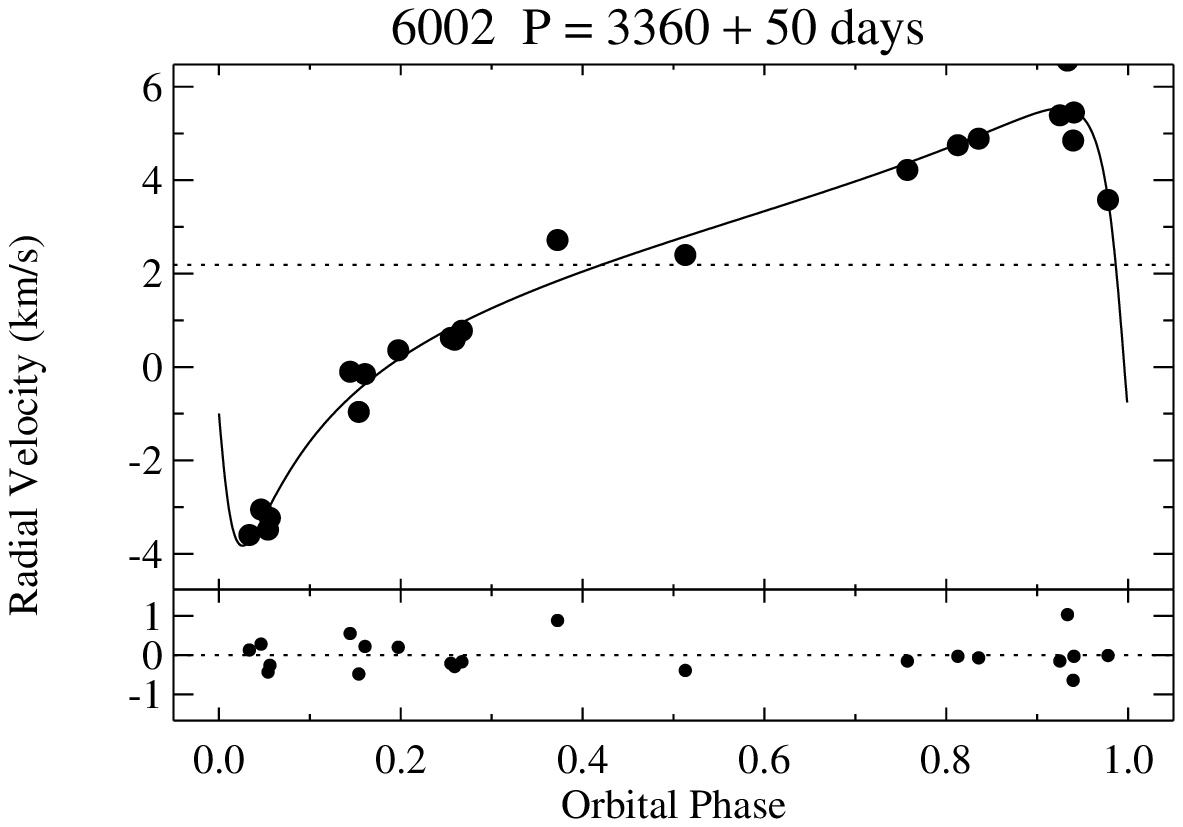}
\includegraphics[width=0.3\linewidth]{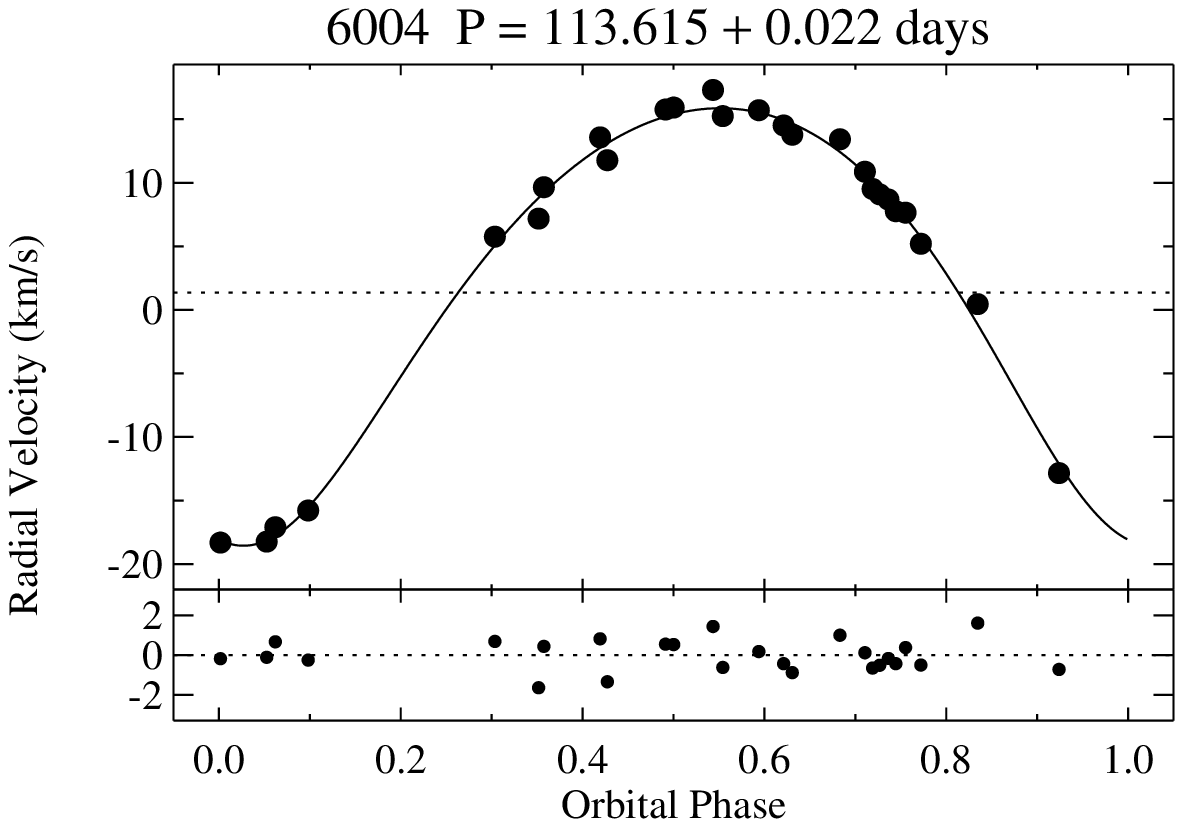}
\includegraphics[width=0.3\linewidth]{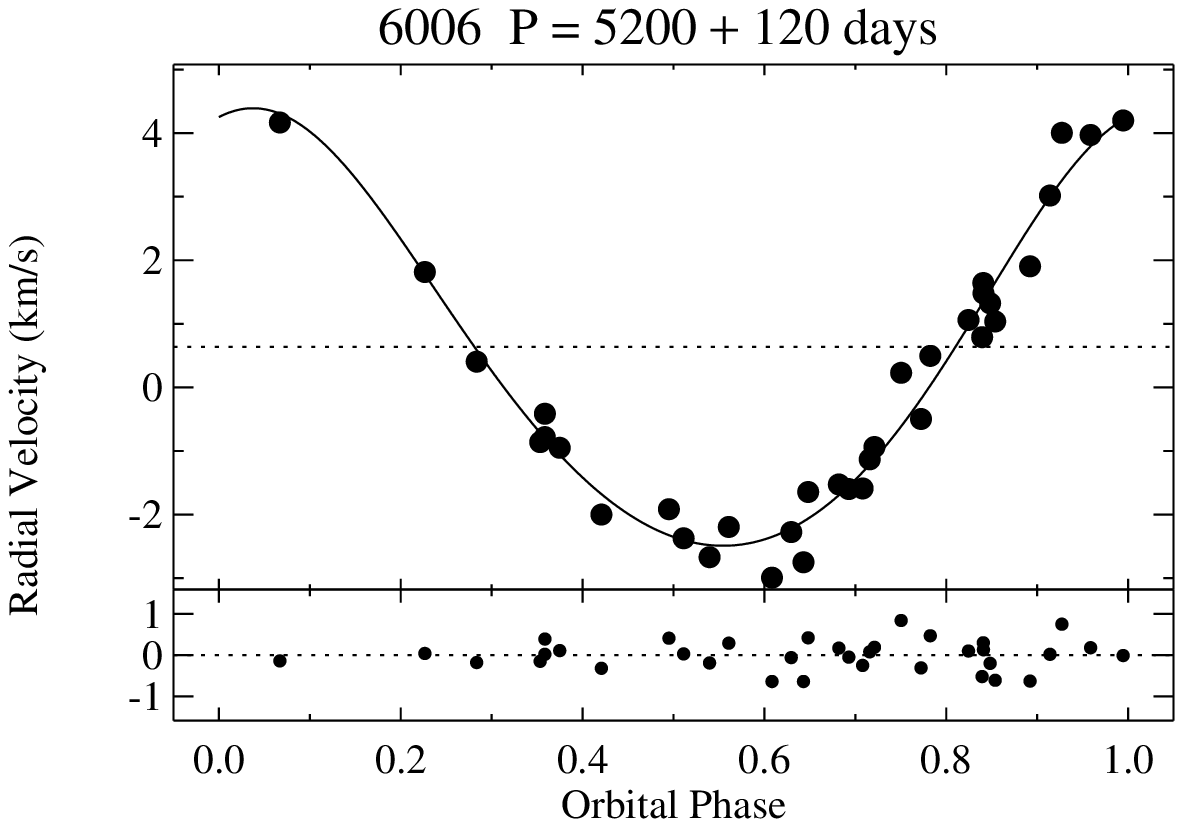}
\includegraphics[width=0.3\linewidth]{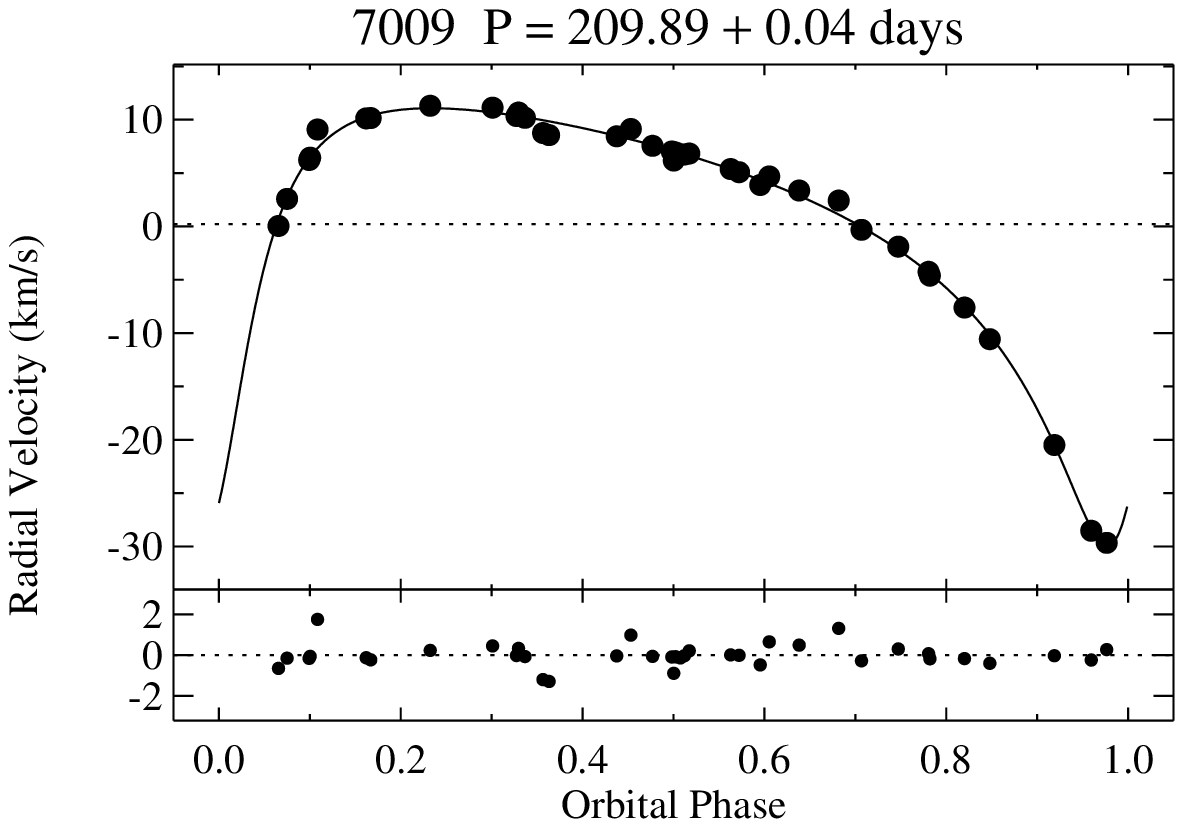}
\includegraphics[width=0.3\linewidth]{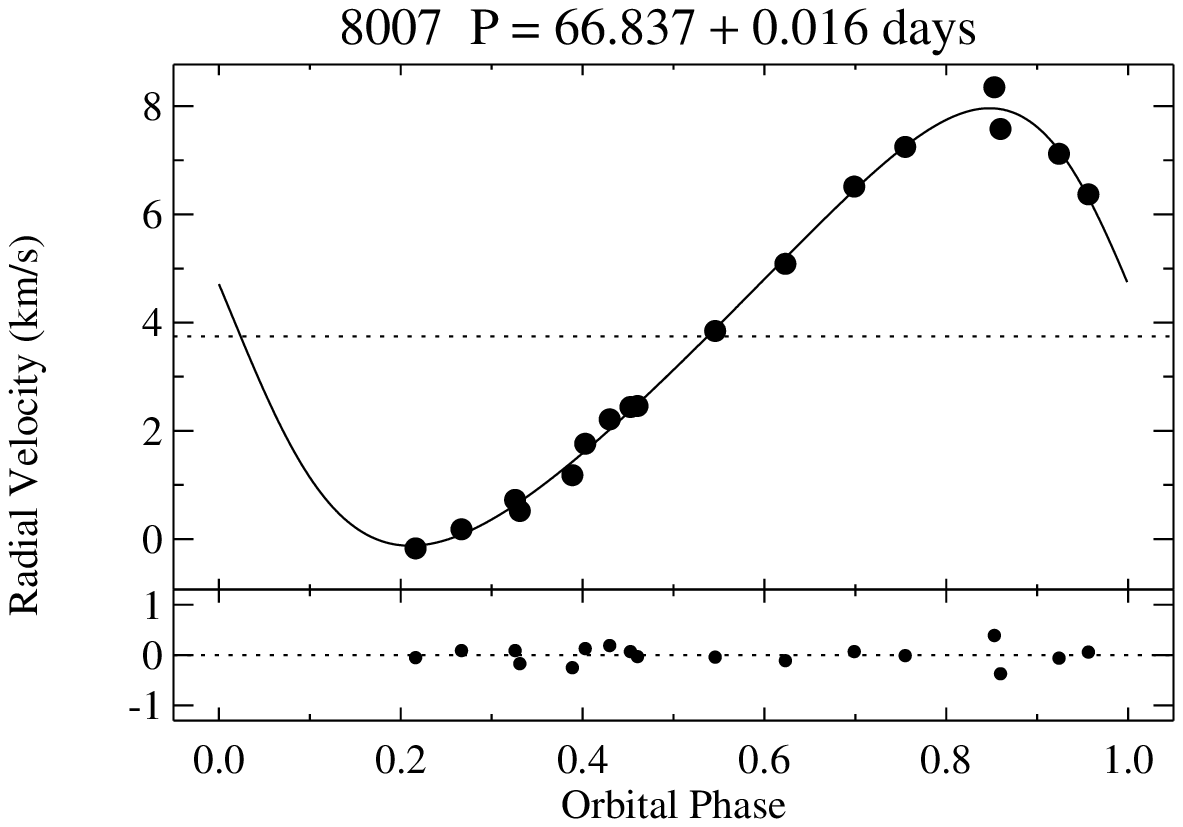}
\includegraphics[width=0.3\linewidth]{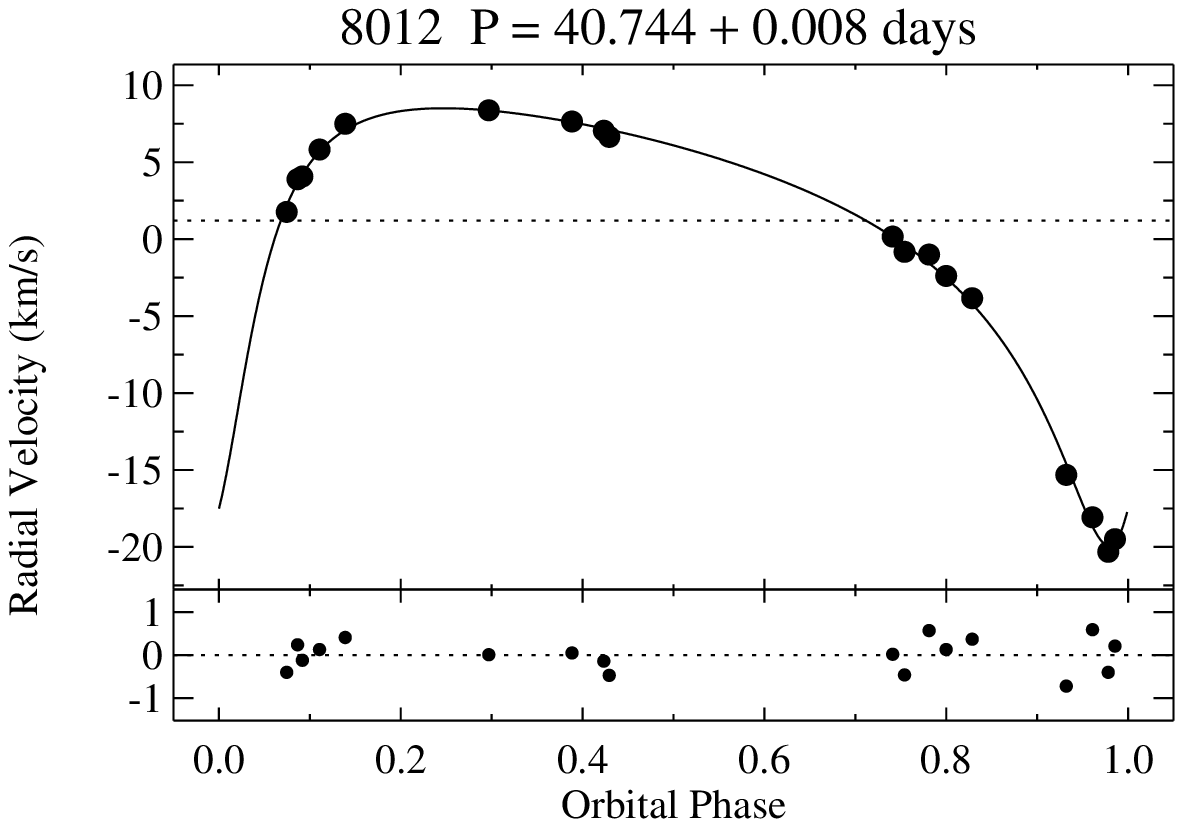}
\includegraphics[width=0.3\linewidth]{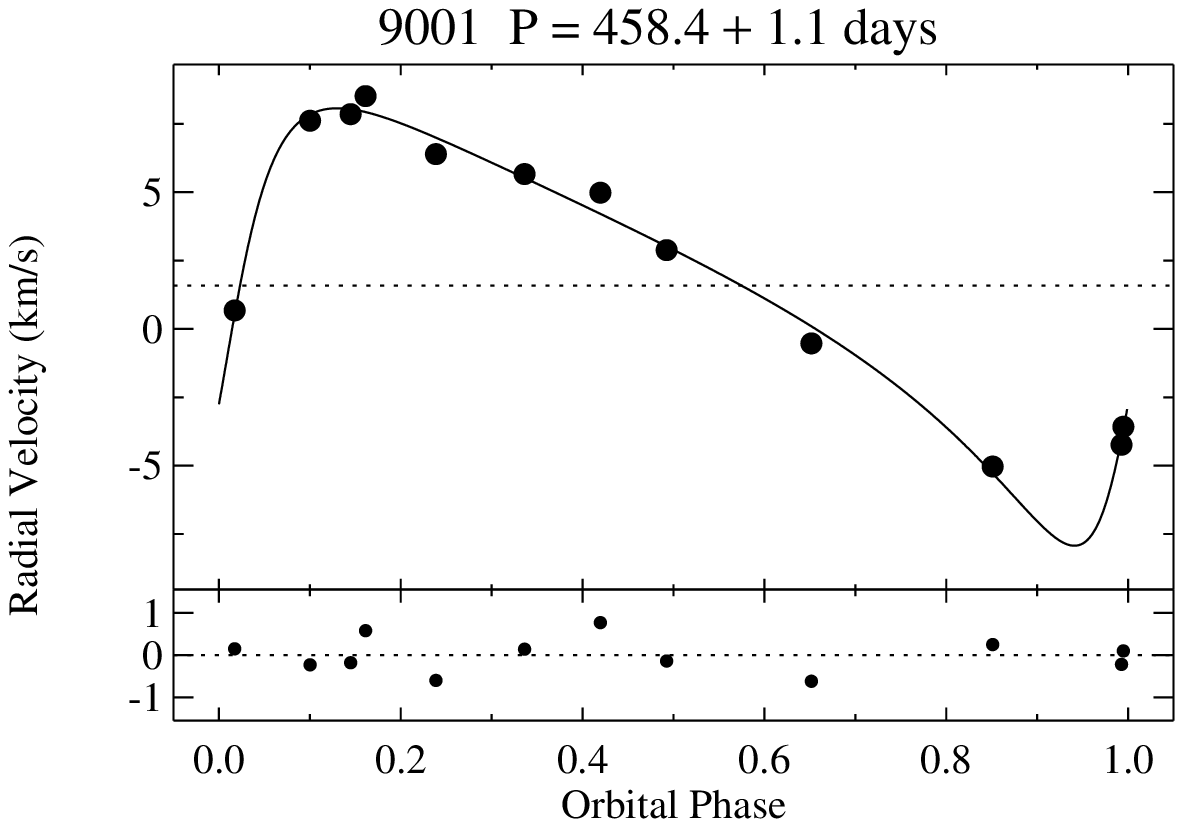}
\includegraphics[width=0.3\linewidth]{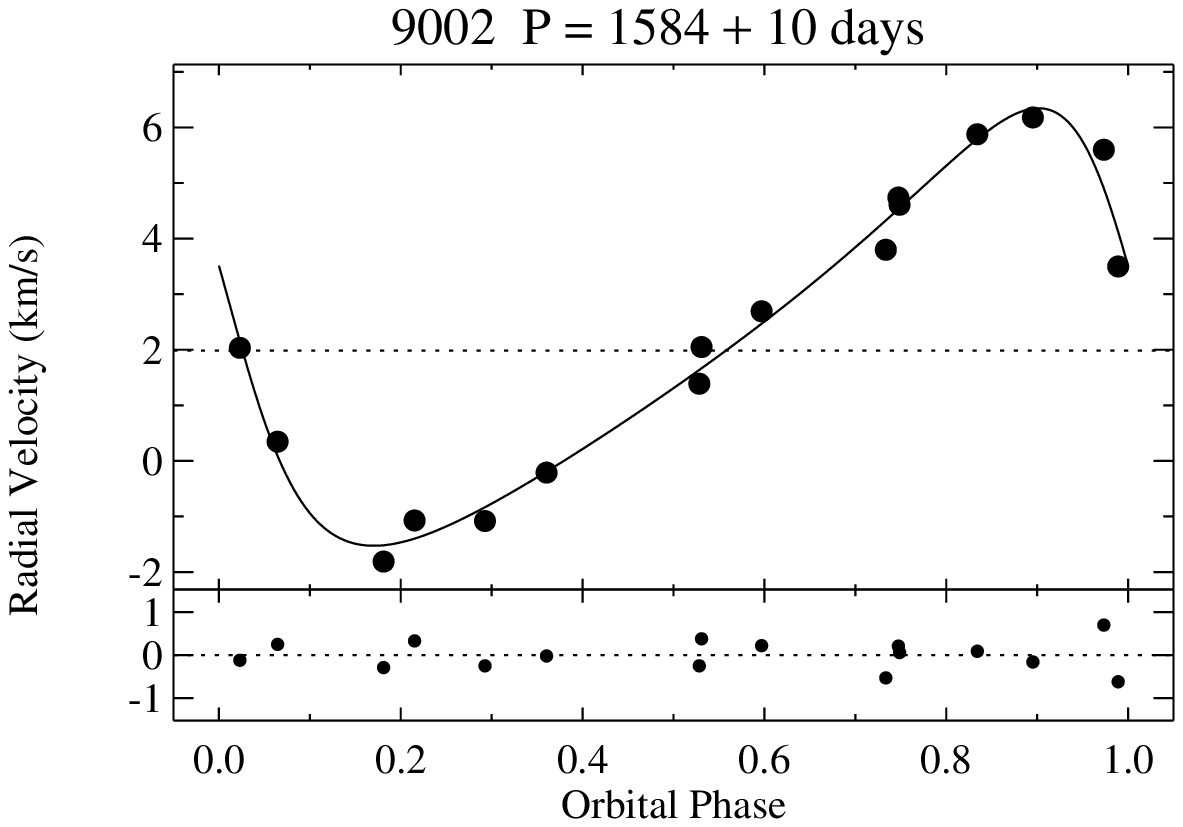}
\includegraphics[width=0.3\linewidth]{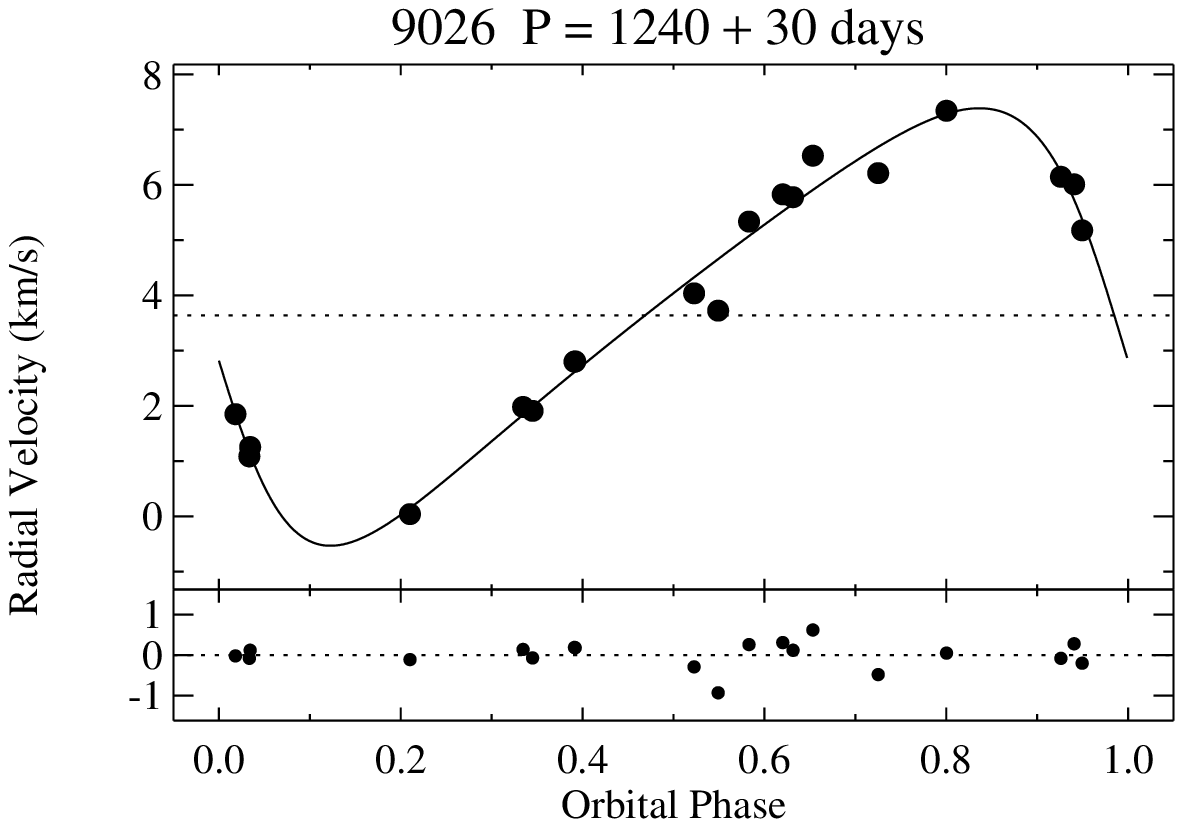}
\caption{\footnotesize NGC 6819 SB1 orbit plots.  For each binary, we plot RV against orbital phase, showing the data points with filled circles and the orbital fit to the data with the solid line. The dotted line marks the $\gamma$-velocity.  Beneath each orbital plot, we show the residuals from the fit.  Above each plot, we give the binary WOCS ID and orbital period.\normalsize}
\label{fig:Sb1plots}
\end{center}
\end{figure*}

For each single-lined spectroscopic binary we use the data given in Table 1 to solve for the orbital solution. In Figure~\ref{fig:Sb1plots} we have plotted the orbital solutions; for each binary the top panel contains the orbit solution and the bottom panel contains the residuals, in the sense of observed minus computed RVs. In Table~\ref{SB1tab} we provide the orbital elements for each binary in two rows: the first row contains the binary ID, the orbital period ($P$), the number of orbital cycles observed, the center-of-mass RV ($\gamma$), the orbital amplitude ($K$), the eccentricity ($e$), the longitude of periastron ($\omega$), a Julian Date of periastron passage ($T_{\circ}$), the projected semi-major axis ($a$ sin $i$), the mass function ($f(m)$), the rms residual velocity from the orbital solution ($\sigma$), and the number of RV measurements ($N$). The second row contains the respective 1$\sigma$ errors on each of these values where appropriate. 
\\
\\

\subsection{Double-lined Orbital Solutions}
\label{sec:SB2}
We used the TODCOR technique formulated by \cite{Zucker1994} to find the RVs for the primary and secondary stars of a given double-lined spectroscopic binary. TODCOR uses two template spectra to derive the two RVs of an SB2 binary simultaneously, greatly increasing our ability to recover reliable RVs even for observations that appear highly blended in a one-dimensional CCF.
We used the same solar template that we use to derive RVs for all single stars and SB1 binaries for both template spectra used in TODCOR. (We found that for most of the stars studied here, substituting a library of synthetic spectra that varied in temperature and gravities for the template spectra only has a minor impact on the RV output from TODCOR relative to our other measurement errors.)

We provide the orbital solutions in Figure~\ref{fig:Sb2plots}. The top panel of each orbital solution shows the measured RVs for the primary (filled circles) and secondary (open circles), fitted orbital solution for the primary (solid line) and secondary (dashed line), and $\gamma$ velocity (dotted line). The bottom panel in each orbital solution shows the observed minus computed values for the primary RVs (filled circles) and secondary RVs (open circles). 
The orbital elements are listed in Table~\ref{SB2tab}. The first row contains the binary ID, the orbital period ($P$), the number of orbital cycles observed, the center-of-mass RV ($\gamma$), the orbital amplitude of the primary ($K$), the eccentricity ($e$), the longitude of periastron ($\omega$), a Julian Date of periastron passage ($T_{\circ}$), the projected primary semi-major axis ($a$ sin $i$), $m$ sin$^{3}$ $i$, the mass ratio ($q$), the rms residual velocity for the primary from the orbital solution ($\sigma$), and the number of RV measurements ($N$). The second row contains the respective errors on each of these values where appropriate. The third and fourth rows are the amplitude ($K$), projected semi-major axis ($a$ sin $i$), $m$ sin$^{3}$ $i$, the rms residual velocity from the orbital solution ($\sigma$), and the number of RV measurements ($N$) for the secondary star and the respective errors.
\LongTables
\begin{deluxetable*}{l r c r r r r r r r c c}
\tabletypesize{\footnotesize}
\tablewidth{0pt}
\centering
\tablecolumns{12}
\tablecaption{Orbital Parameters For NGC 6819 Single-lined Binaries\label{SB1tab}}
\tablehead{\colhead{ID} & \colhead{$P$} & \colhead{Orbital} & \colhead{$\gamma$} & \colhead{$K$} & \colhead{$e$} & \colhead{$\omega$} & \colhead{$T_\circ$} & \colhead{$a \sin i$} & \colhead{$f(m)$} & \colhead{$\sigma$} & \colhead{$N$} \\
\colhead{} & \colhead{(days)} & \colhead{Cycles} & \colhead{(\kms)} & \colhead{(\kms)} & \colhead{} & \colhead{(deg)} & \colhead{(HJD-2400000 d)} & \colhead{(10$^6$ km)} & \colhead{(\Msolar)} & \colhead{(\kms)} & \colhead{}}
\startdata
    1001 &           653.4 &    6.1 &            1.43 &           21.49 &           0.226 &              88 &           48784 &           188.1 &         6.21$e-$1 &  0.65 &   23 \\
         &       $\pm$ 0.5 &       &      $\pm$ 0.17 &      $\pm$ 0.23 &     $\pm$ 0.011 &         $\pm$ 3 &         $\pm$ 4 &       $\pm$ 2.0 &    $\pm$ 2.0$e-$2 &       &      \\
    1006 &            1524 &    5.5 &            2.73 &            5.37 &            0.24 &             131 &           49770 &             109 &         2.25$e-$2 &  0.49 &   18 \\
         &         $\pm$ 5 &       &      $\pm$ 0.13 &      $\pm$ 0.18 &      $\pm$ 0.04 &        $\pm$ 10 &        $\pm$ 40 &         $\pm$ 4 &    $\pm$ 2.2$e-$3 &       &      \\
    1010 &            1144 &    7.7 &             1.1 &             7.2 &            0.55 &             107 &           53065 &              95 &          2.6$e-$2 &  1.56 &   32 \\
         &         $\pm$ 3 &       &       $\pm$ 0.4 &       $\pm$ 0.7 &      $\pm$ 0.04 &        $\pm$ 11 &        $\pm$ 22 &         $\pm$ 9 &    $\pm$ 0.7$e-$2 &       &      \\
    2008 &            2379 &    3.3 &            1.79 &            5.76 &            0.24 &             143 &           52040 &             183 &          4.3$e-$2 &  0.51 &   41 \\
         &         $\pm$ 9 &       &      $\pm$ 0.14 &      $\pm$ 0.16 &      $\pm$ 0.04 &         $\pm$ 5 &        $\pm$ 30 &         $\pm$ 5 &    $\pm$ 0.4$e-$2 &       &      \\
    2012 &            2920 &    1.8 &            2.64 &            6.01 &            0.39 &              19 &           51860 &             223 &          5.1$e-$2 &  0.41 &   19 \\
         &        $\pm$ 30 &       &      $\pm$ 0.16 &      $\pm$ 0.20 &      $\pm$ 0.03 &         $\pm$ 4 &        $\pm$ 30 &         $\pm$ 8 &    $\pm$ 0.5$e-$2 &       &      \\
    3002 &         17.6978 &  414.5 &             2.2 &            42.7 &           0.022 &             130 &         48961.8 &           10.38 &         1.42$e-$1 &  2.02 &   37 \\
         &    $\pm$ 0.0003 &       &       $\pm$ 0.3 &       $\pm$ 0.5 &     $\pm$ 0.012 &        $\pm$ 30 &       $\pm$ 1.4 &      $\pm$ 0.12 &    $\pm$ 0.5$e-$2 &       &      \\
    3011 &            7369 &    1.1 &            3.02 &            4.28 &           0.608 &             345 &           54980 &             344 &         2.99$e-$2 &  0.32 &   26 \\
         &        $\pm$ 19 &       &      $\pm$ 0.11 &      $\pm$ 0.11 &     $\pm$ 0.012 &         $\pm$ 4 &        $\pm$ 30 &        $\pm$ 10 &    $\pm$ 2.3$e-$3 &       &      \\
    4008 &            1449 &    4.5 &            1.22 &            5.13 &            0.06 &             180 &           50710 &             102 &         2.01$e-$2 &  0.50 &   33 \\
         &         $\pm$ 4 &       &      $\pm$ 0.09 &      $\pm$ 0.13 &      $\pm$ 0.03 &        $\pm$ 30 &       $\pm$ 130 &         $\pm$ 3 &    $\pm$ 1.5$e-$3 &       &      \\
    6001 &           323.1 &    6.0 &             2.5 &             7.0 &            0.43 &             319 &           55442 &              28 &          9.0$e-$3 &  0.79 &   16 \\
         &       $\pm$ 2.1 &       &       $\pm$ 0.3 &       $\pm$ 0.7 &      $\pm$ 0.05 &         $\pm$ 9 &         $\pm$ 7 &         $\pm$ 3 &    $\pm$ 0.3$e-$2 &       &      \\
    6002 &            3360 &    1.6 &            2.19 &             4.7 &            0.72 &             113 &           48249 &             151 &          1.2$e-$2 &  0.49 &   21 \\
         &        $\pm$ 50 &       &      $\pm$ 0.20 &       $\pm$ 0.6 &      $\pm$ 0.06 &         $\pm$ 4 &        $\pm$ 24 &        $\pm$ 23 &    $\pm$ 0.5$e-$2 &       &      \\
    6004 &         113.615 &   35.8 &            1.36 &            17.2 &           0.162 &             166 &         49453.3 &            26.5 &          5.8$e-$2 &  0.89 &   26 \\
         &     $\pm$ 0.022 &       &      $\pm$ 0.21 &       $\pm$ 0.3 &     $\pm$ 0.019 &         $\pm$ 6 &       $\pm$ 1.9 &       $\pm$ 0.4 &    $\pm$ 0.3$e-$2 &       &      \\
    6006 &            5200 &    1.6 &            0.64 &            3.44 &            0.10 &             344 &           56200 &             245 &          2.2$e-$2 &  0.40 &   35 \\
         &       $\pm$ 120 &       &      $\pm$ 0.12 &      $\pm$ 0.14 &      $\pm$ 0.04 &        $\pm$ 21 &       $\pm$ 300 &        $\pm$ 10 &    $\pm$ 0.3$e-$2 &       &      \\
    7009 &          209.89 &   23.1 &            0.22 &            20.5 &           0.585 &           216.5 &         49988.9 &            48.0 &         1.00$e-$1 &  0.61 &   38 \\
         &      $\pm$ 0.04 &       &      $\pm$ 0.11 &       $\pm$ 0.3 &     $\pm$ 0.007 &       $\pm$ 1.1 &       $\pm$ 0.6 &       $\pm$ 0.8 &    $\pm$ 0.5$e-$2 &       &      \\
    8007 &          66.837 &   82.6 &            3.74 &            4.04 &            0.22 &              79 &         51623.2 &            3.62 &          4.2$e-$4 &  0.21 &   17 \\
         &     $\pm$ 0.016 &       &      $\pm$ 0.10 &      $\pm$ 0.12 &      $\pm$ 0.03 &         $\pm$ 6 &       $\pm$ 1.2 &      $\pm$ 0.11 &    $\pm$ 0.4$e-$4 &       &      \\
    8012 &          40.744 &   50.0 &            1.21 &           14.21 &           0.586 &           213.9 &        52417.46 &            6.45 &          6.5$e-$3 &  0.45 &   18 \\
         &     $\pm$ 0.008 &       &      $\pm$ 0.12 &      $\pm$ 0.18 &     $\pm$ 0.009 &       $\pm$ 1.4 &      $\pm$ 0.13 &      $\pm$ 0.10 &    $\pm$ 0.3$e-$3 &       &      \\
    9001 &           458.4 &    4.1 &            1.58 &             8.0 &            0.54 &             249 &           55368 &              43 &          1.5$e-$2 &  0.57 &   12 \\
         &       $\pm$ 1.1 &       &      $\pm$ 0.20 &       $\pm$ 0.4 &      $\pm$ 0.03 &         $\pm$ 5 &         $\pm$ 3 &         $\pm$ 3 &    $\pm$ 0.3$e-$2 &       &      \\
    9002 &            1584 &    2.7 &            1.99 &            3.94 &            0.39 &              74 &           50031 &              79 &          7.9$e-$3 &  0.43 &   16 \\
         &        $\pm$ 10 &       &      $\pm$ 0.12 &      $\pm$ 0.24 &      $\pm$ 0.06 &         $\pm$ 7 &        $\pm$ 22 &         $\pm$ 5 &    $\pm$ 1.5$e-$3 &       &      \\
    9026 &            1240 &    1.8 &            3.64 &             4.0 &            0.35 &              99 &           51700 &              63 &          6.6$e-$3 &  0.39 &   19 \\
         &        $\pm$ 30 &       &      $\pm$ 0.12 &       $\pm$ 0.4 &      $\pm$ 0.06 &        $\pm$ 10 &        $\pm$ 30 &         $\pm$ 6 &    $\pm$ 1.8$e-$3 &       &      \\
   11006 &            7810 &    1.1 &            2.35 &             3.8 &            0.67 &             172 &           48470 &             310 &          1.9$e-$2 &  0.70 &   32 \\
         &        $\pm$ 60 &       &      $\pm$ 0.22 &       $\pm$ 0.3 &      $\pm$ 0.04 &         $\pm$ 7 &        $\pm$ 70 &        $\pm$ 30 &    $\pm$ 0.4$e-$2 &       &      \\
   13001 &         4.24079 &  871.0 &             2.7 &            55.1 &           0.047 &             205 &        54002.05 &            3.21 &          7.3$e-$2 &  2.43 &   16 \\
         &   $\pm$ 0.00004 &       &       $\pm$ 0.6 &       $\pm$ 1.0 &     $\pm$ 0.020 &        $\pm$ 20 &      $\pm$ 0.24 &      $\pm$ 0.06 &    $\pm$ 0.4$e-$2 &       &      \\
   13007 &             511 &    5.5 &            1.92 &             1.7 &            0.47 &             353 &           52927 &              10 &          1.7$e-$4 &  0.45 &   15 \\
         &         $\pm$ 3 &       &      $\pm$ 0.13 &       $\pm$ 0.4 &      $\pm$ 0.12 &        $\pm$ 21 &        $\pm$ 20 &         $\pm$ 3 &    $\pm$ 1.2$e-$4 &       &      \\
   14008 &        1.719530 &  860.0 &             2.6 &            45.3 &            0.02 &             300 &         52959.5 &            1.07 &         1.66$e-$2 &  2.93 &   14 \\
         &  $\pm$ 0.000024 &       &       $\pm$ 0.9 &       $\pm$ 1.3 &      $\pm$ 0.03 &        $\pm$ 80 &       $\pm$ 0.4 &      $\pm$ 0.03 &    $\pm$ 1.5$e-$3 &       &      \\
   14009 &         174.464 &   23.4 &            2.63 &            15.7 &           0.691 &              84 &         53620.3 &            27.2 &          2.6$e-$2 &  0.72 &   32 \\
         &     $\pm$ 0.020 &       &      $\pm$ 0.23 &       $\pm$ 0.8 &     $\pm$ 0.017 &         $\pm$ 4 &       $\pm$ 0.5 &       $\pm$ 1.6 &    $\pm$ 0.4$e-$2 &       &      \\
   14012 &             762 &    5.7 &             3.6 &             7.0 &            0.13 &             193 &           55610 &              73 &          2.7$e-$2 &  1.37 &   42 \\
         &         $\pm$ 4 &       &       $\pm$ 0.3 &       $\pm$ 0.5 &      $\pm$ 0.05 &        $\pm$ 24 &        $\pm$ 50 &         $\pm$ 5 &    $\pm$ 0.6$e-$2 &       &      \\
   16006 &           620.5 &    8.8 &            2.12 &            10.5 &           0.714 &           211.9 &         53378.9 &              63 &          2.5$e-$2 &  0.77 &   33 \\
         &       $\pm$ 0.6 &       &      $\pm$ 0.17 &       $\pm$ 0.4 &     $\pm$ 0.019 &       $\pm$ 2.2 &       $\pm$ 2.3 &         $\pm$ 3 &    $\pm$ 0.3$e-$2 &       &      \\
   16011 &         209.446 &    6.9 &             3.1 &              18 &            0.92 &             273 &         55232.0 &              20 &          7.7$e-$3 &  0.26 &   15 \\
         &     $\pm$ 0.019 &       &       $\pm$ 0.5 &        $\pm$ 10 &      $\pm$ 0.07 &        $\pm$ 21 &       $\pm$ 0.7 &        $\pm$ 14 &    $\pm$ 1.4$e-$2 &       &      \\
   17005 &           543.1 &    6.7 &            1.08 &            9.83 &           0.498 &           249.3 &           52892 &            63.7 &         3.49$e-$2 &  0.33 &   21 \\
         &       $\pm$ 0.7 &       &      $\pm$ 0.11 &      $\pm$ 0.17 &     $\pm$ 0.015 &       $\pm$ 1.5 &         $\pm$ 3 &       $\pm$ 1.3 &    $\pm$ 1.9$e-$3 &       &      \\
   17008 &            2230 &    1.3 &            2.99 &            3.54 &            0.16 &              97 &           54710 &             107 &          9.8$e-$3 &  0.46 &   19 \\
         &        $\pm$ 80 &       &      $\pm$ 0.15 &      $\pm$ 0.19 &      $\pm$ 0.07 &        $\pm$ 16 &        $\pm$ 90 &         $\pm$ 6 &    $\pm$ 1.6$e-$3 &       &      \\
   17025 &           135.0 &    8.8 &             0.7 &            10.3 &            0.28 &             300 &           52094 &            18.3 &         1.35$e-$2 &  0.68 &   16 \\
         &       $\pm$ 0.3 &       &       $\pm$ 0.4 &       $\pm$ 0.3 &      $\pm$ 0.03 &         $\pm$ 8 &         $\pm$ 4 &       $\pm$ 0.5 &    $\pm$ 1.2$e-$3 &       &      \\
   18019 &           142.3 &   13.3 &            1.84 &            4.78 &            0.32 &             304 &         52045.1 &             8.9 &         1.38$e-$3 &  0.34 &   19 \\
         &       $\pm$ 0.3 &       &      $\pm$ 0.12 &      $\pm$ 0.17 &      $\pm$ 0.03 &         $\pm$ 5 &       $\pm$ 1.8 &       $\pm$ 0.3 &    $\pm$ 1.5$e-$4 &       &      \\
   20002 &          141.74 &   30.4 &             2.8 &             7.1 &            0.52 &             197 &           52665 &            11.8 &          3.3$e-$3 &  1.00 &   17 \\
         &      $\pm$ 0.22 &       &       $\pm$ 0.4 &       $\pm$ 1.3 &      $\pm$ 0.09 &         $\pm$ 8 &         $\pm$ 4 &       $\pm$ 2.3 &    $\pm$ 1.9$e-$3 &       &      \\
   20003 &            3540 &    1.3 &             2.3 &            3.17 &            0.08 &              40 &           53300 &             154 &          1.1$e-$2 &  0.64 &   16 \\
         &       $\pm$ 140 &       &       $\pm$ 0.3 &      $\pm$ 0.25 &      $\pm$ 0.12 &        $\pm$ 80 &       $\pm$ 900 &        $\pm$ 12 &    $\pm$ 0.3$e-$2 &       &      \\
   20008 &            2540 &    1.6 &             3.3 &             7.9 &            0.37 &             283 &           55630 &             258 &         1.06$e-$1 &  1.26 &   15 \\
         &       $\pm$ 160 &       &       $\pm$ 0.5 &       $\pm$ 0.5 &      $\pm$ 0.07 &        $\pm$ 14 &        $\pm$ 80 &        $\pm$ 19 &    $\pm$ 2.2$e-$2 &       &      \\
   20010 &           51.00 &   33.1 &             3.1 &            15.8 &            0.53 &             135 &         53102.8 &             9.4 &          1.3$e-$2 &  1.50 &   15 \\
         &      $\pm$ 0.07 &       &       $\pm$ 0.5 &       $\pm$ 2.4 &      $\pm$ 0.06 &         $\pm$ 9 &       $\pm$ 1.2 &       $\pm$ 1.5 &    $\pm$ 0.6$e-$2 &       &      \\
   20013 &         121.570 &   15.1 &            2.23 &            15.6 &           0.454 &             255 &         52296.6 &            23.2 &         3.36$e-$2 &  0.26 &   15 \\
         &     $\pm$ 0.017 &       &      $\pm$ 0.20 &       $\pm$ 0.3 &     $\pm$ 0.008 &         $\pm$ 3 &       $\pm$ 0.6 &       $\pm$ 0.4 &    $\pm$ 1.7$e-$3 &       &      \\
   21007 &          23.921 &   49.6 &             2.3 &            16.2 &            0.40 &             248 &         52123.3 &            4.87 &          8.0$e-$3 &  0.88 &   15 \\
         &     $\pm$ 0.017 &       &       $\pm$ 0.5 &       $\pm$ 0.4 &      $\pm$ 0.03 &         $\pm$ 6 &       $\pm$ 0.4 &      $\pm$ 0.13 &    $\pm$ 0.6$e-$3 &       &      \\
   22006 &            2390 &    1.5 &            2.55 &             5.5 &            0.59 &             311 &           51960 &             147 &          2.2$e-$2 &  0.82 &   16 \\
         &        $\pm$ 40 &       &      $\pm$ 0.24 &       $\pm$ 0.4 &      $\pm$ 0.04 &         $\pm$ 9 &        $\pm$ 50 &        $\pm$ 12 &    $\pm$ 0.5$e-$2 &       &      \\
   22020 &           81.68 &   13.7 &            0.46 &            10.2 &            0.14 &             352 &           52125 &            11.4 &          8.8$e-$3 &  0.69 &   12 \\
         &      $\pm$ 0.13 &       &      $\pm$ 0.23 &       $\pm$ 0.3 &      $\pm$ 0.03 &        $\pm$ 14 &         $\pm$ 3 &       $\pm$ 0.3 &    $\pm$ 0.8$e-$3 &       &      \\
   23009 &             771 &    7.2 &            2.22 &            7.04 &           0.302 &             291 &           53411 &            71.1 &         2.41$e-$2 &  0.65 &   32 \\
         &         $\pm$ 3 &       &      $\pm$ 0.14 &      $\pm$ 0.20 &     $\pm$ 0.024 &         $\pm$ 5 &        $\pm$ 12 &       $\pm$ 2.1 &    $\pm$ 2.1$e-$3 &       &      \\
   24004 &           83.33 &   13.2 &            1.08 &            20.4 &           0.501 &             172 &         55281.1 &            20.2 &          4.8$e-$2 &  0.62 &   12 \\
         &      $\pm$ 0.11 &       &      $\pm$ 0.22 &       $\pm$ 0.4 &     $\pm$ 0.010 &         $\pm$ 3 &       $\pm$ 0.5 &       $\pm$ 0.4 &    $\pm$ 0.3$e-$2 &       &      \\
   24012 &         25.5266 &  114.6 &            1.69 &            19.1 &           0.004 &             120 &           52137 &            6.69 &         1.83$e-$2 &  0.53 &   18 \\
         &    $\pm$ 0.0013 &       &      $\pm$ 0.15 &       $\pm$ 0.3 &     $\pm$ 0.011 &       $\pm$ 170 &        $\pm$ 12 &      $\pm$ 0.09 &    $\pm$ 0.7$e-$3 &       &      \\
   24031 &           91.20 &   16.2 &            2.15 &            16.9 &           0.323 &              14 &         54884.0 &            20.1 &         3.90$e-$2 &  0.69 &   18 \\
         &      $\pm$ 0.04 &       &      $\pm$ 0.19 &       $\pm$ 0.3 &     $\pm$ 0.016 &         $\pm$ 3 &       $\pm$ 0.7 &       $\pm$ 0.4 &    $\pm$ 2.4$e-$3 &       &      \\
   25004 &          21.284 &  138.3 &             1.2 &           17.53 &           0.417 &           297.4 &        52029.50 &            4.66 &          8.9$e-$3 &  0.48 &   21 \\
         &     $\pm$ 0.003 &       &       $\pm$ 0.3 &      $\pm$ 0.23 &     $\pm$ 0.016 &       $\pm$ 2.1 &      $\pm$ 0.21 &      $\pm$ 0.07 &    $\pm$ 0.4$e-$3 &       &      \\
   25011 &        1.574695 & 1947.6 &            1.47 &            31.3 &           0.009 &             140 &         54467.8 &           0.677 &         4.99$e-$3 &  1.36 &   33 \\
         &  $\pm$ 0.000005 &       &      $\pm$ 0.24 &       $\pm$ 0.4 &     $\pm$ 0.010 &        $\pm$ 90 &       $\pm$ 0.4 &     $\pm$ 0.008 &    $\pm$ 1.7$e-$4 &       &      \\
   26007 &          46.693 &   87.7 &            2.91 &            24.9 &           0.277 &             297 &         53070.2 &           15.35 &          6.6$e-$2 &  0.74 &   13 \\
         &     $\pm$ 0.007 &       &      $\pm$ 0.24 &       $\pm$ 0.4 &     $\pm$ 0.011 &         $\pm$ 4 &       $\pm$ 0.5 &      $\pm$ 0.23 &    $\pm$ 0.3$e-$2 &       &      \\
   27011 &            1340 &    2.8 &             2.0 &             4.4 &            0.77 &              83 &           52240 &              52 &          3.0$e-$3 &  0.67 &   20 \\
         &        $\pm$ 30 &       &       $\pm$ 0.3 &       $\pm$ 1.1 &      $\pm$ 0.11 &        $\pm$ 13 &        $\pm$ 30 &        $\pm$ 17 &    $\pm$ 0.3$e-$2 &       &      \\
   27020 &             583 &    4.2 &            2.44 &             5.3 &            0.26 &             268 &           52267 &            41.1 &          8.2$e-$3 &  0.52 &   16 \\
         &         $\pm$ 3 &       &      $\pm$ 0.15 &       $\pm$ 0.3 &      $\pm$ 0.05 &         $\pm$ 9 &        $\pm$ 13 &       $\pm$ 2.1 &    $\pm$ 1.2$e-$3 &       &      \\
   28004 &             414 &    3.7 &             2.2 &            13.0 &            0.71 &             119 &           55639 &              52 &          3.4$e-$2 &  0.61 &   13 \\
         &         $\pm$ 3 &       &       $\pm$ 0.4 &       $\pm$ 1.8 &      $\pm$ 0.08 &         $\pm$ 3 &         $\pm$ 4 &         $\pm$ 9 &    $\pm$ 1.5$e-$2 &       &      \\
   29021 &           370.3 &    4.6 &             1.2 &            12.8 &            0.11 &             118 &           54967 &              65 &          7.9$e-$2 &  0.92 &   16 \\
         &       $\pm$ 2.1 &       &       $\pm$ 0.4 &       $\pm$ 0.7 &      $\pm$ 0.04 &        $\pm$ 16 &        $\pm$ 16 &         $\pm$ 3 &    $\pm$ 1.3$e-$2 &       &      \\
   30008 &           93.16 &   41.8 &            1.14 &            4.43 &            0.46 &             297 &           55174 &             5.0 &          5.9$e-$4 &  0.44 &   14 \\
         &      $\pm$ 0.13 &       &      $\pm$ 0.20 &      $\pm$ 0.23 &      $\pm$ 0.04 &         $\pm$ 8 &         $\pm$ 3 &       $\pm$ 0.3 &    $\pm$ 1.0$e-$4 &       &      \\
   31004 &          22.857 &   84.3 &             2.8 &              14 &            0.39 &             307 &         55742.2 &             4.0 &          5.0$e-$3 &  0.42 &   14 \\
         &     $\pm$ 0.003 &       &       $\pm$ 1.3 &         $\pm$ 3 &      $\pm$ 0.07 &         $\pm$ 8 &       $\pm$ 1.1 &       $\pm$ 0.8 &    $\pm$ 0.3$e-$2 &       &      \\
   32006 &         1.84716 & 2623.5 &             2.2 &            18.2 &            0.05 &             330 &        52305.65 &           0.462 &         1.15$e-$3 &  1.08 &   20 \\
         &   $\pm$ 0.00004 &       &       $\pm$ 0.3 &       $\pm$ 0.3 &      $\pm$ 0.03 &        $\pm$ 30 &      $\pm$ 0.14 &     $\pm$ 0.009 &    $\pm$ 0.7$e-$4 &       &      \\
   33002 &         10.9025 &  232.3 &             3.1 &            40.5 &           0.022 &              10 &         54194.1 &            6.07 &          7.5$e-$2 &  1.56 &   18 \\
         &    $\pm$ 0.0006 &       &       $\pm$ 0.5 &       $\pm$ 0.5 &     $\pm$ 0.016 &        $\pm$ 30 &       $\pm$ 1.0 &      $\pm$ 0.08 &    $\pm$ 0.3$e-$2 &       &      \\
   33005 &             839 &    4.4 &             2.8 &            20.3 &            0.26 &             165 &           51882 &             226 &          6.6$e-$1 &  1.73 &   19 \\
         &         $\pm$ 5 &       &       $\pm$ 0.5 &       $\pm$ 0.8 &      $\pm$ 0.03 &         $\pm$ 8 &        $\pm$ 19 &        $\pm$ 10 &    $\pm$ 0.8$e-$1 &       &      \\
   34006 &           60.48 &   20.2 &             1.3 &            14.2 &            0.24 &             226 &         55353.8 &            11.4 &         1.63$e-$2 &  0.89 &   15 \\
         &      $\pm$ 0.04 &       &       $\pm$ 0.3 &       $\pm$ 0.4 &      $\pm$ 0.03 &         $\pm$ 7 &       $\pm$ 1.2 &       $\pm$ 0.3 &    $\pm$ 1.4$e-$3 &       &      \\
   35020 &          24.737 &   45.4 &            1.40 &            12.4 &            0.36 &             136 &        52042.16 &            3.93 &          4.0$e-$3 &  0.58 &   15 \\
         &     $\pm$ 0.005 &       &      $\pm$ 0.17 &       $\pm$ 0.5 &      $\pm$ 0.03 &         $\pm$ 4 &      $\pm$ 0.19 &      $\pm$ 0.15 &    $\pm$ 0.4$e-$3 &       &      \\
   35025 &         4.64123 &  199.4 &            2.45 &            39.2 &           0.208 &           117.2 &        51999.66 &            2.45 &         2.71$e-$2 &  0.74 &   17 \\
         &   $\pm$ 0.00013 &       &      $\pm$ 0.22 &       $\pm$ 0.5 &     $\pm$ 0.012 &       $\pm$ 2.4 &      $\pm$ 0.03 &      $\pm$ 0.03 &    $\pm$ 1.0$e-$3 &       &      \\
   36012 &           176.4 &   20.6 &             3.3 &             5.8 &            0.65 &             299 &           54410 &            10.6 &          1.5$e-$3 &  0.69 &   23 \\
         &       $\pm$ 0.6 &       &       $\pm$ 0.4 &       $\pm$ 0.7 &      $\pm$ 0.04 &         $\pm$ 8 &         $\pm$ 5 &       $\pm$ 1.4 &    $\pm$ 0.6$e-$3 &       &      \\
   36043 &           71.70 &   13.3 &             3.3 &            14.2 &            0.09 &             348 &           56244 &           13.97 &         2.11$e-$2 &  0.47 &   12 \\
         &      $\pm$ 0.04 &       &       $\pm$ 0.4 &       $\pm$ 0.3 &      $\pm$ 0.03 &        $\pm$ 15 &         $\pm$ 3 &      $\pm$ 0.25 &    $\pm$ 1.1$e-$3 &       &      \\
   37028 &           288.5 &   11.5 &            1.32 &            4.05 &            0.31 &             250 &           54638 &            15.3 &          1.7$e-$3 &  0.56 &   19 \\
         &       $\pm$ 1.7 &       &      $\pm$ 0.17 &      $\pm$ 0.21 &      $\pm$ 0.06 &        $\pm$ 12 &         $\pm$ 9 &       $\pm$ 0.8 &    $\pm$ 0.3$e-$3 &       &      \\
   39013 &          36.708 &   99.2 &             2.7 &            17.1 &            0.06 &             240 &           54427 &             8.6 &          1.9$e-$2 &  2.00 &   18 \\
         &     $\pm$ 0.010 &       &       $\pm$ 0.6 &       $\pm$ 0.8 &      $\pm$ 0.05 &        $\pm$ 40 &         $\pm$ 4 &       $\pm$ 0.4 &    $\pm$ 0.3$e-$2 &       &      \\
   40006 &            2070 &    1.8 &            2.09 &             4.4 &            0.38 &             111 &           51560 &             116 &          1.5$e-$2 &  0.81 &   17 \\
         &        $\pm$ 60 &       &      $\pm$ 0.22 &       $\pm$ 0.4 &      $\pm$ 0.08 &        $\pm$ 13 &        $\pm$ 90 &        $\pm$ 10 &    $\pm$ 0.4$e-$2 &       &      \\
   40017 &           370.5 &   11.7 &             2.5 &             6.1 &            0.16 &             332 &           53502 &            30.5 &          8.3$e-$3 &  0.85 &   24 \\
         &       $\pm$ 0.7 &       &       $\pm$ 0.3 &       $\pm$ 0.4 &      $\pm$ 0.05 &        $\pm$ 20 &        $\pm$ 20 &       $\pm$ 2.0 &    $\pm$ 1.6$e-$3 &       &      \\
   46013 &         14.2111 &  105.4 &             2.3 &            10.8 &            0.53 &             256 &        51878.89 &            1.79 &         1.13$e-$3 &  1.41 &   26 \\
         &    $\pm$ 0.0015 &       &       $\pm$ 0.3 &       $\pm$ 0.5 &      $\pm$ 0.04 &         $\pm$ 5 &      $\pm$ 0.11 &      $\pm$ 0.10 &    $\pm$ 1.8$e-$4 &       &      \\
   49002 &        1.615916 & 1573.2 &             3.8 &            53.7 &           0.022 &             240 &         53897.1 &            1.19 &         2.59$e-$2 &  3.18 &   20 \\
         &  $\pm$ 0.000010 &       &       $\pm$ 0.8 &       $\pm$ 1.2 &     $\pm$ 0.020 &        $\pm$ 60 &       $\pm$ 0.3 &      $\pm$ 0.03 &    $\pm$ 1.7$e-$3 &       &      \\
   49004 &             646 &    7.9 &             0.5 &              16 &            0.68 &             129 &           53207 &             100 &         1.09$e-$1 &  0.75 &   23 \\
         &         $\pm$ 3 &       &       $\pm$ 1.0 &         $\pm$ 7 &      $\pm$ 0.12 &        $\pm$ 11 &         $\pm$ 5 &        $\pm$ 50 &   $\pm$  1.1$e-$2 &       &      \\
   52015 &          247.34 &   12.2 &            1.36 &           14.03 &           0.053 &             326 &           53291 &            47.6 &          7.1$e-$2 &  0.69 &   18 \\
         &      $\pm$ 0.18 &       &      $\pm$ 0.21 &      $\pm$ 0.25 &     $\pm$ 0.018 &        $\pm$ 21 &        $\pm$ 15 &       $\pm$ 0.8 &    $\pm$ 0.4$e-$2 &       &      \\
   53003 &           42.54 &   23.7 &             1.9 &             5.4 &            0.39 &              77 &         55107.7 &             2.9 &          5.0$e-$4 &  0.73 &   16 \\
         &      $\pm$ 0.10 &       &       $\pm$ 0.5 &       $\pm$ 1.0 &      $\pm$ 0.14 &        $\pm$ 14 &       $\pm$ 2.0 &       $\pm$ 0.6 &    $\pm$ 0.3$e-$3 &       &      \\
   54027 &          73.027 &   31.1 &            3.09 &            21.2 &            0.55 &            41.3 &         55589.2 &            17.9 &         4.25$e-$2 &  0.49 &   13 \\
         &     $\pm$ 0.019 &       &      $\pm$ 0.22 &       $\pm$ 0.3 &      $\pm$ 0.03 &       $\pm$ 2.2 &       $\pm$ 0.7 &       $\pm$ 0.4 &    $\pm$ 2.2$e-$3 &       &      \\
   56010 &            2600 &    1.5 &             2.6 &               9 &            0.23 &              90 &           54510 &             320 &          2.0$e-$1 &  0.77 &   12 \\
         &        $\pm$ 40 &       &       $\pm$ 1.9 &         $\pm$ 3 &      $\pm$ 0.19 &        $\pm$ 30 &       $\pm$ 150 &       $\pm$ 110 &    $\pm$ 1.9$e-$1 &       &      \\
   57004 &          9.1537 &  128.4 &            3.40 &           40.03 &           0.123 &            59.3 &        56052.14 &            5.00 &         5.95$e-$2 &  0.47 &   12 \\
         &    $\pm$ 0.0004 &       &      $\pm$ 0.18 &      $\pm$ 0.22 &     $\pm$ 0.006 &       $\pm$ 2.3 &      $\pm$ 0.06 &      $\pm$ 0.03 &    $\pm$ 1.0$e-$3 &       &      \\
   59003 &          21.368 &   99.7 &             2.5 &            31.6 &           0.275 &           187.1 &        52336.13 &            8.93 &         6.22$e-$2 &  0.80 &   17 \\
         &     $\pm$ 0.003 &       &       $\pm$ 0.3 &       $\pm$ 0.4 &     $\pm$ 0.016 &       $\pm$ 2.3 &      $\pm$ 0.16 &      $\pm$ 0.11 &    $\pm$ 2.2$e-$3 &       &      \\
   60006 &          7.8431 &  153.2 &             1.9 &           33.14 &           0.019 &             319 &         51728.8 &           3.574 &         2.96$e-$2 &  0.66 &   18 \\
         &    $\pm$ 0.0003 &       &       $\pm$ 0.3 &      $\pm$ 0.22 &     $\pm$ 0.011 &        $\pm$ 21 &       $\pm$ 0.5 &     $\pm$ 0.024 &    $\pm$ 0.6$e-$3 &       &      \\
   62010 &            1332 &    3.3 &            2.12 &             7.0 &            0.70 &             338 &           54382 &              91 &         1.72$e-$2 &  0.79 &   23 \\
         &         $\pm$ 7 &       &      $\pm$ 0.20 &       $\pm$ 0.3 &      $\pm$ 0.03 &         $\pm$ 5 &        $\pm$ 17 &         $\pm$ 6 &    $\pm$ 2.5$e-$3 &       &      \\
   66004 &        2.277936 & 1324.8 &             2.9 &            38.6 &           0.016 &             150 &         53178.6 &           1.208 &         1.35$e-$2 &  1.80 &   17 \\
         &  $\pm$ 0.000015 &       &       $\pm$ 0.5 &       $\pm$ 0.7 &     $\pm$ 0.017 &        $\pm$ 70 &       $\pm$ 0.5 &     $\pm$ 0.023 &    $\pm$ 0.8$e-$3 &       &      \\
   66016 &          59.383 &   49.9 &             2.2 &            19.8 &           0.190 &             329 &         54456.9 &            15.9 &         4.51$e-$2 &  0.87 &   16 \\
         &     $\pm$ 0.018 &       &       $\pm$ 0.3 &       $\pm$ 0.3 &     $\pm$ 0.023 &         $\pm$ 5 &       $\pm$ 0.8 &       $\pm$ 0.3 &    $\pm$ 2.2$e-$3 &       &      \\
   69005 &             640 &    4.9 &             3.9 &              15 &            0.64 &              15 &           54836 &             100 &          1.1$e-$1 &  0.77 &   18 \\
         &         $\pm$ 5 &       &       $\pm$ 0.9 &         $\pm$ 6 &      $\pm$ 0.09 &         $\pm$ 3 &         $\pm$ 4 &        $\pm$ 40 &    $\pm$ 1.2$e-$1 &       &      \\
   77007 &            1560 &    2.8 &            1.56 &             3.9 &            0.48 &             298 &           53500 &              74 &          6.6$e-$3 &  0.83 &   21 \\
         &        $\pm$ 30 &       &      $\pm$ 0.22 &       $\pm$ 0.4 &      $\pm$ 0.10 &        $\pm$ 11 &        $\pm$ 50 &         $\pm$ 9 &    $\pm$ 2.2$e-$3 &       &      
\enddata
\end{deluxetable*}

\begin{figure*}
\begin{center}
\includegraphics[width=0.3\linewidth]{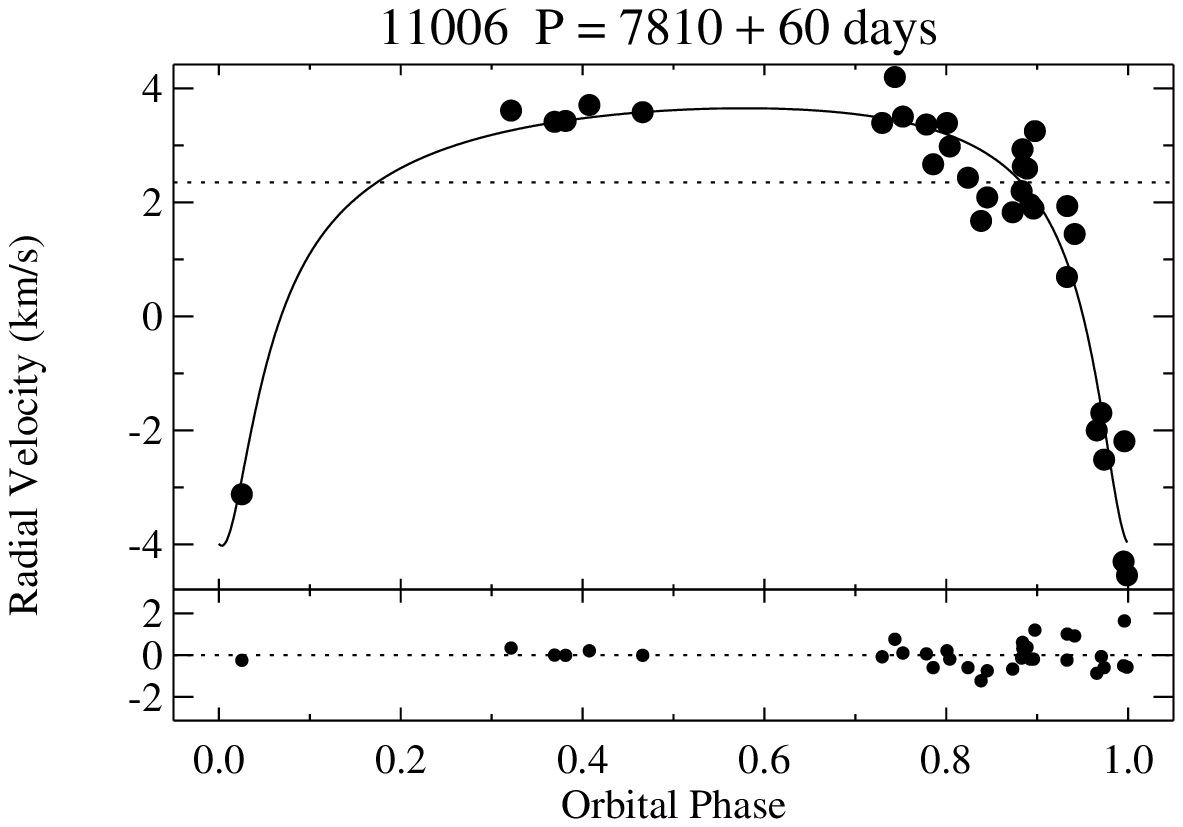}
\includegraphics[width=0.3\linewidth]{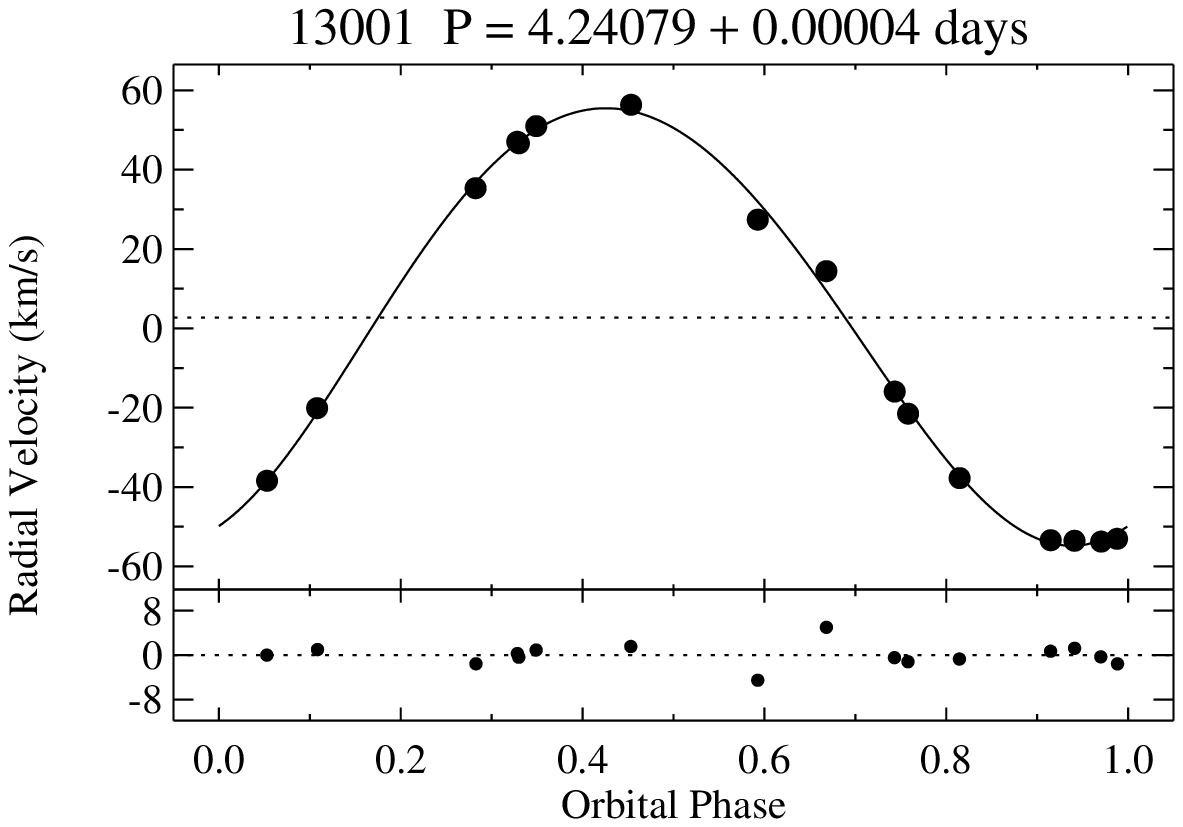}
\includegraphics[width=0.3\linewidth]{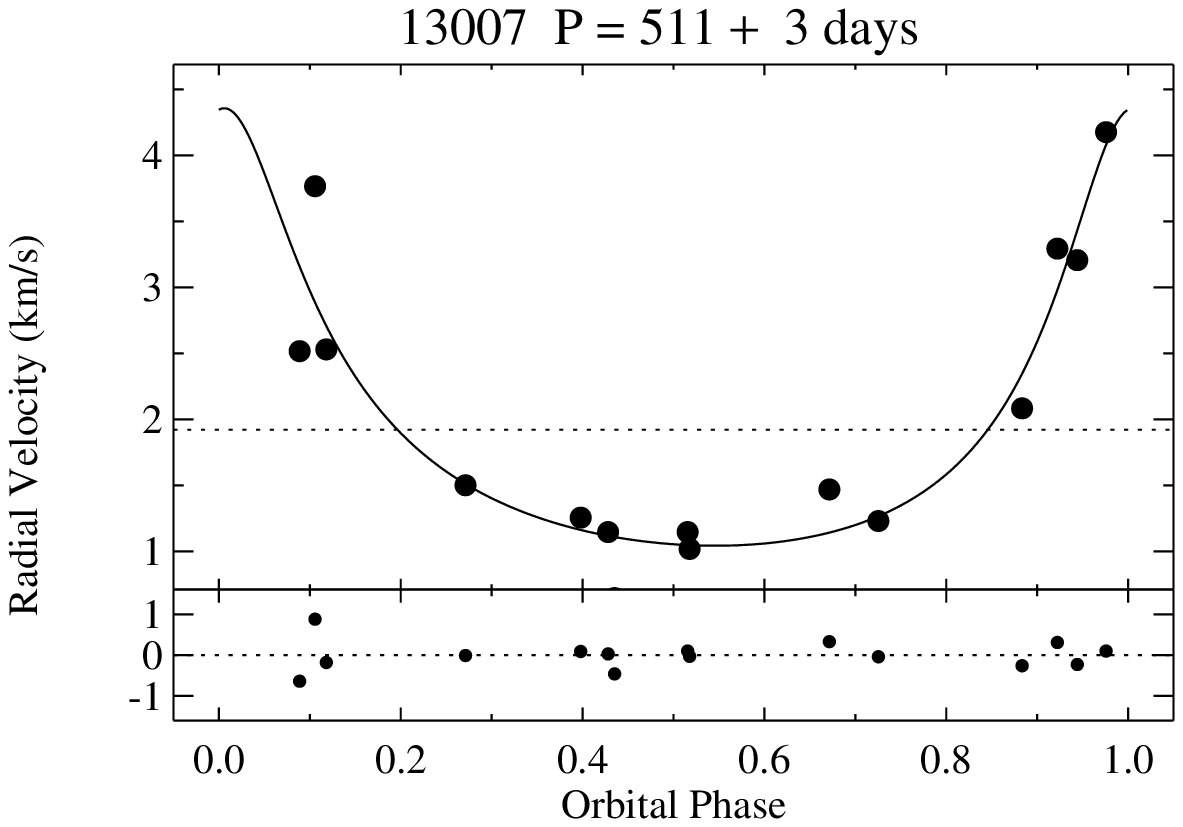}
\includegraphics[width=0.3\linewidth]{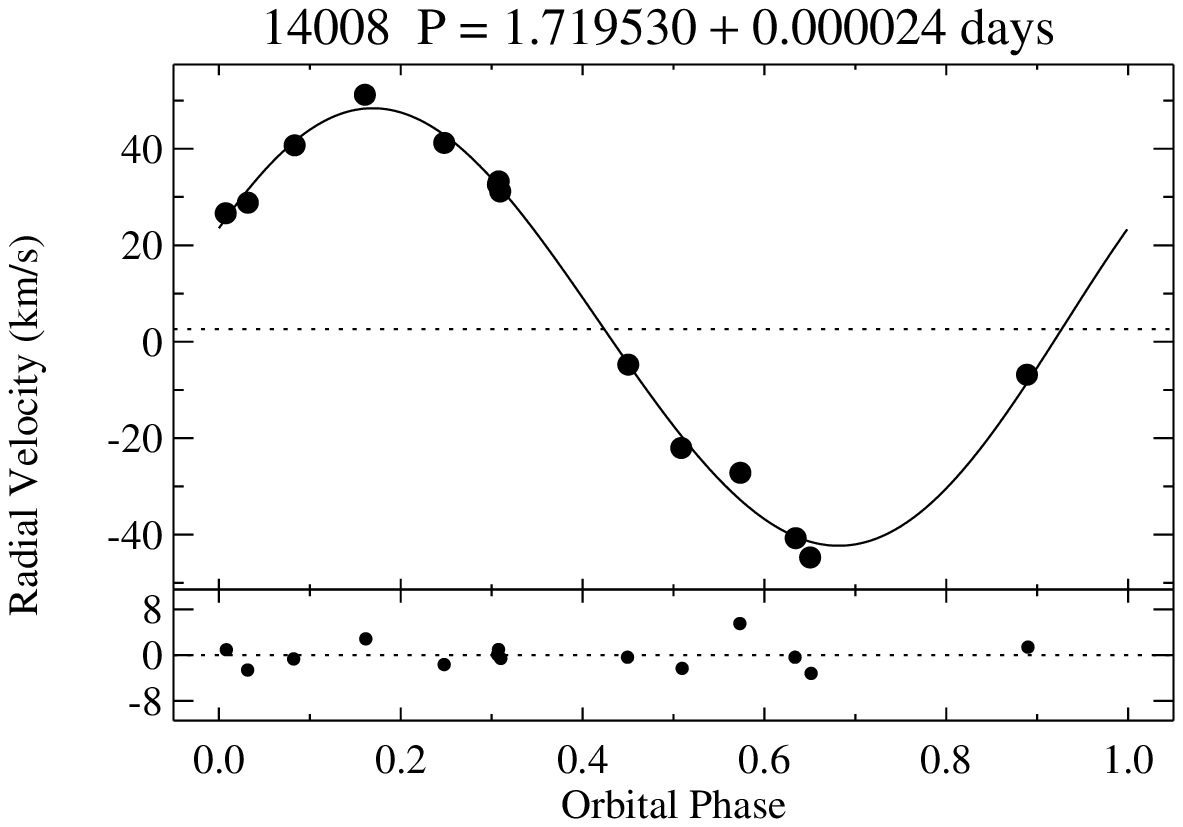}
\includegraphics[width=0.3\linewidth]{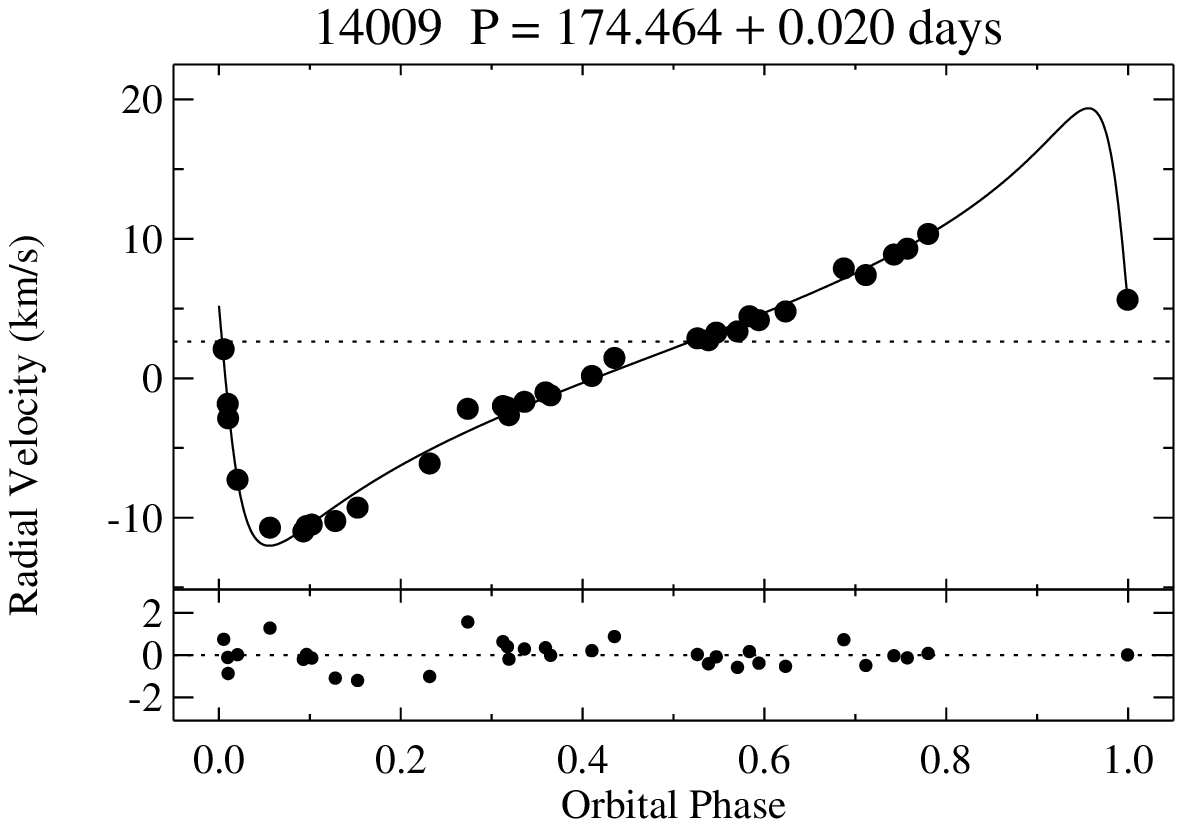}
\includegraphics[width=0.3\linewidth]{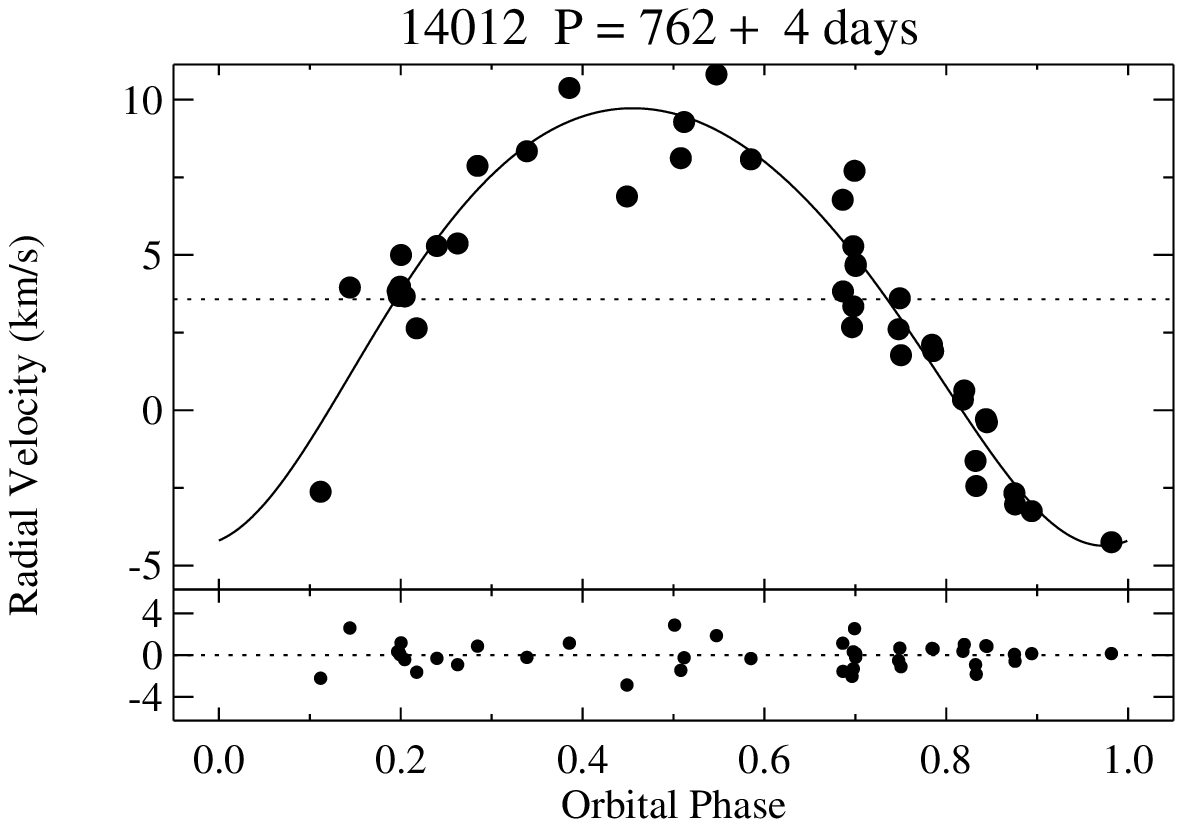}
\includegraphics[width=0.3\linewidth]{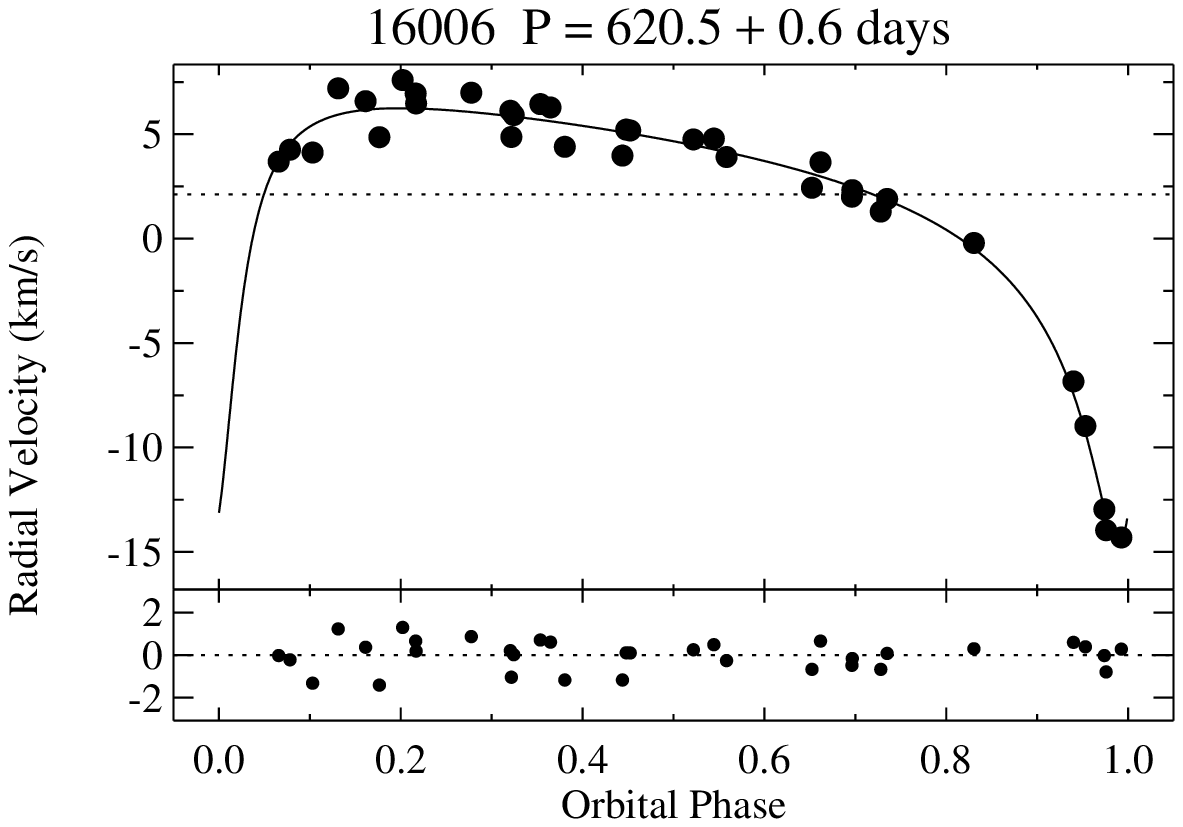}
\includegraphics[width=0.3\linewidth]{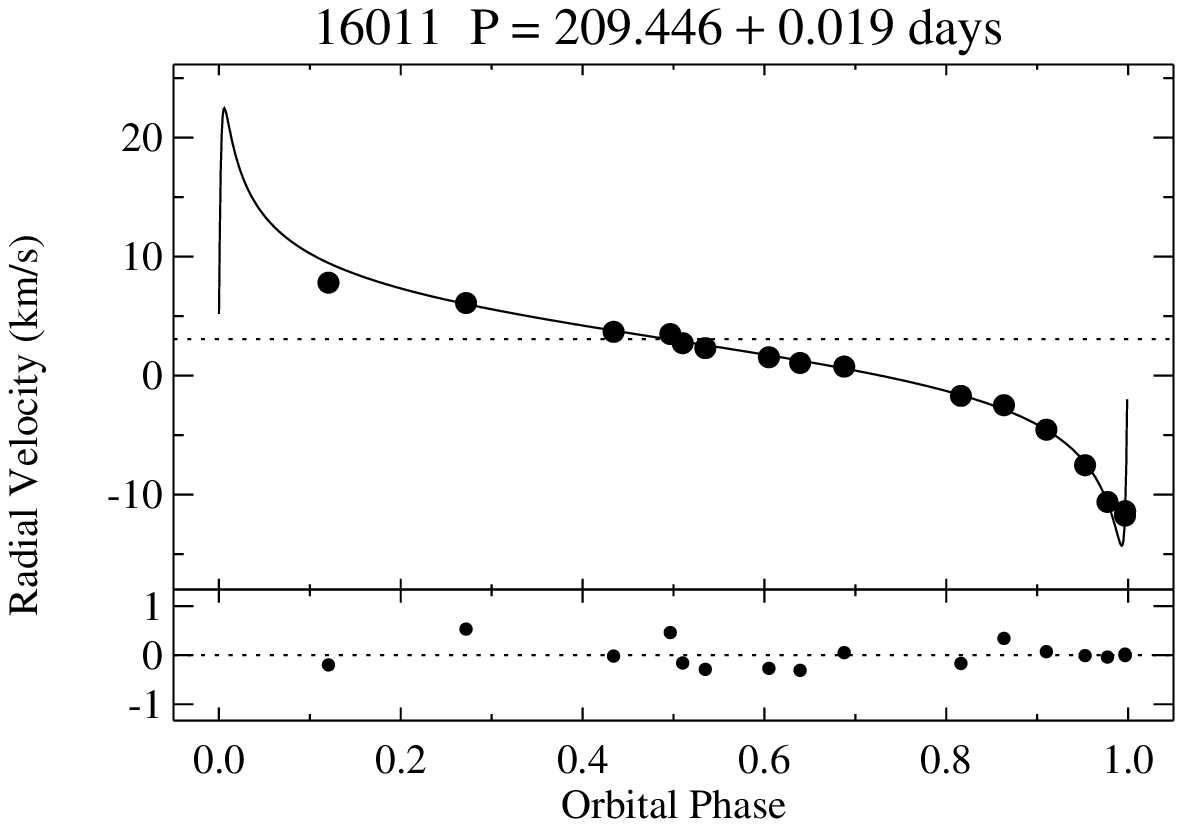}
\includegraphics[width=0.3\linewidth]{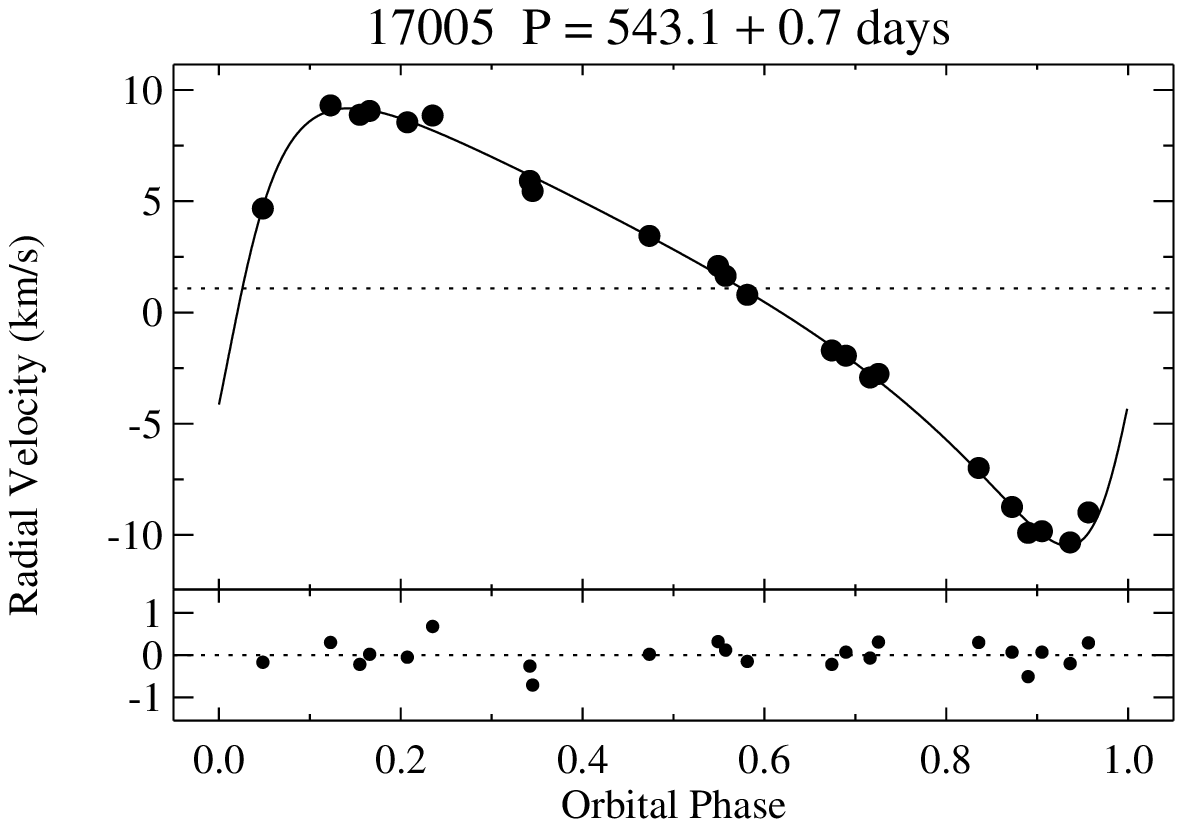}
\includegraphics[width=0.3\linewidth]{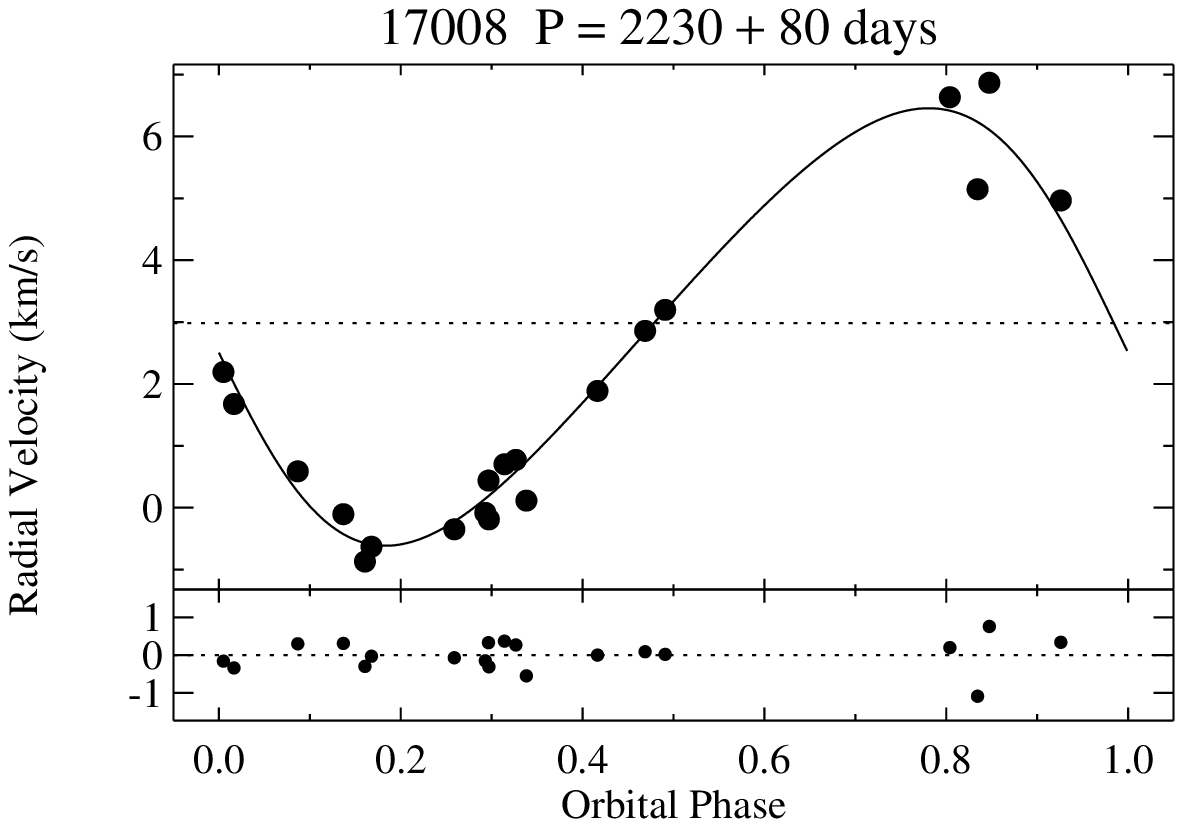}
\includegraphics[width=0.3\linewidth]{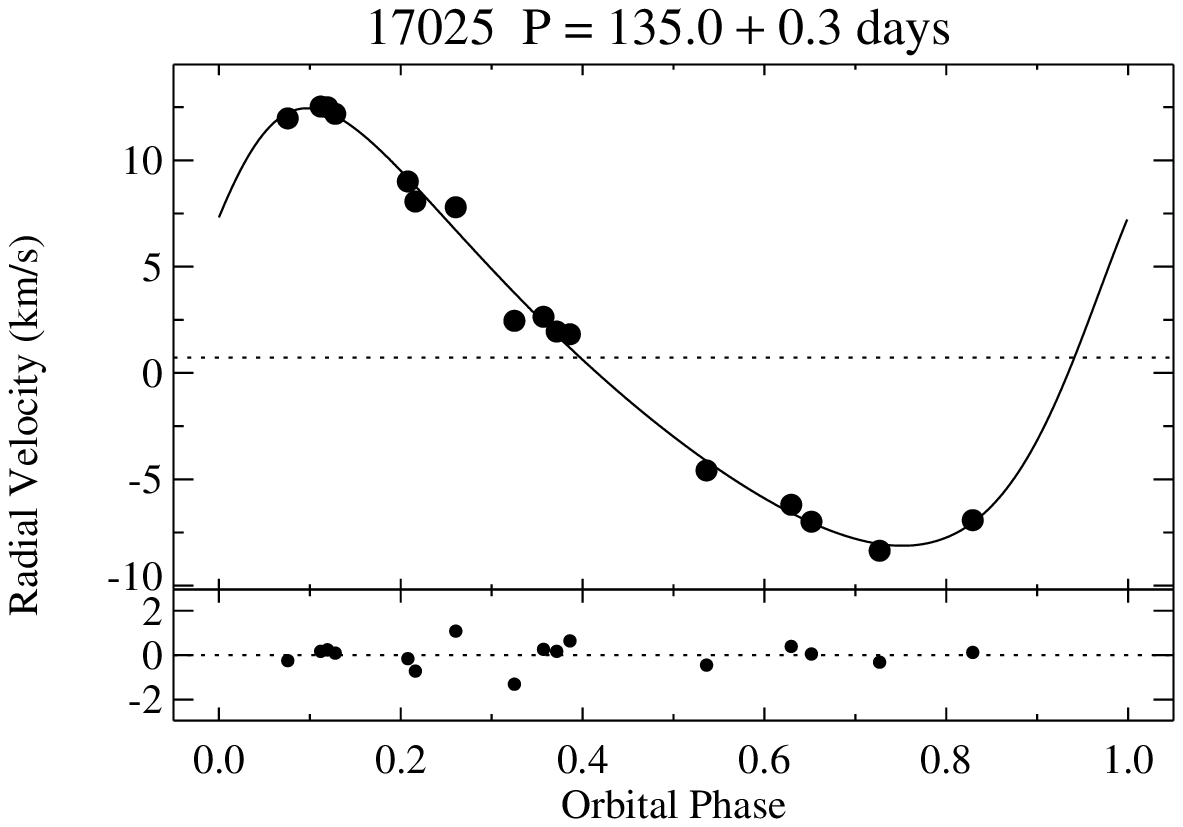}
\includegraphics[width=0.3\linewidth]{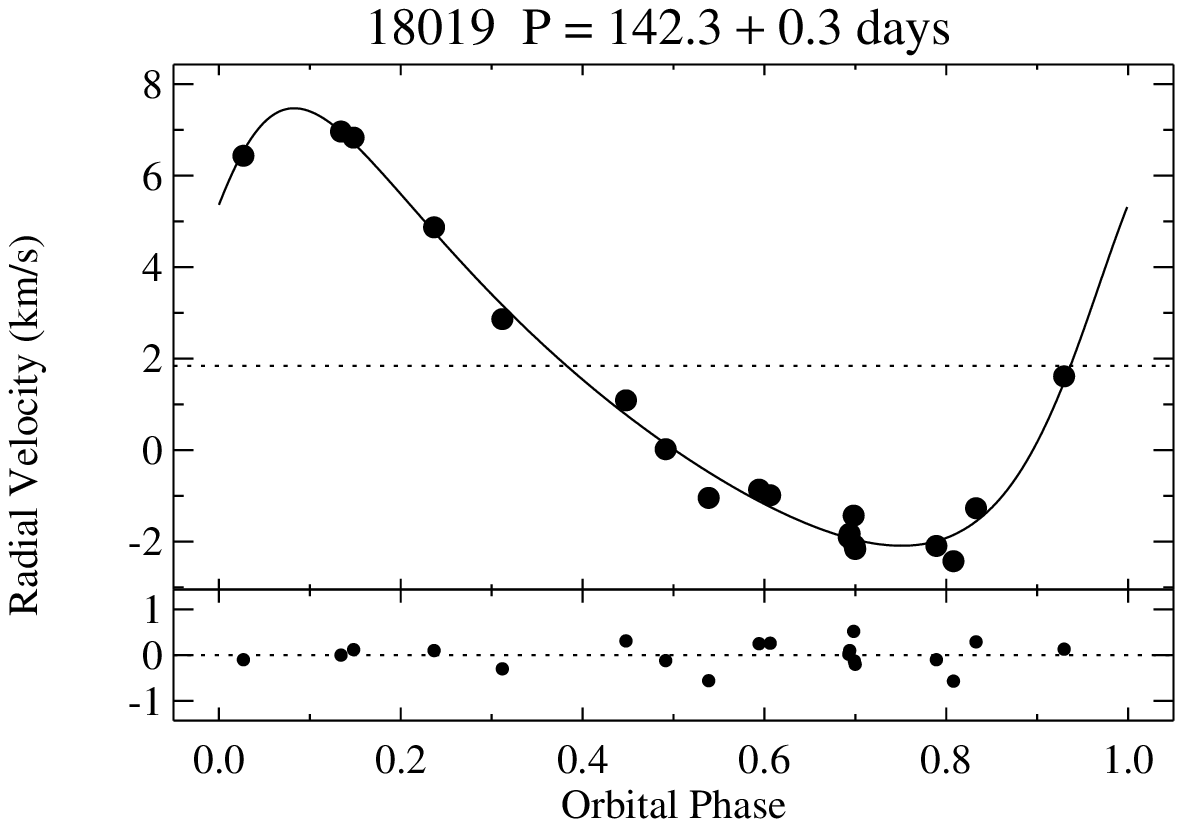}
\includegraphics[width=0.3\linewidth]{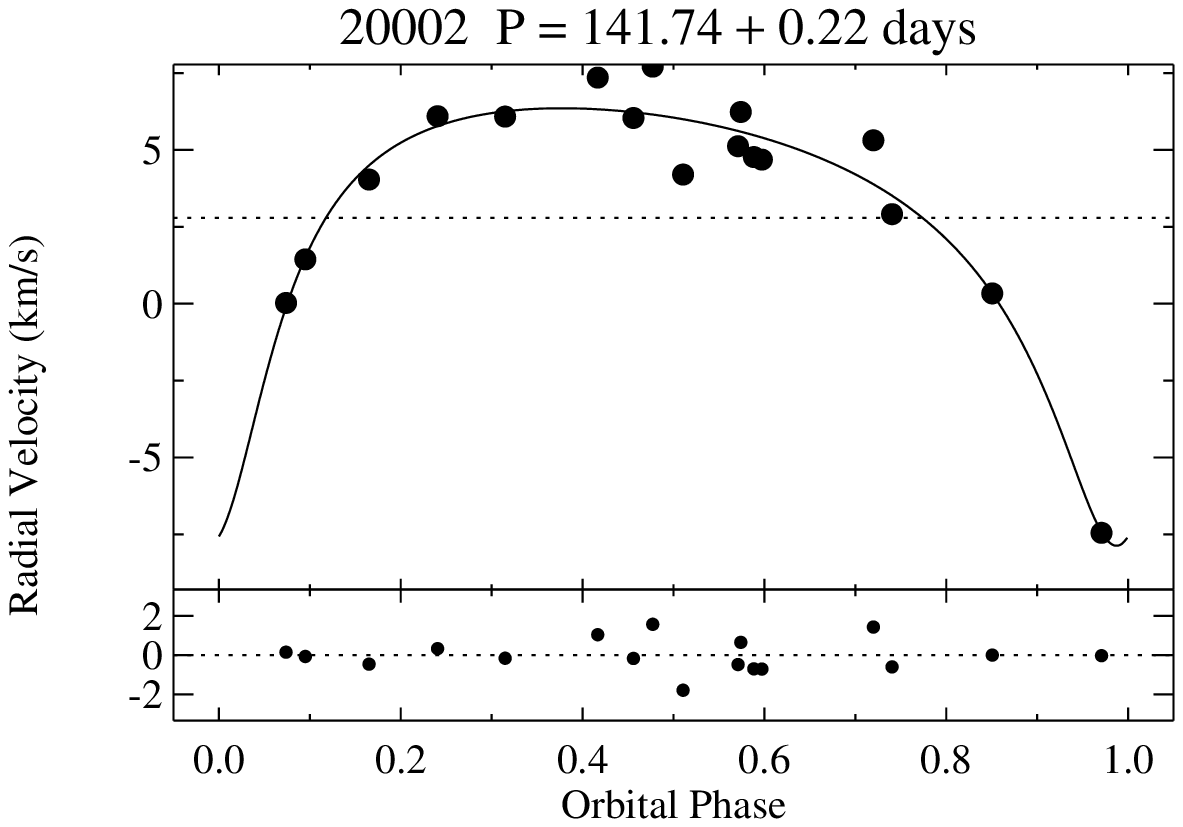}
\includegraphics[width=0.3\linewidth]{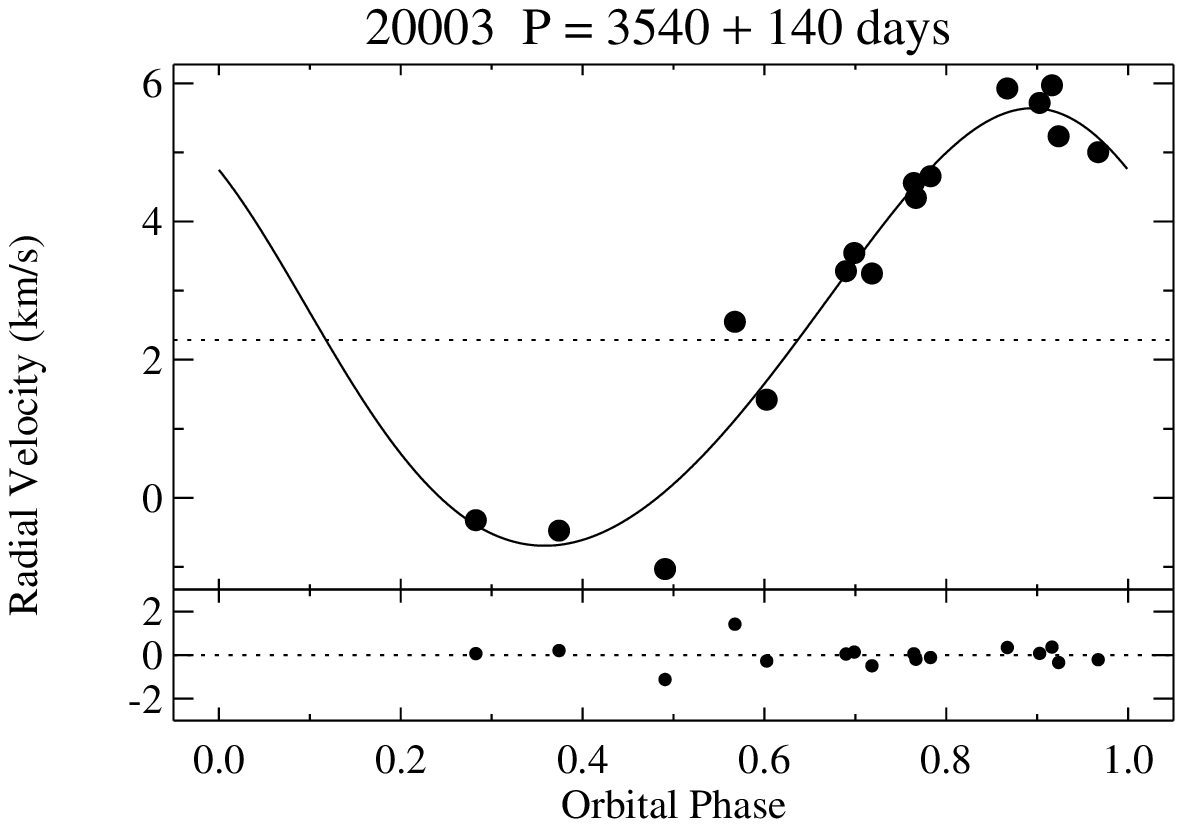}
\includegraphics[width=0.3\linewidth]{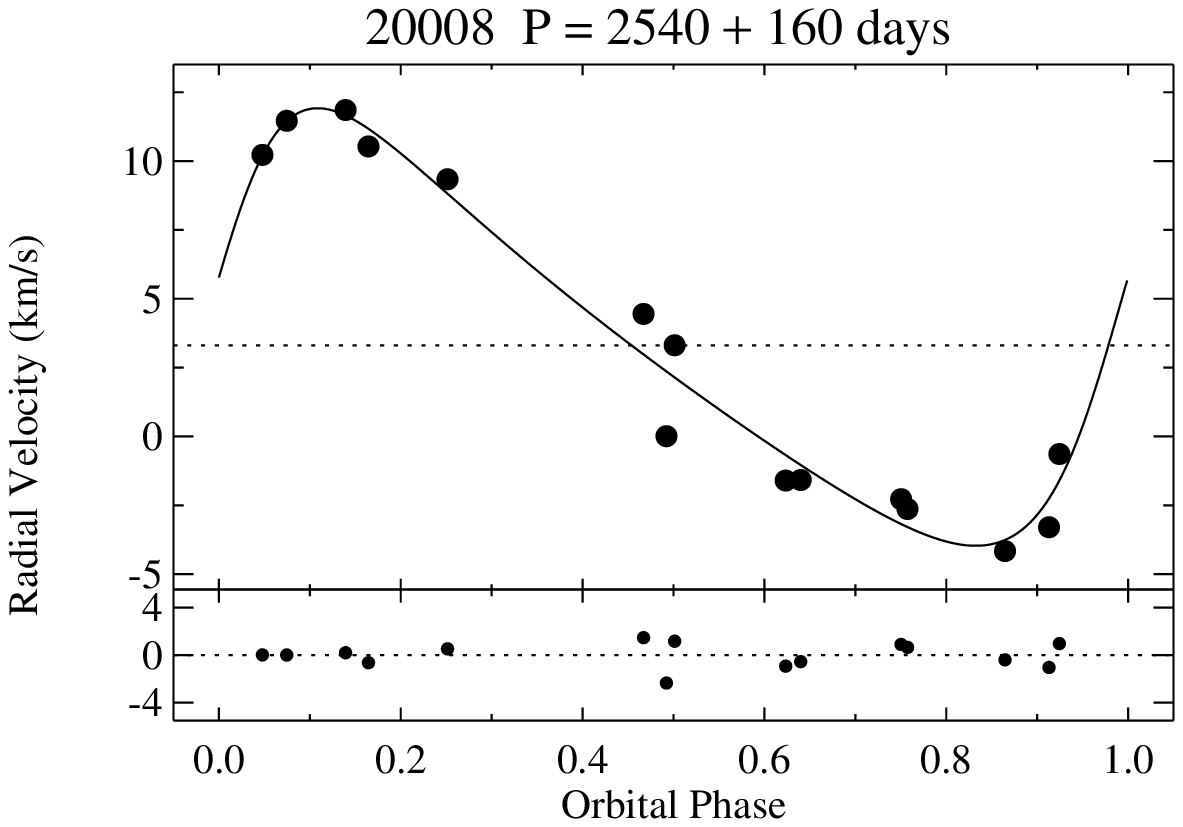}
\includegraphics[width=0.3\linewidth]{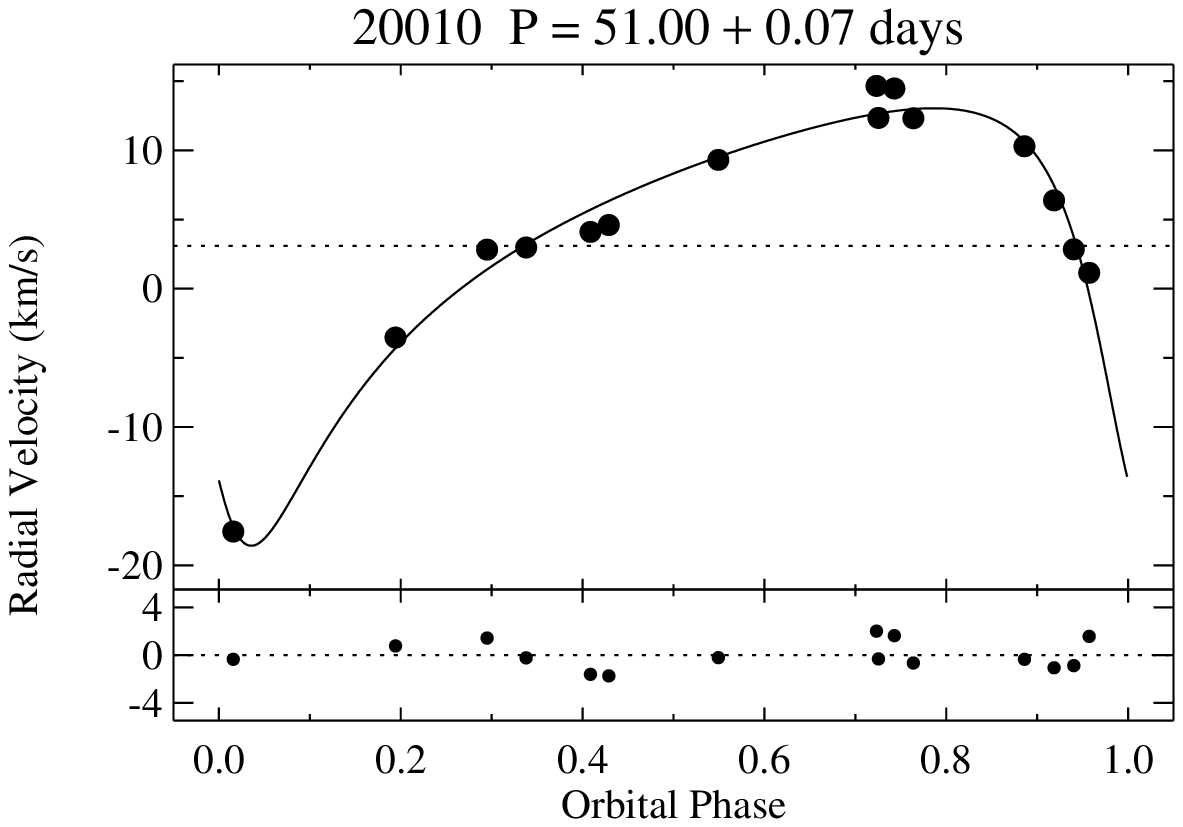}
\includegraphics[width=0.3\linewidth]{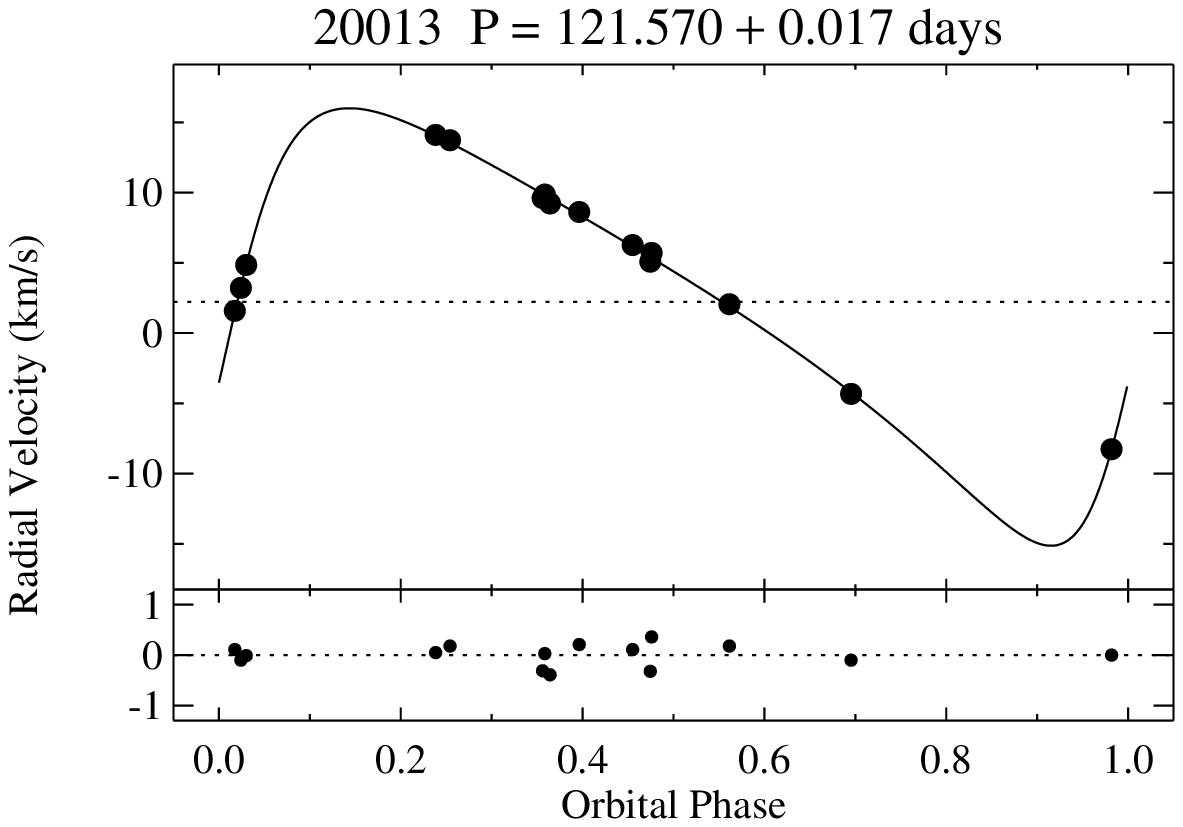}
\includegraphics[width=0.3\linewidth]{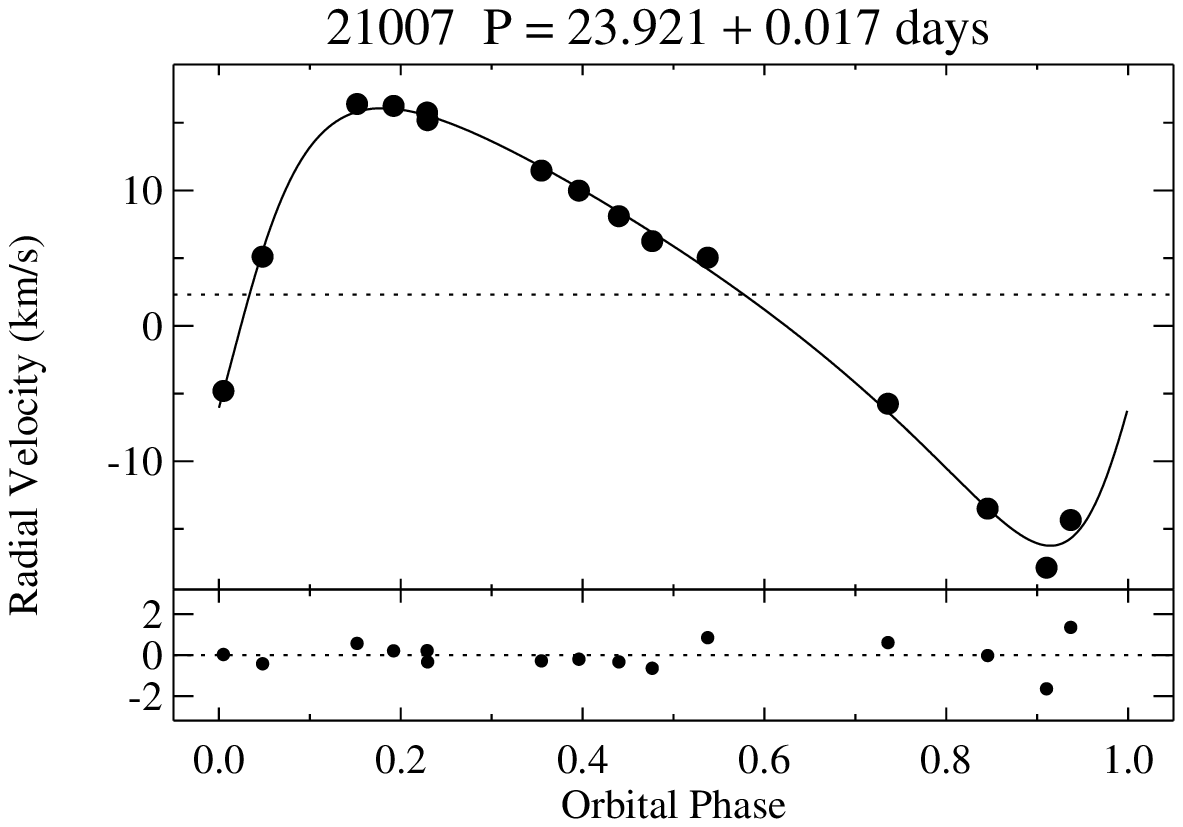}

{\bf{Figure 7.} (Continued)}
\end{center}
\end{figure*}
\begin{figure*}
\begin{center}
\includegraphics[width=0.3\linewidth]{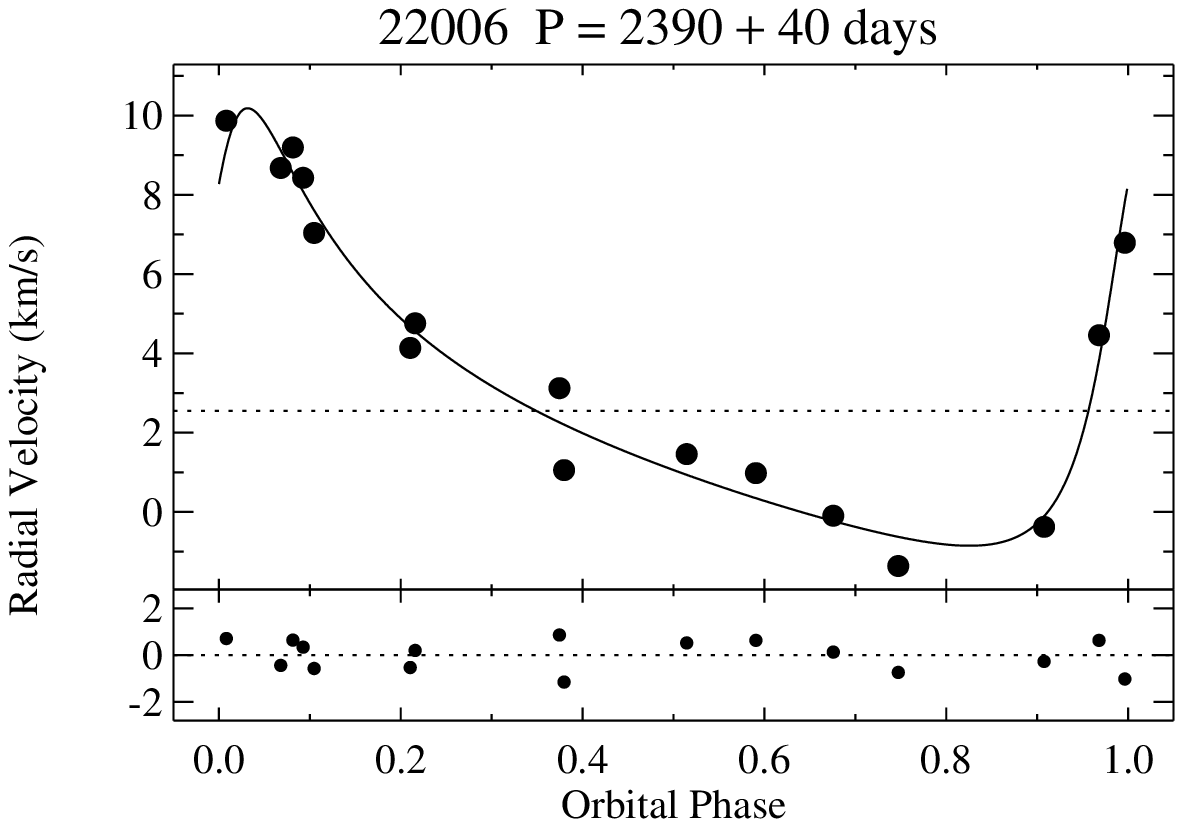}
\includegraphics[width=0.3\linewidth]{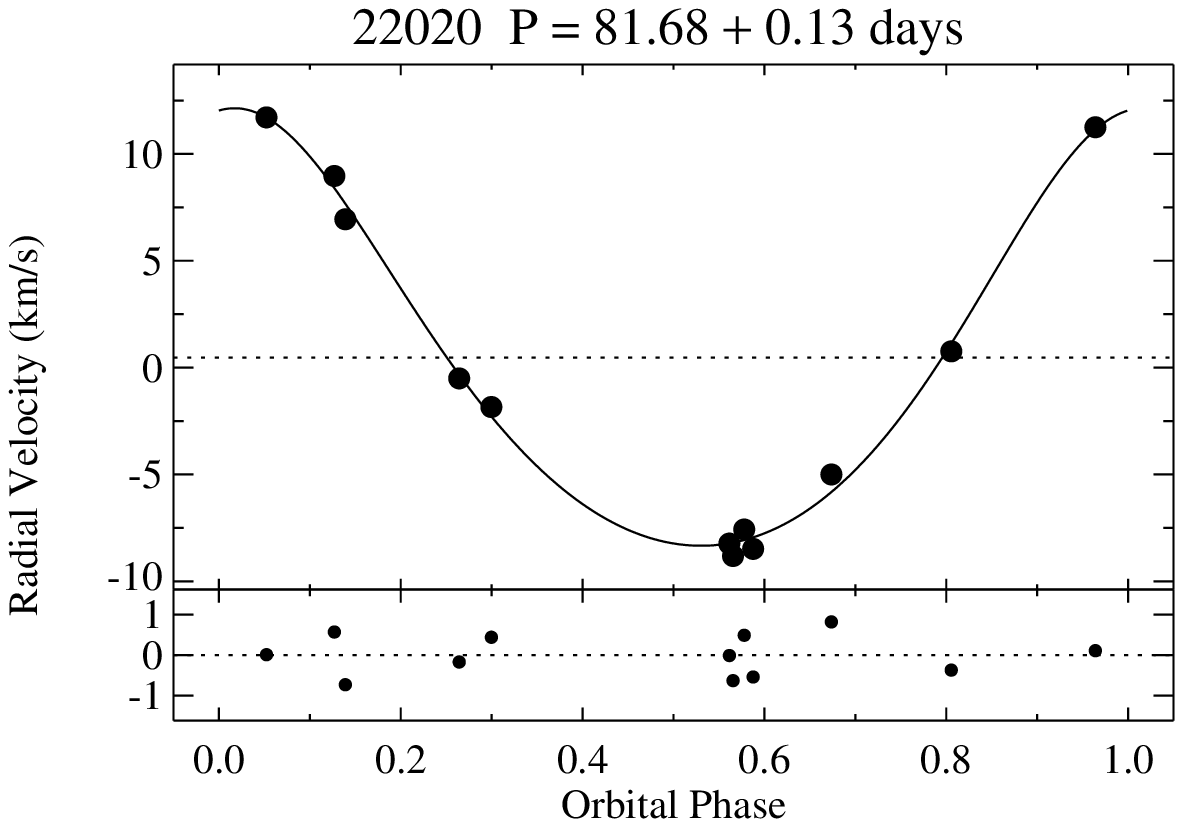}
\includegraphics[width=0.3\linewidth]{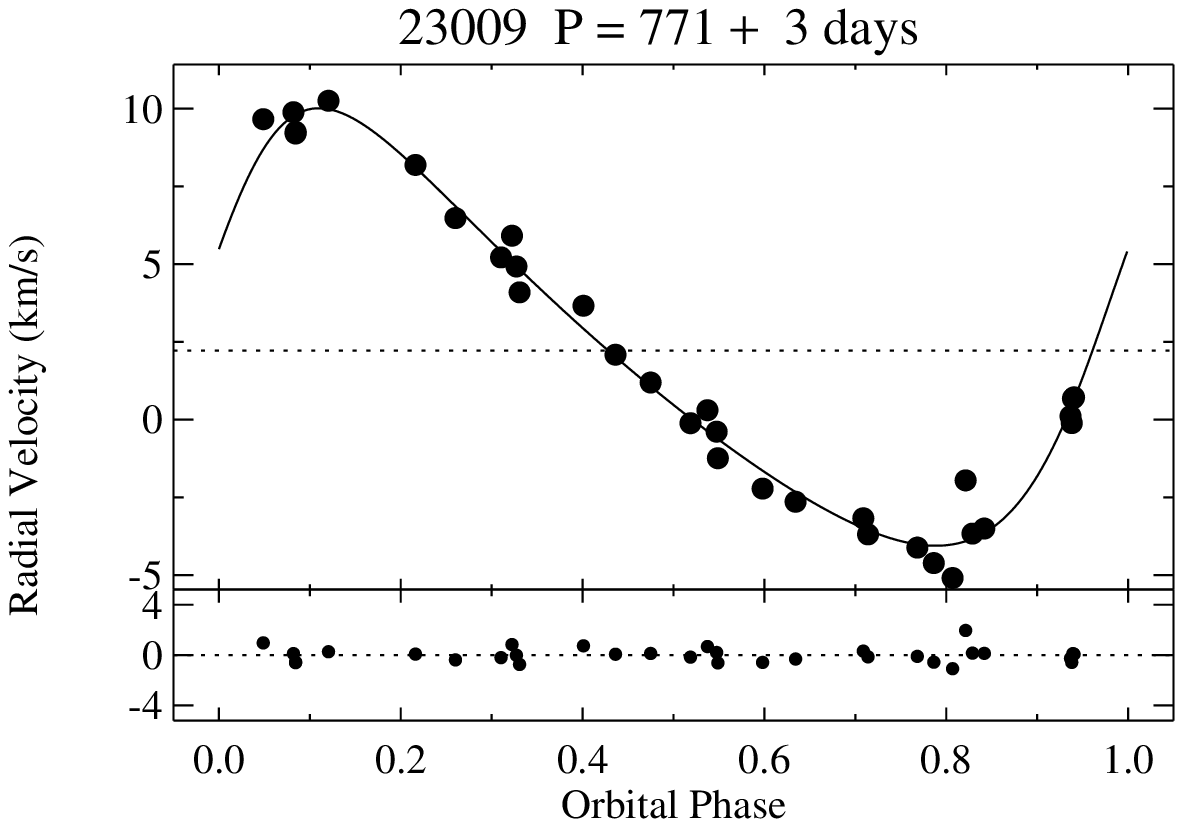}
\includegraphics[width=0.3\linewidth]{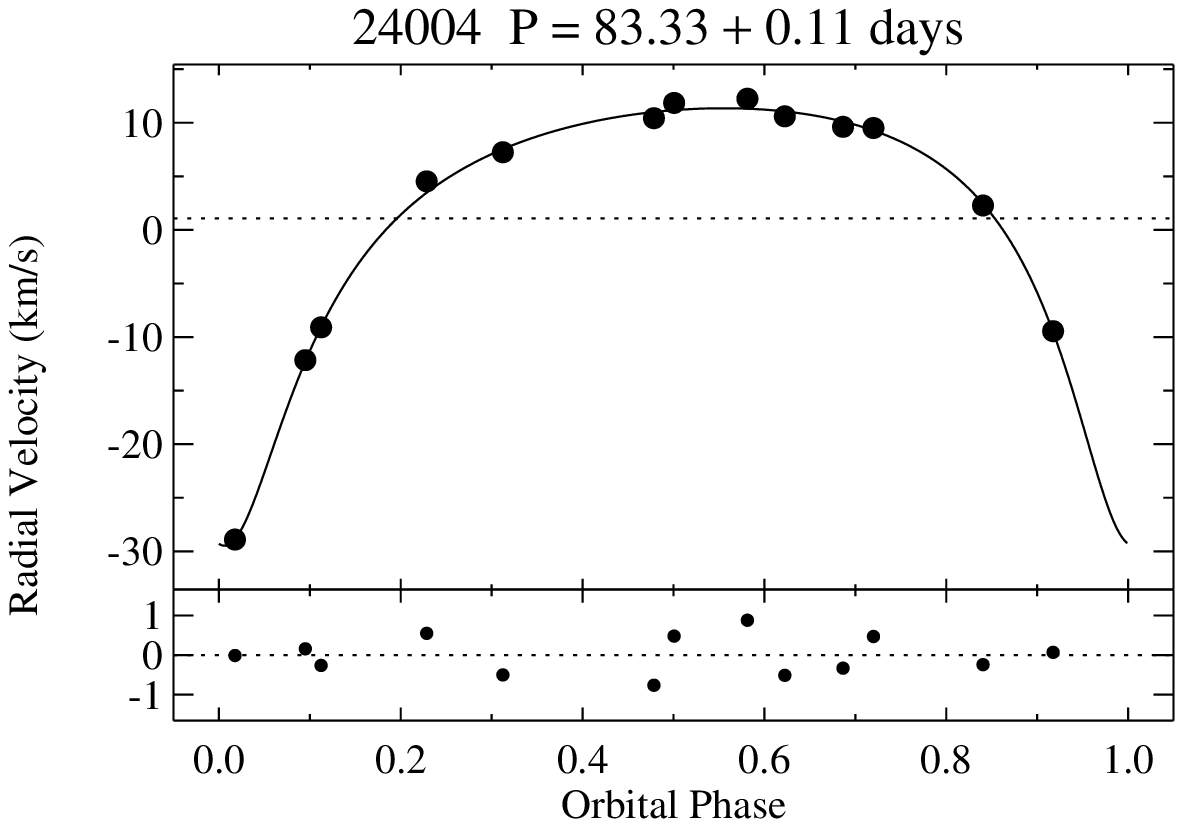}
\includegraphics[width=0.3\linewidth]{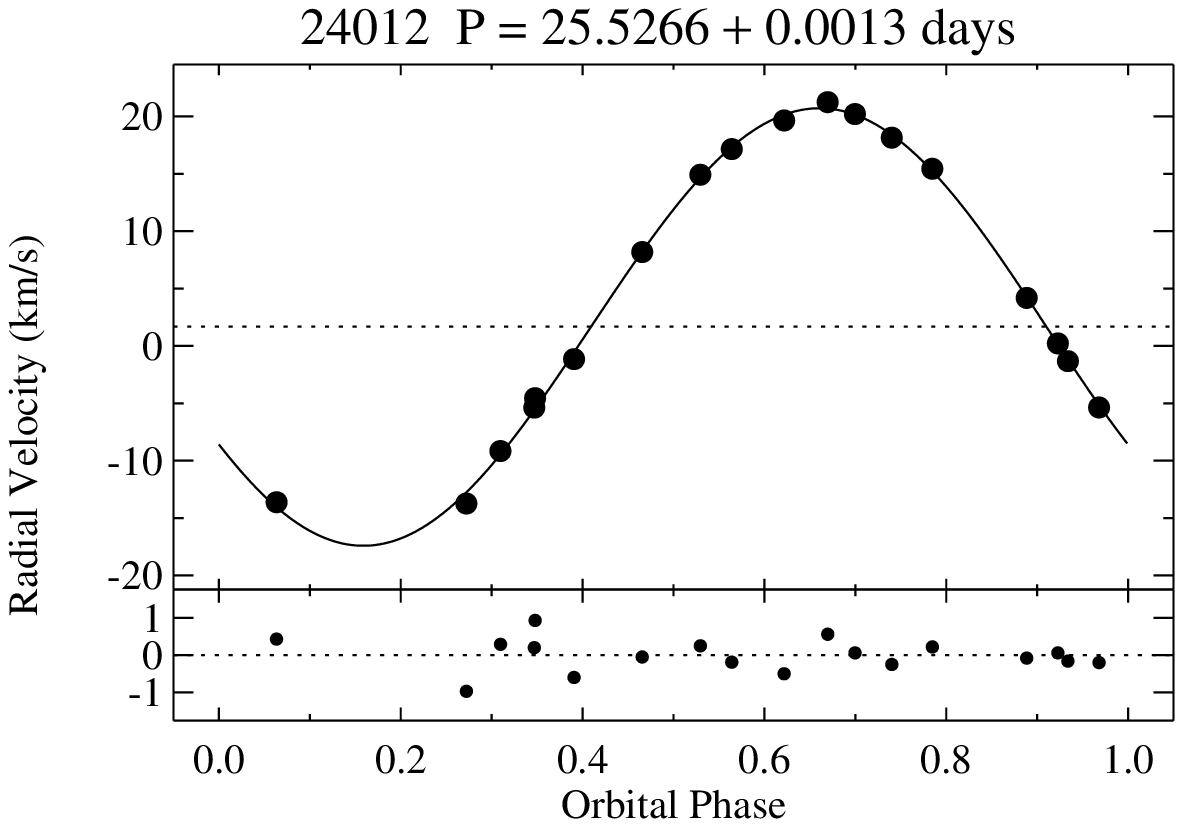}
\includegraphics[width=0.3\linewidth]{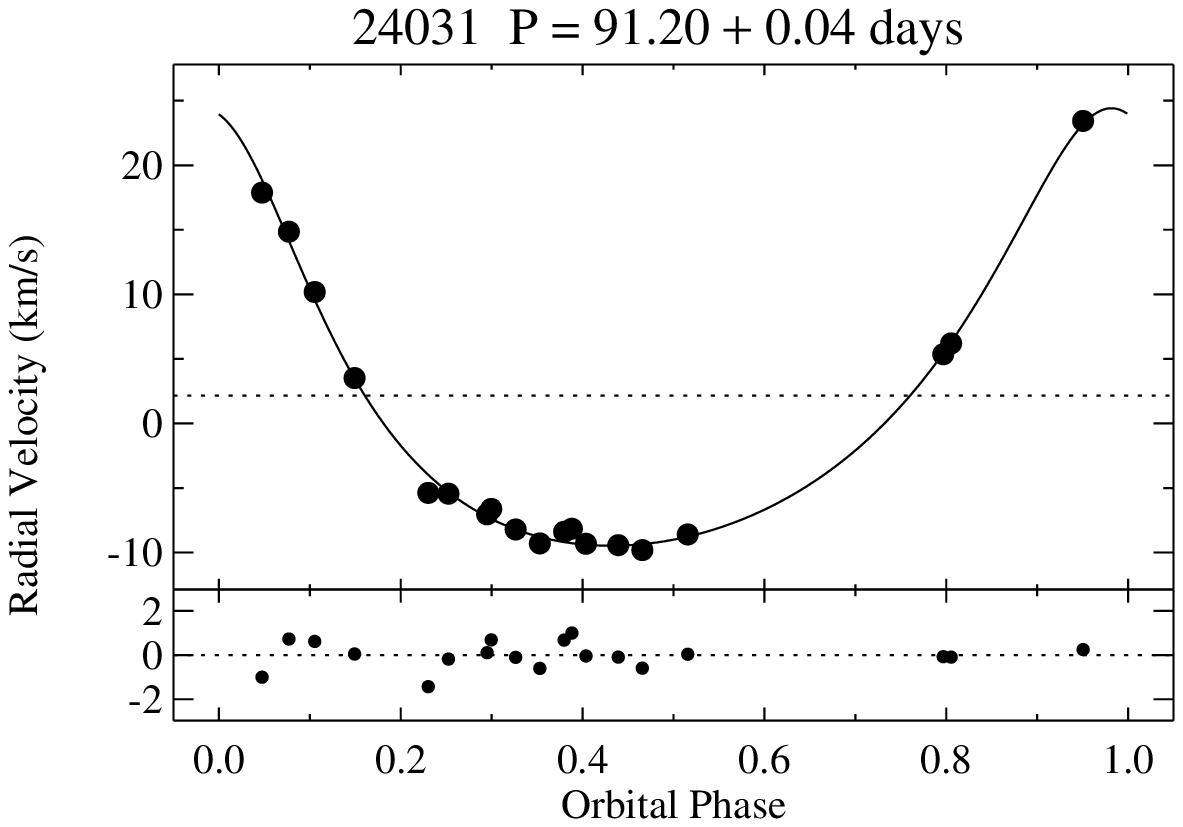}
\includegraphics[width=0.3\linewidth]{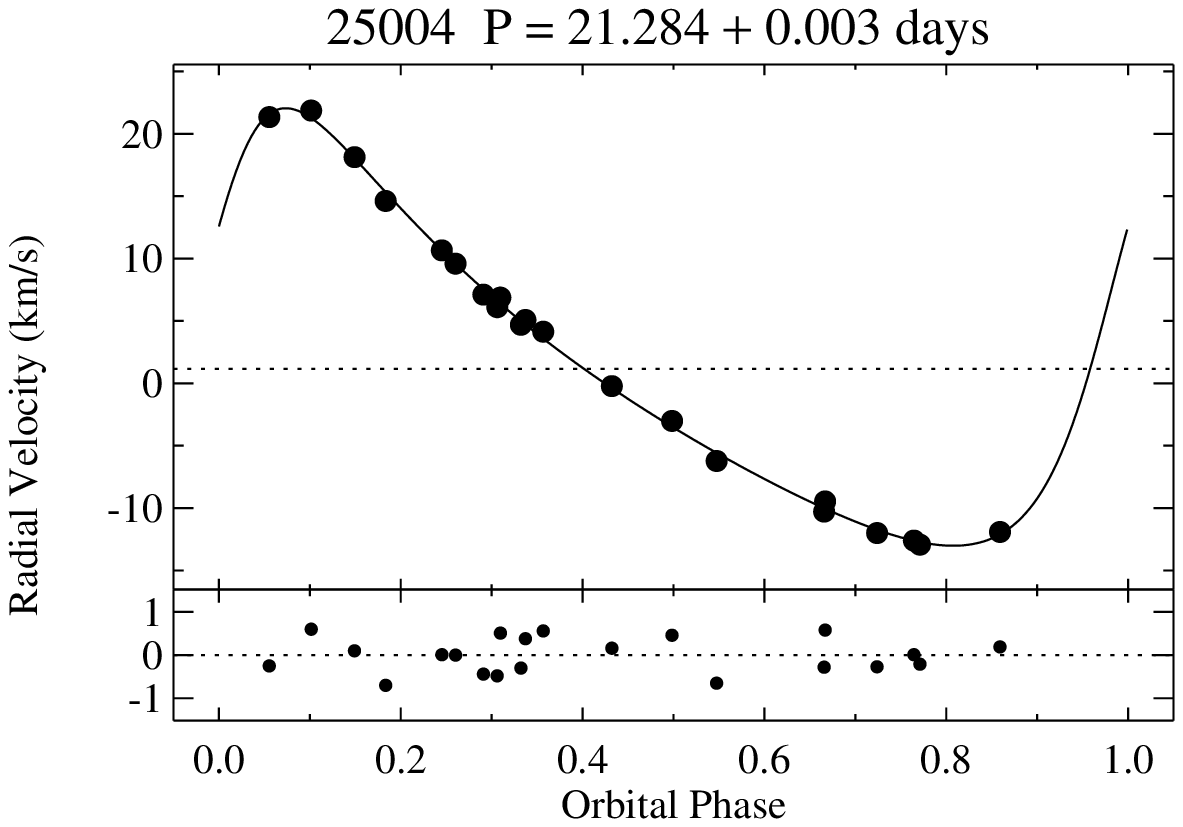}
\includegraphics[width=0.3\linewidth]{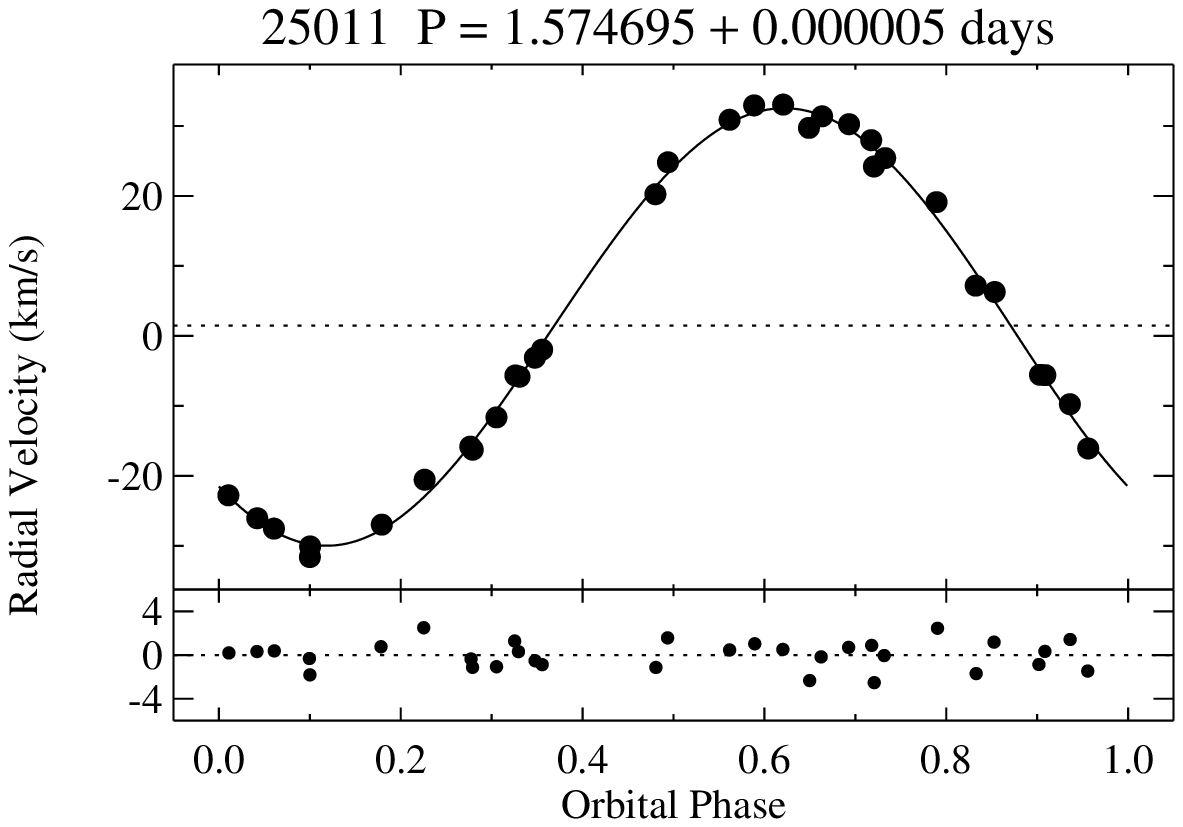}
\includegraphics[width=0.3\linewidth]{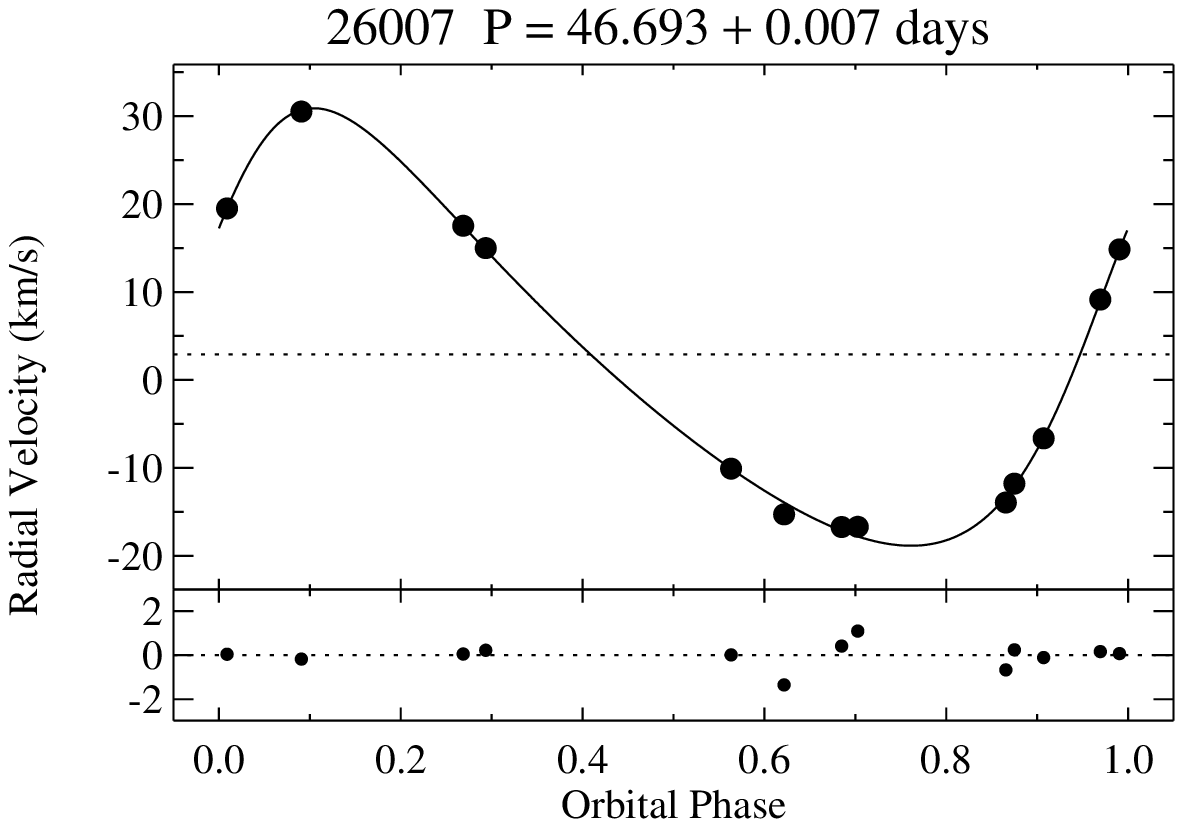}
\includegraphics[width=0.3\linewidth]{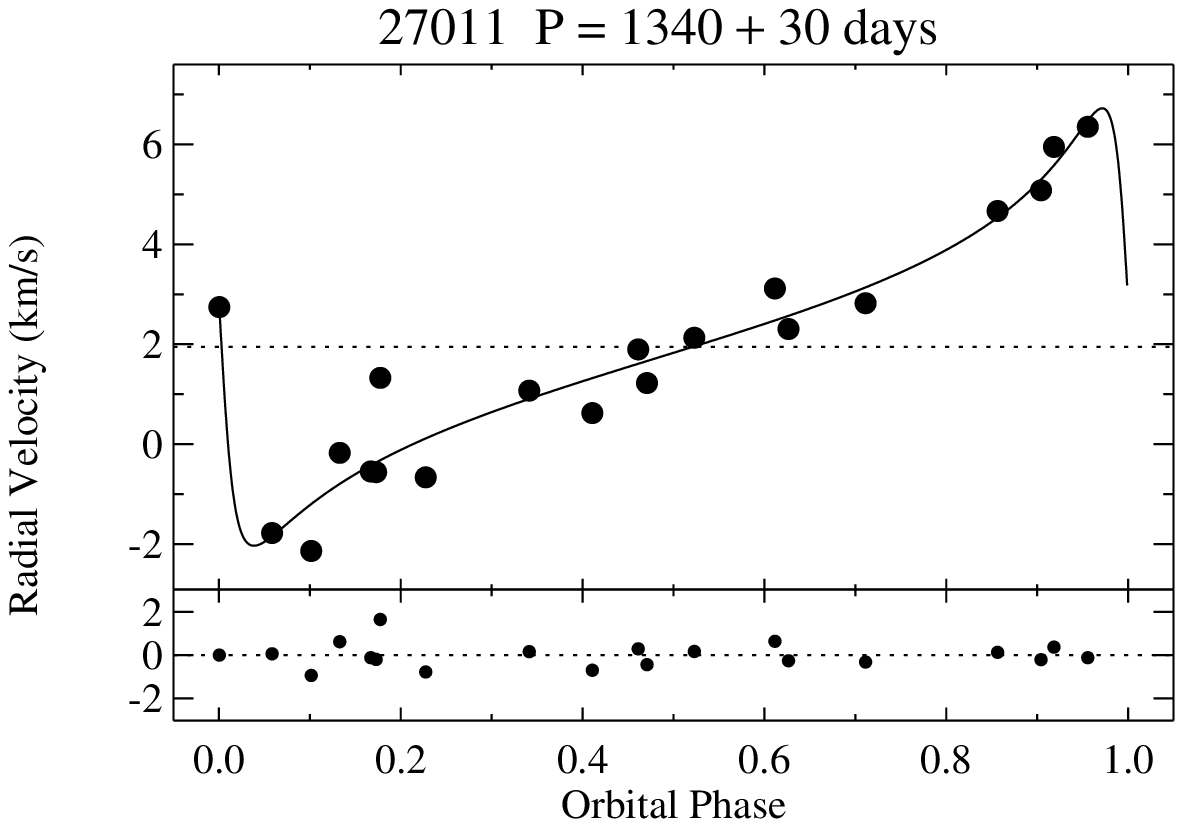}
\includegraphics[width=0.3\linewidth]{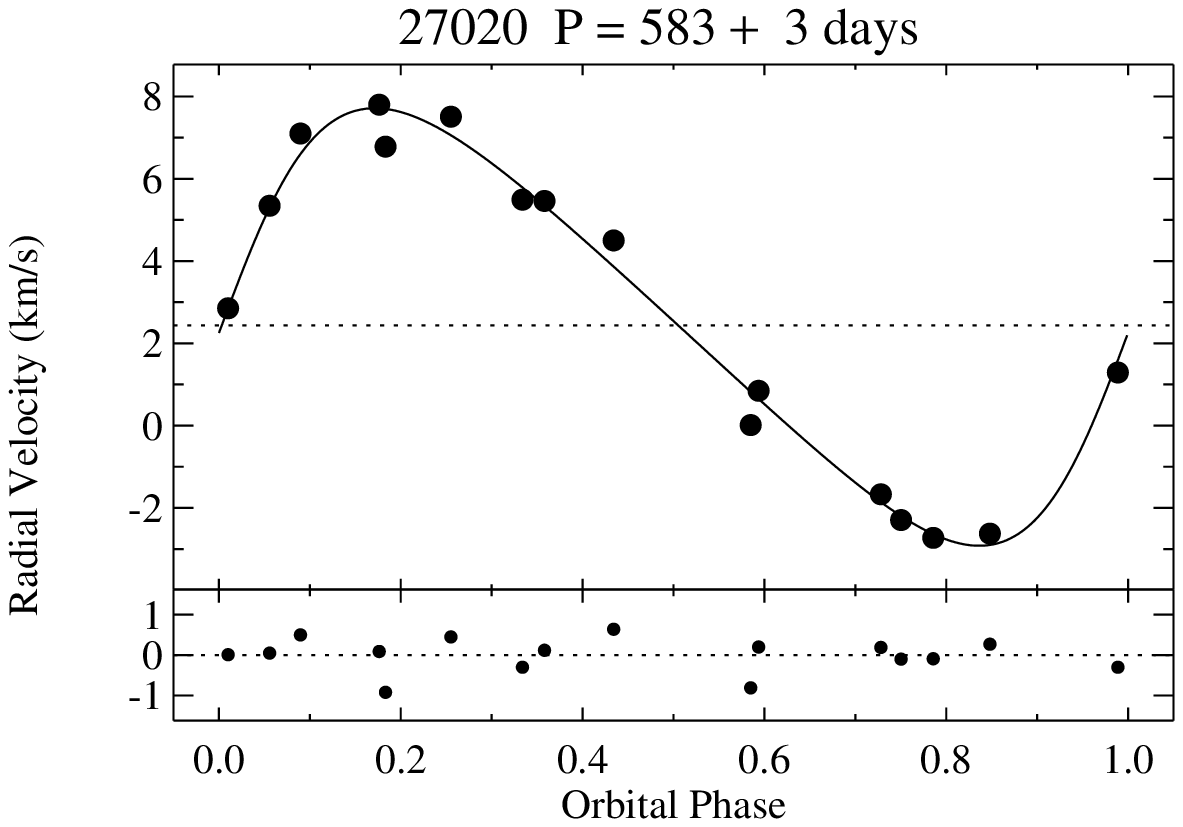}
\includegraphics[width=0.3\linewidth]{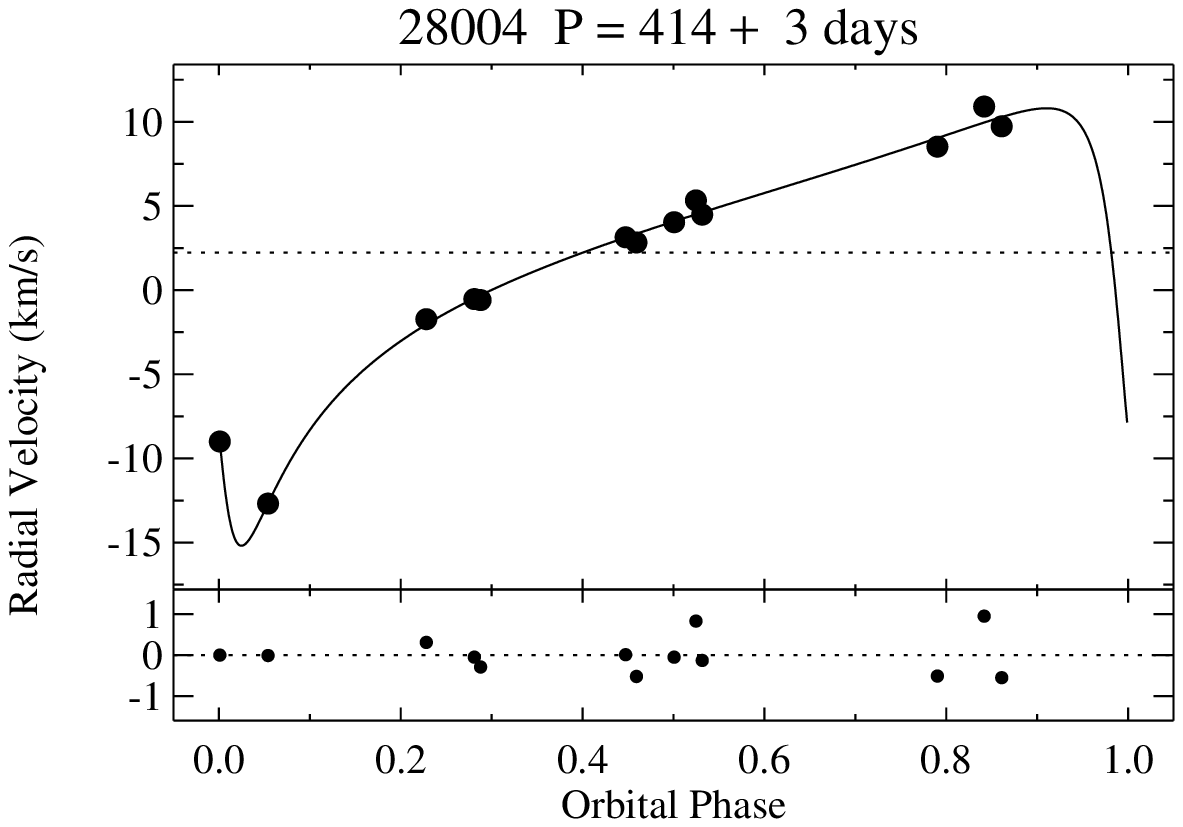}
\includegraphics[width=0.3\linewidth]{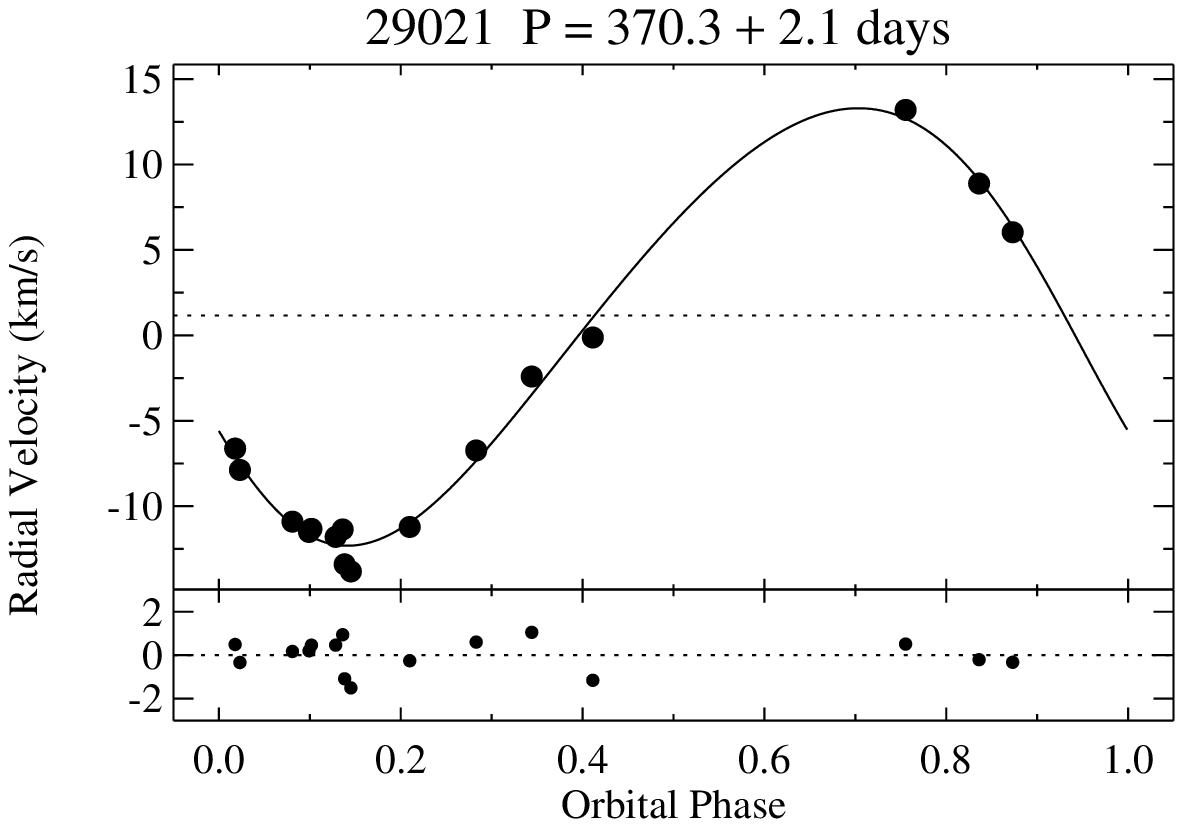}
\includegraphics[width=0.3\linewidth]{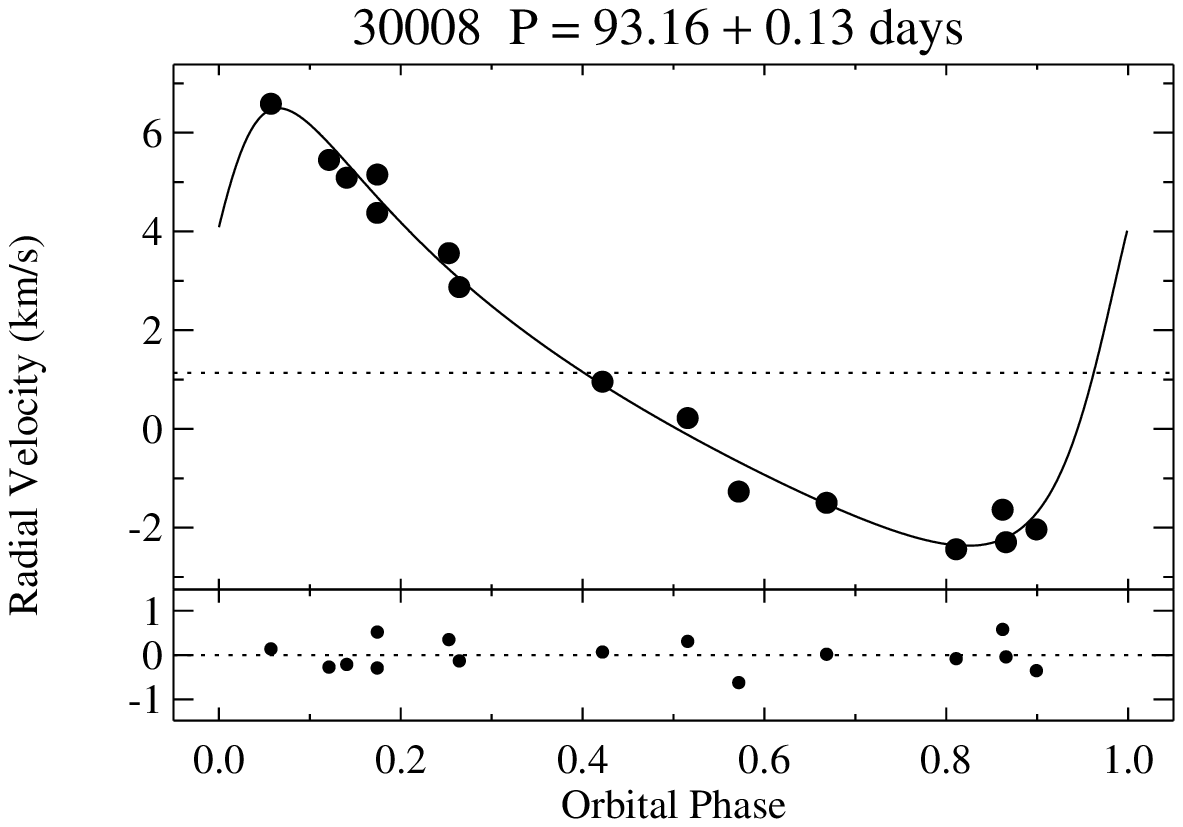}
\includegraphics[width=0.3\linewidth]{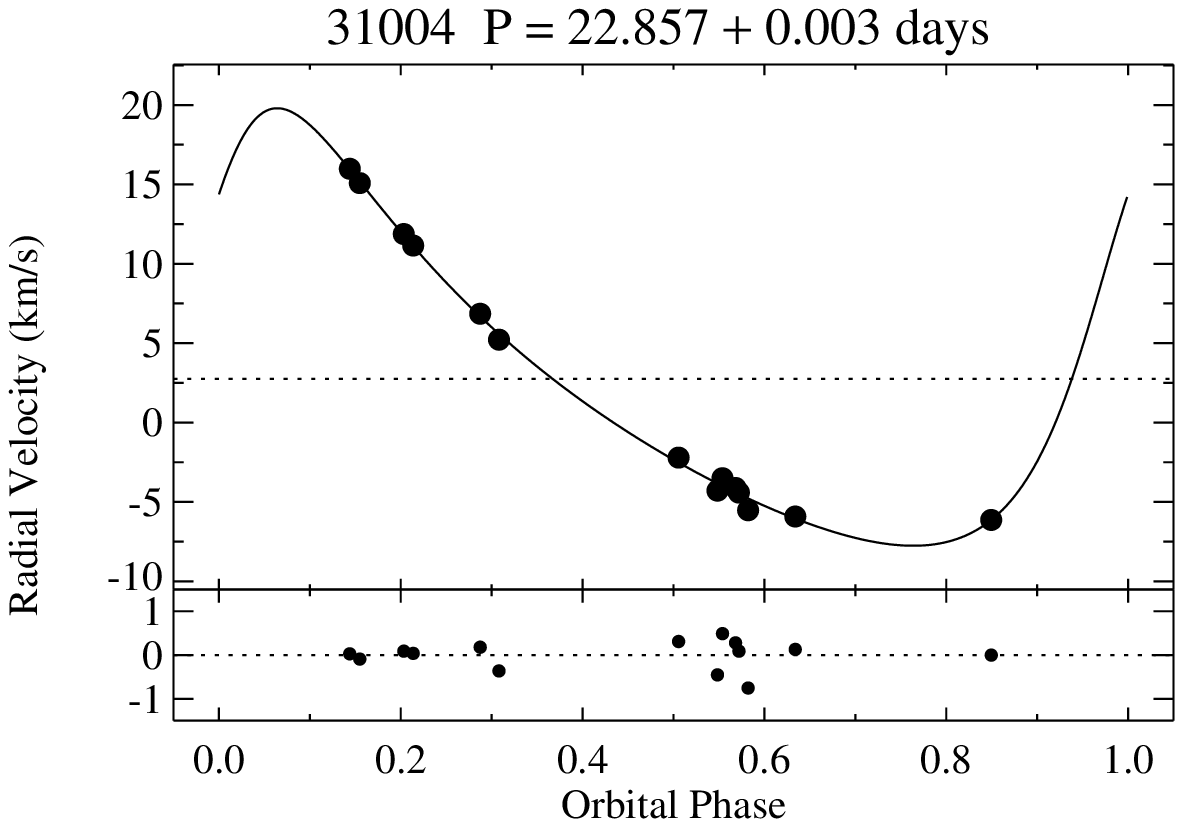}
\includegraphics[width=0.3\linewidth]{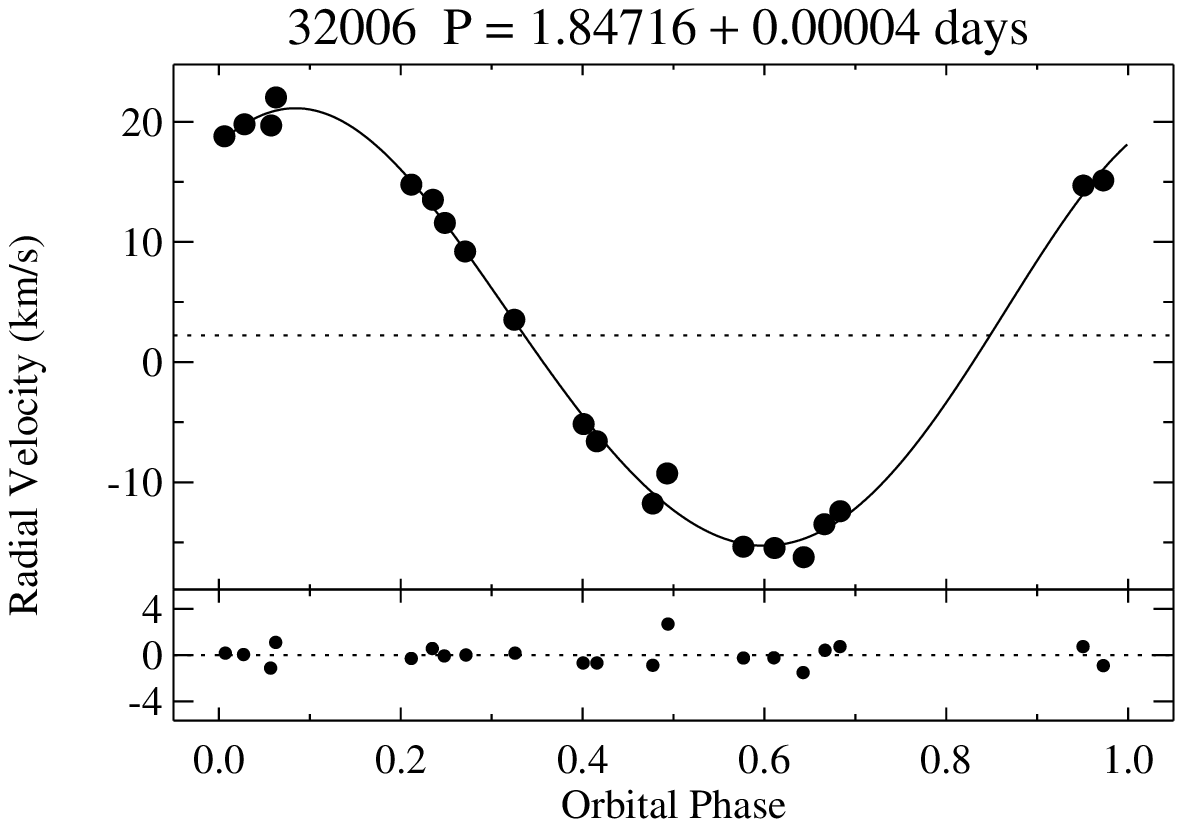}
\includegraphics[width=0.3\linewidth]{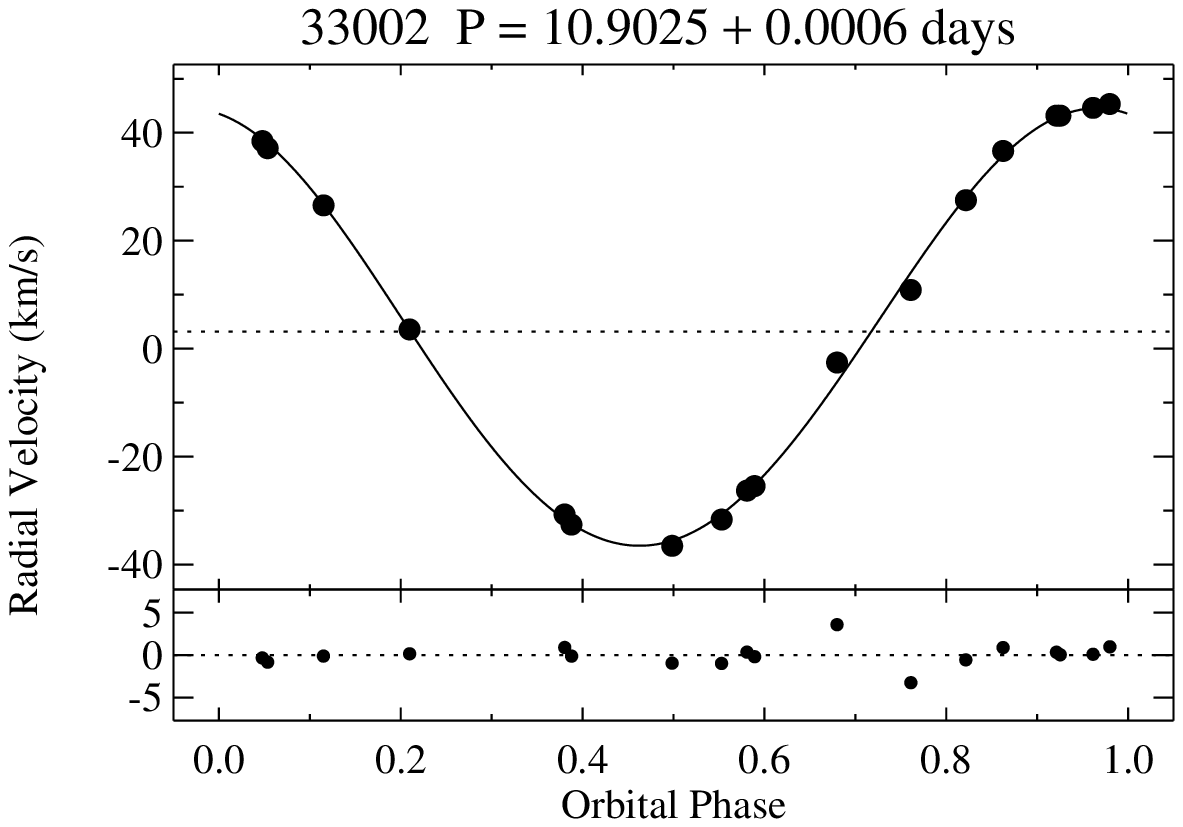}
\includegraphics[width=0.3\linewidth]{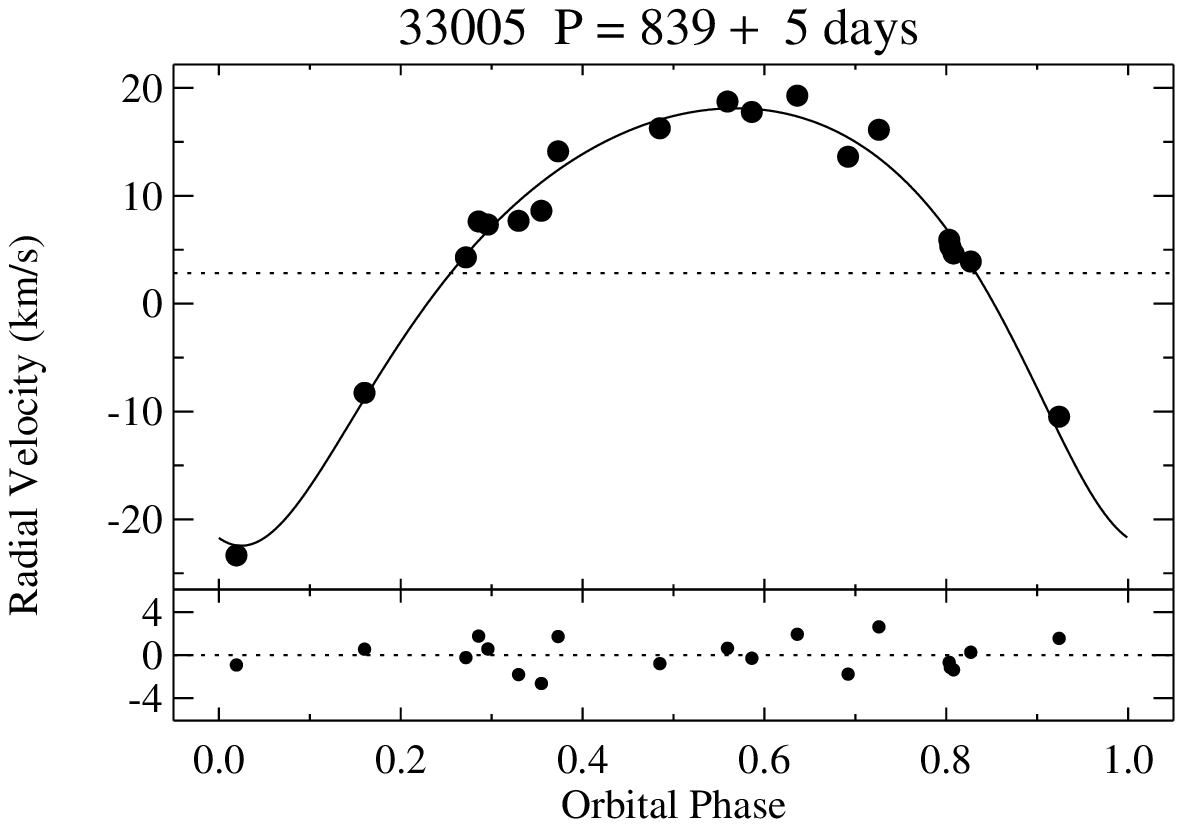}

{\bf{Figure 7.} (Continued)}
\end{center}
\end{figure*}
\begin{figure*}
\begin{center}
\includegraphics[width=0.3\linewidth]{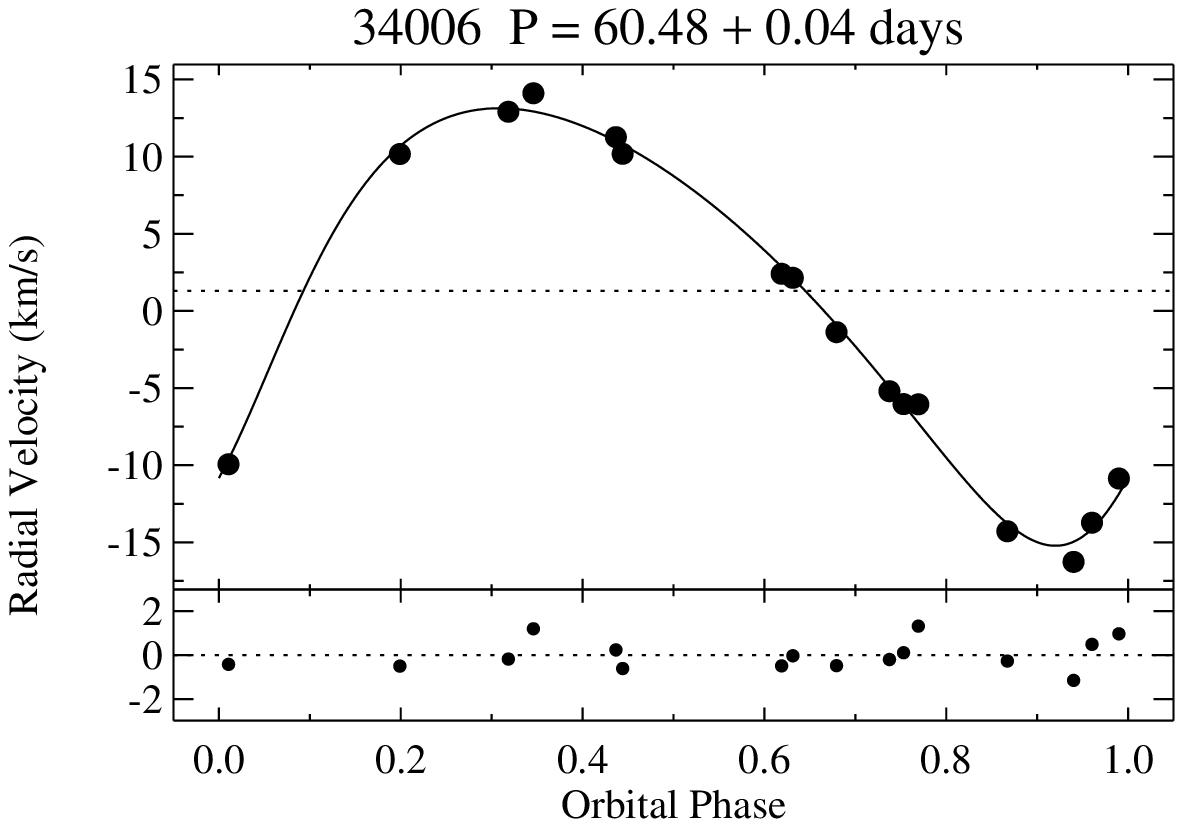}
\includegraphics[width=0.3\linewidth]{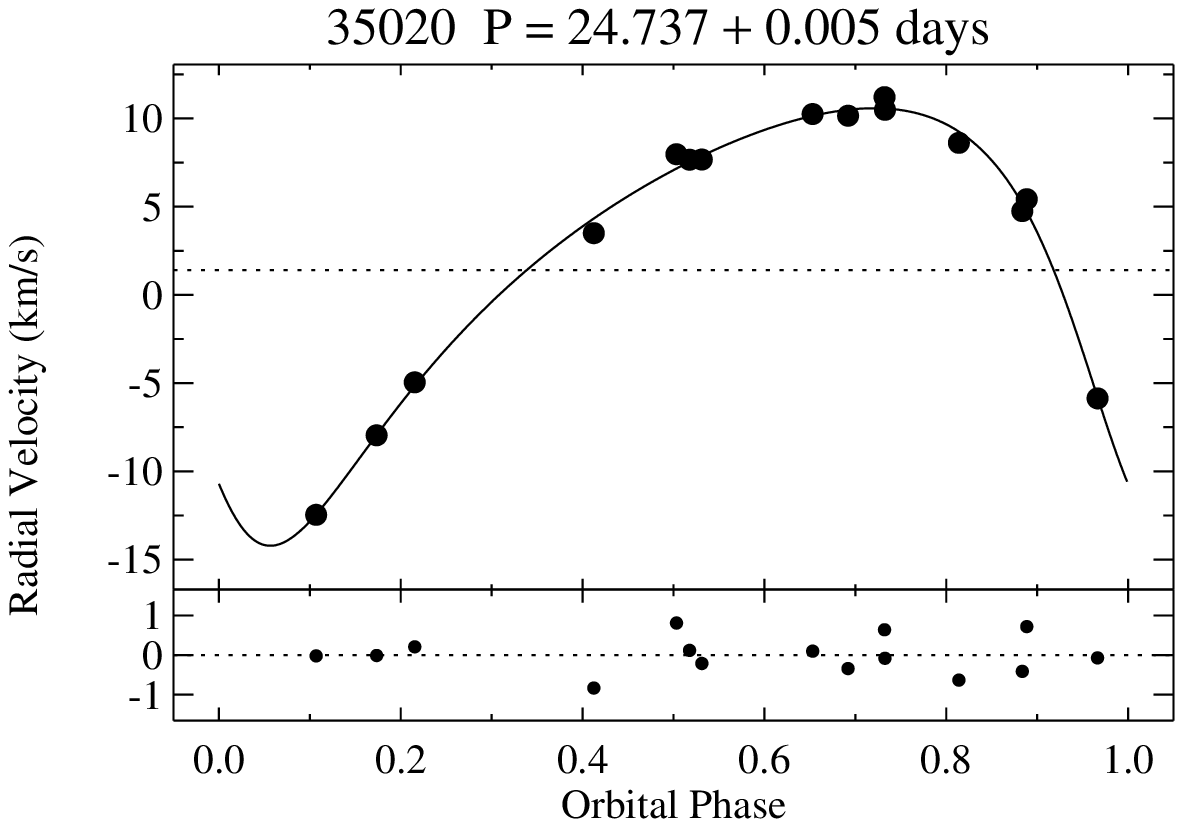}
\includegraphics[width=0.3\linewidth]{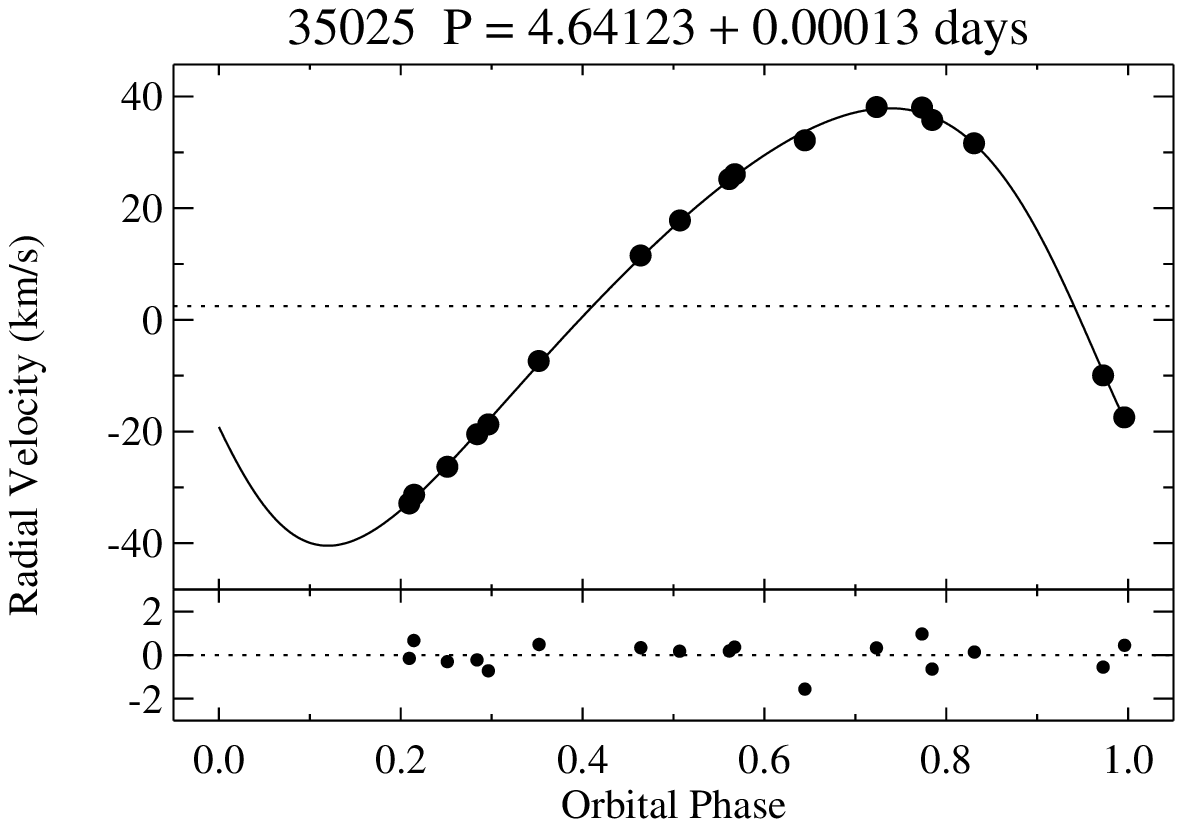}
\includegraphics[width=0.3\linewidth]{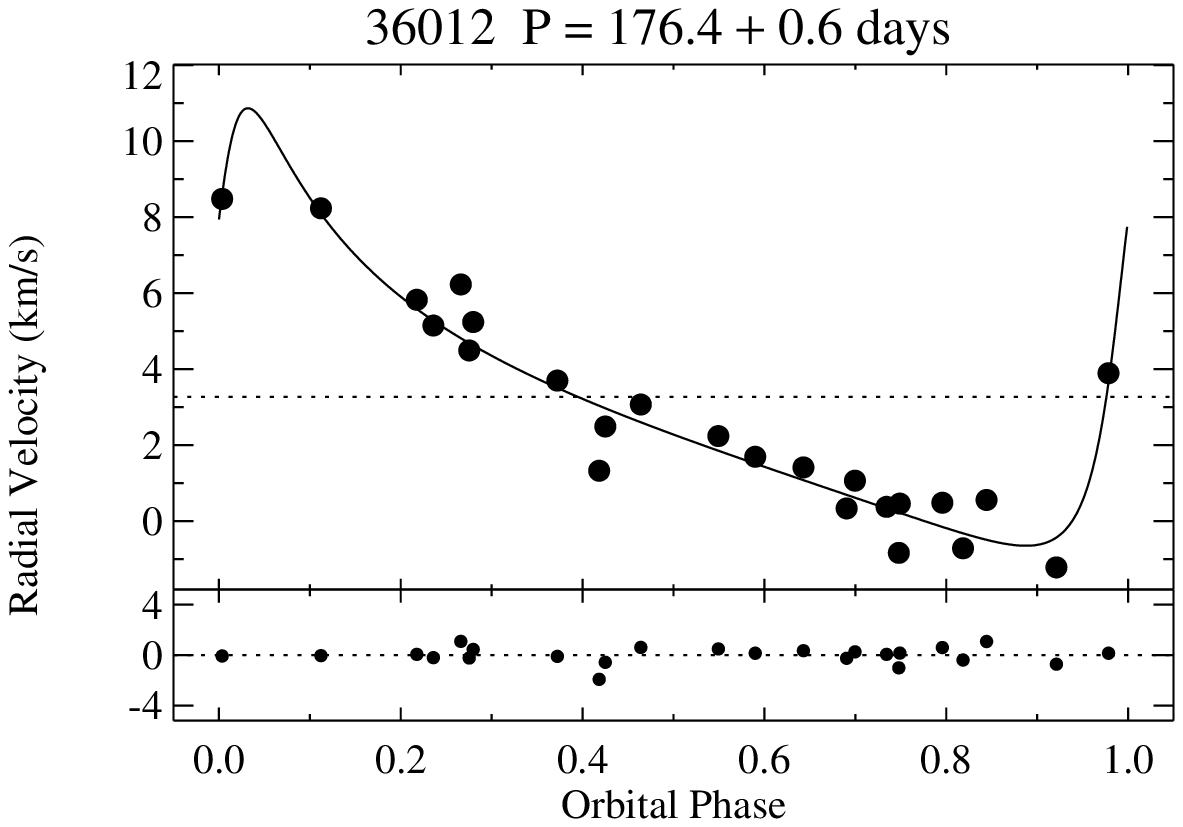}
\includegraphics[width=0.3\linewidth]{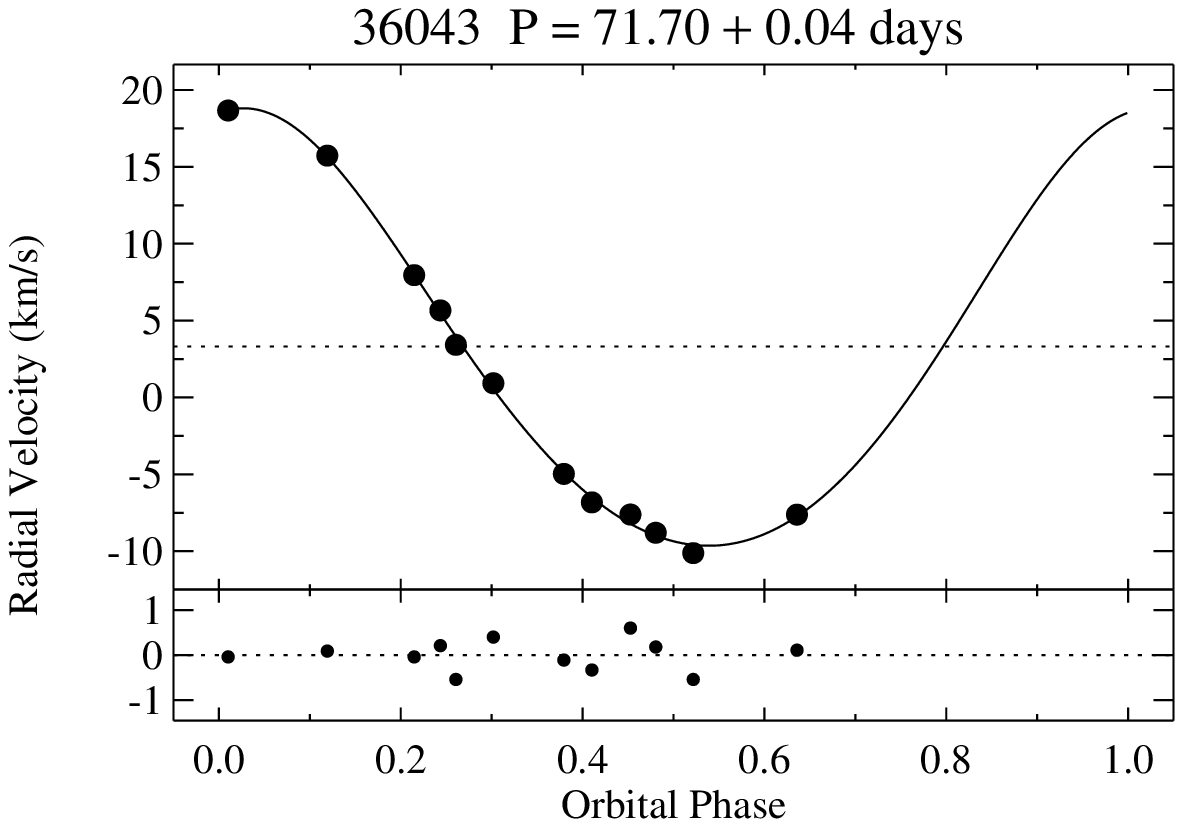}
\includegraphics[width=0.3\linewidth]{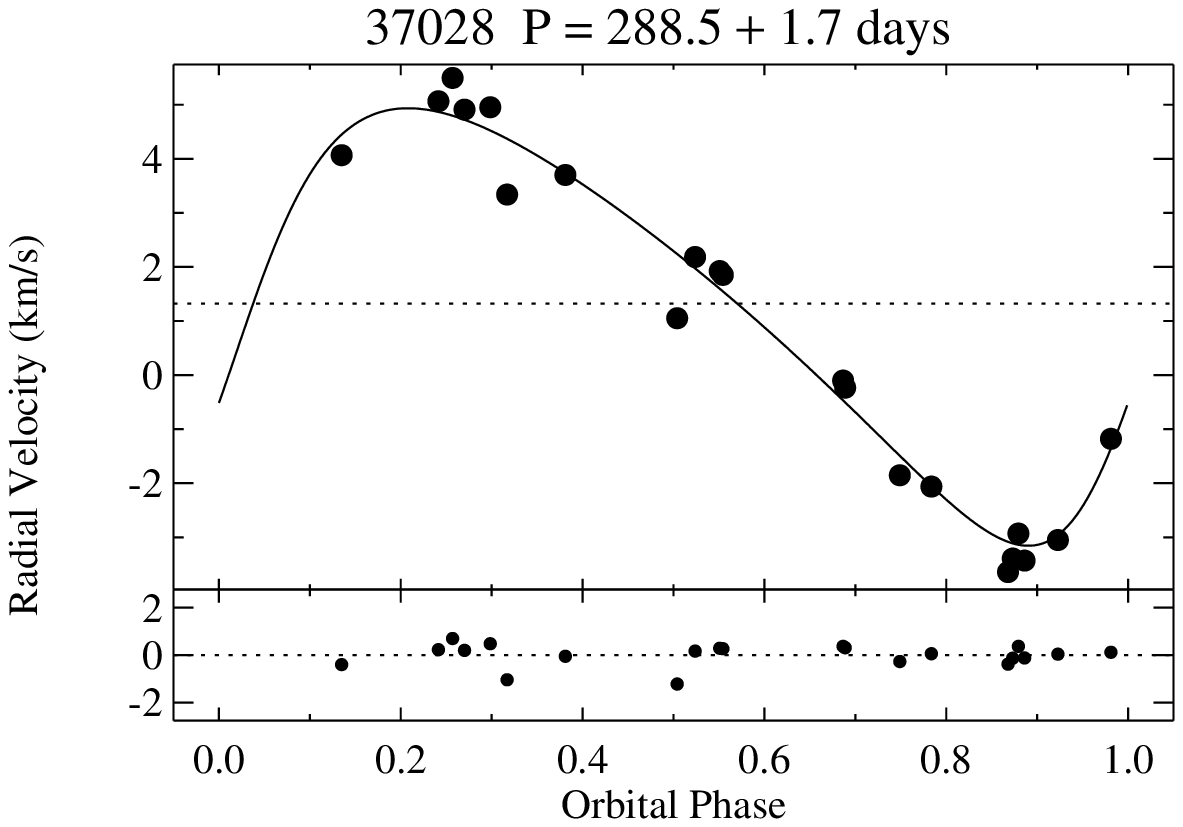}
\includegraphics[width=0.3\linewidth]{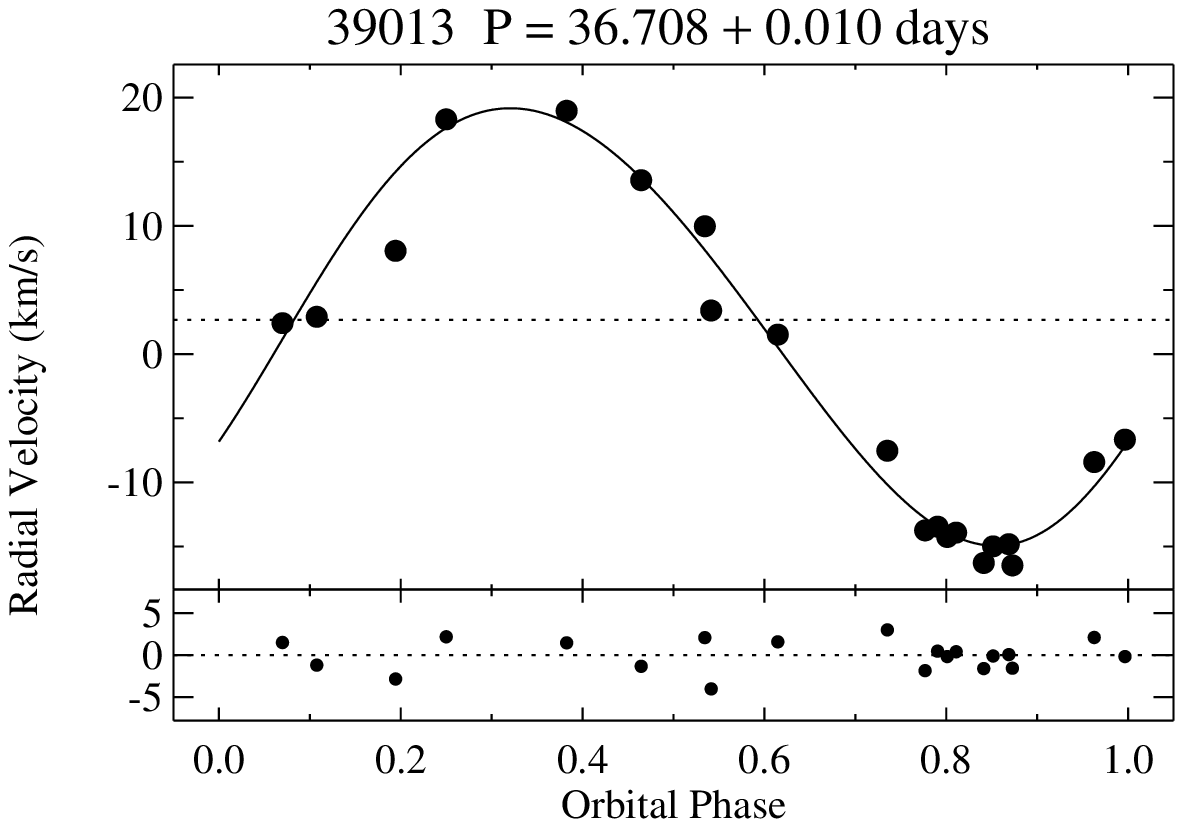}
\includegraphics[width=0.3\linewidth]{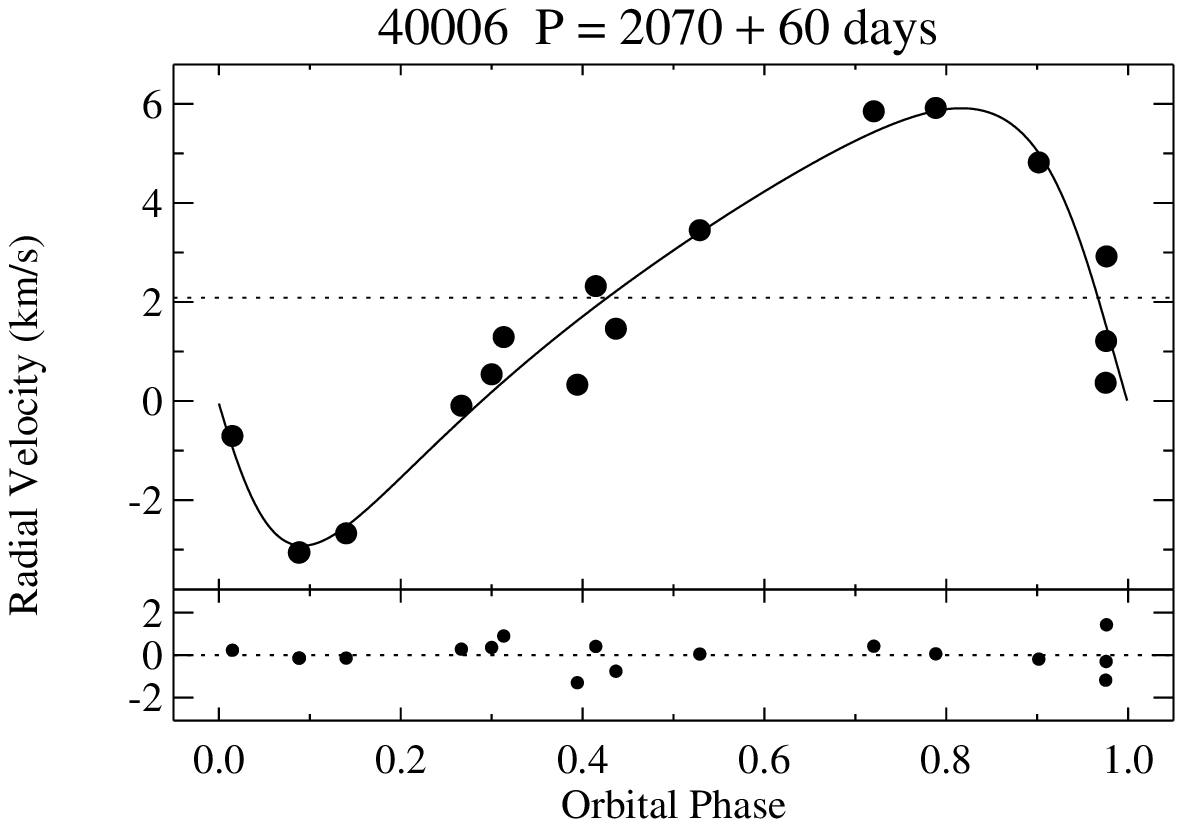}
\includegraphics[width=0.3\linewidth]{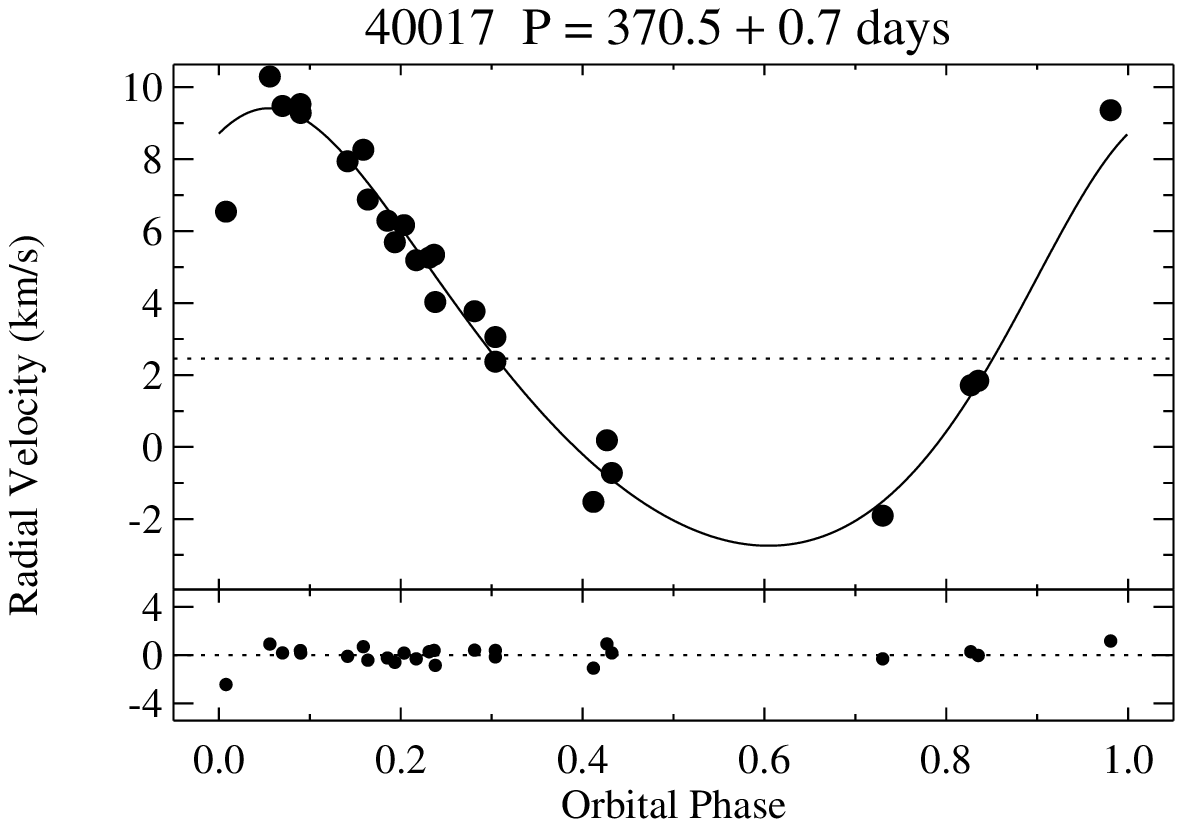}
\includegraphics[width=0.3\linewidth]{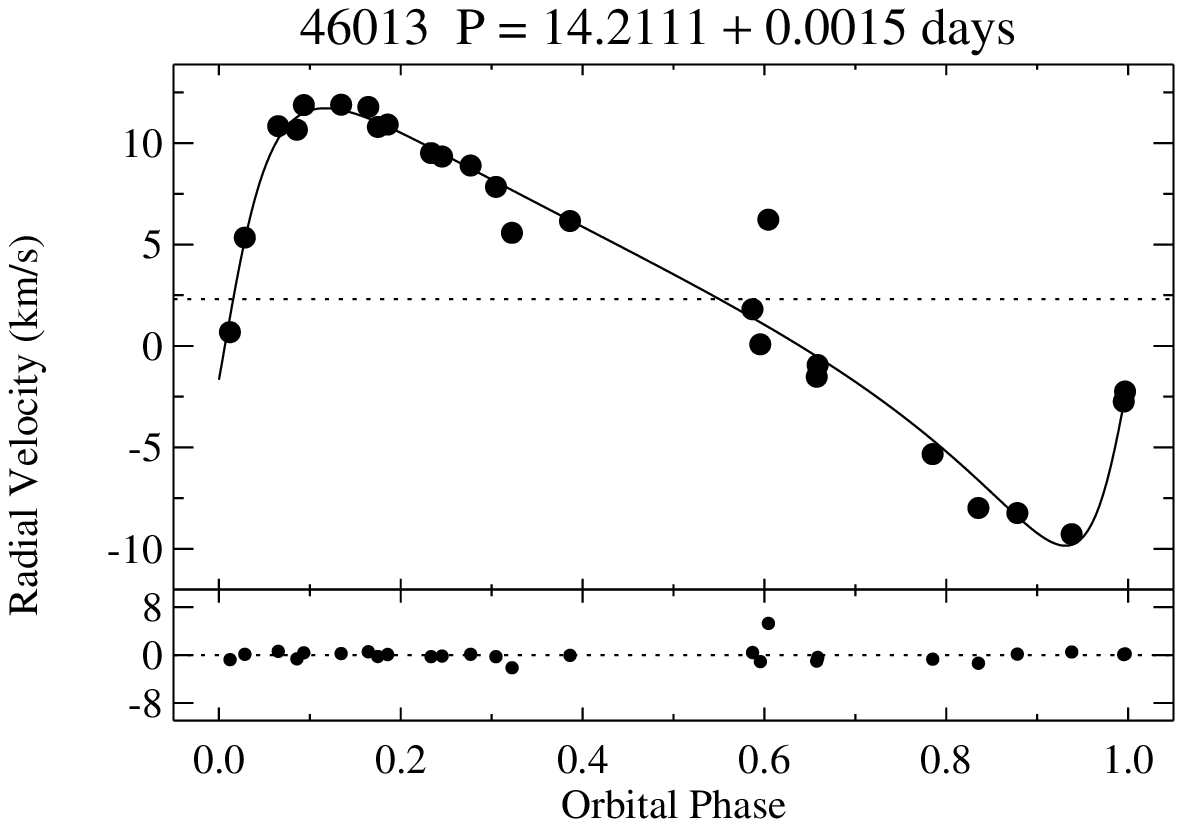}
\includegraphics[width=0.3\linewidth]{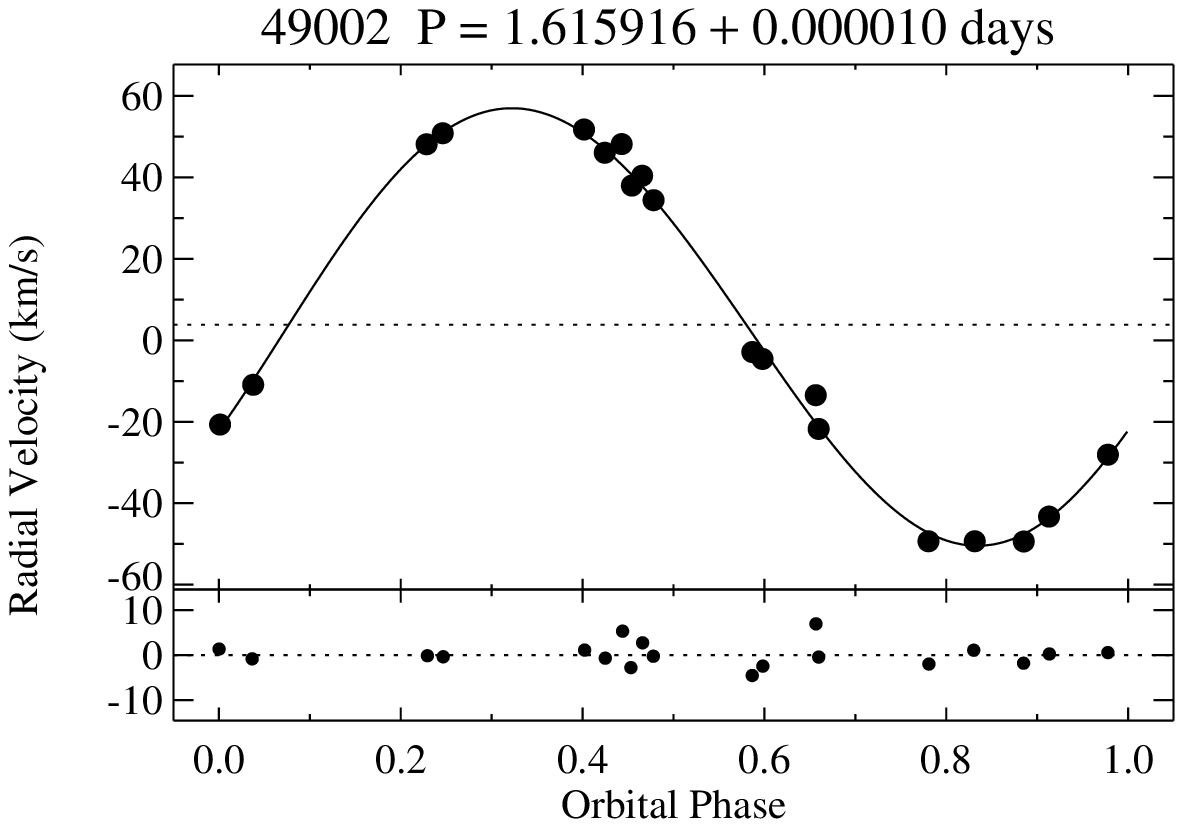}
\includegraphics[width=0.3\linewidth]{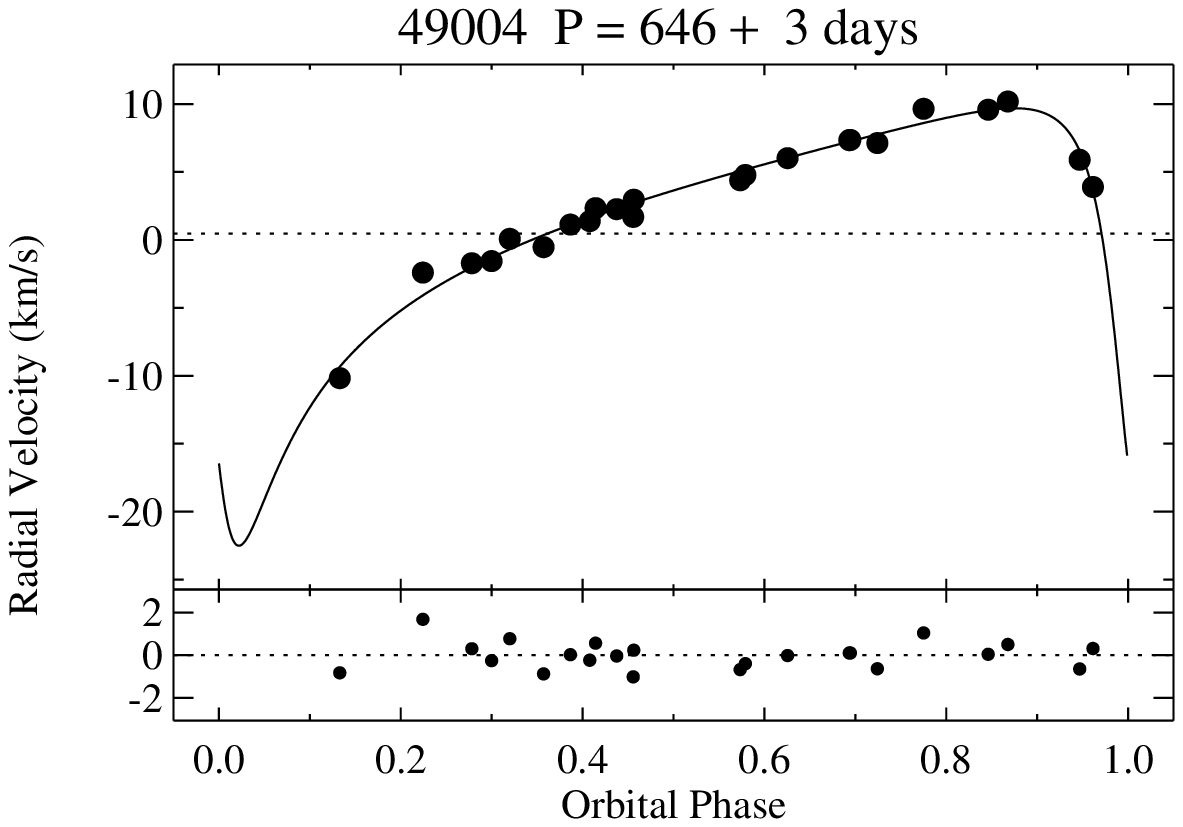}
\includegraphics[width=0.3\linewidth]{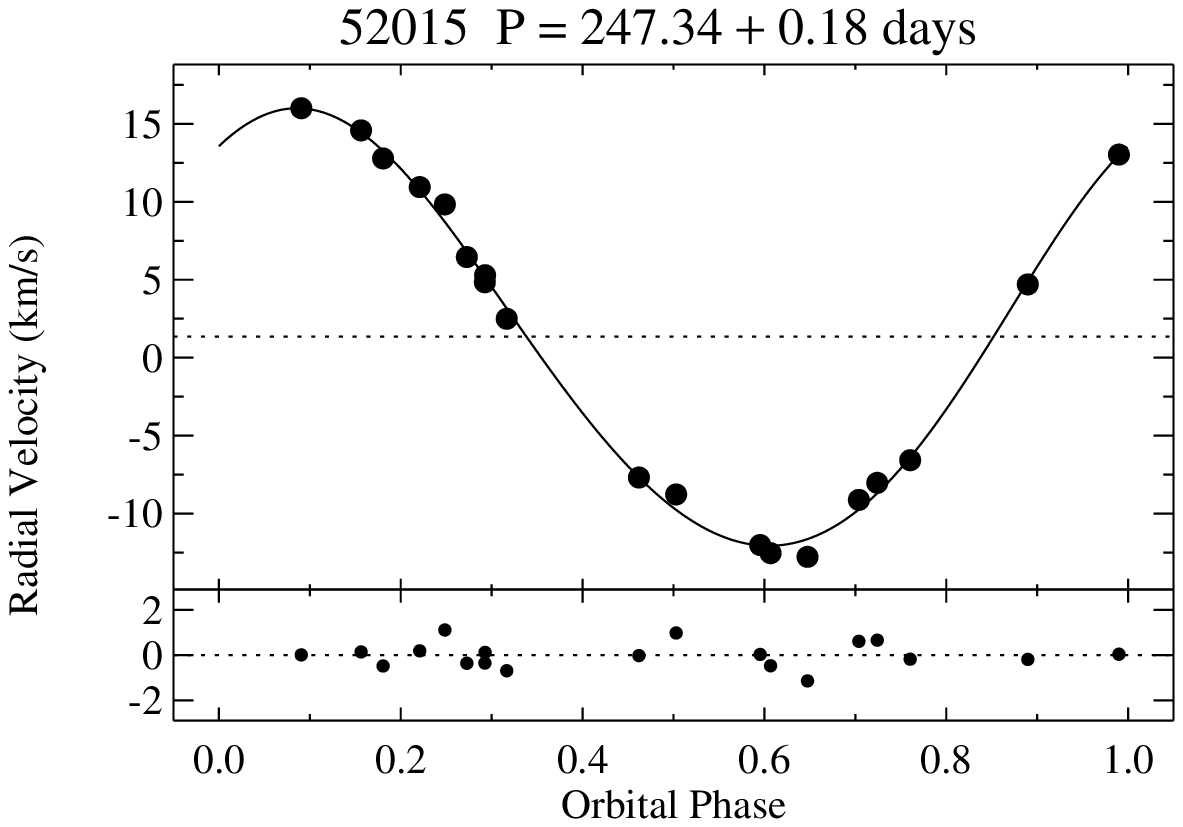}
\includegraphics[width=0.3\linewidth]{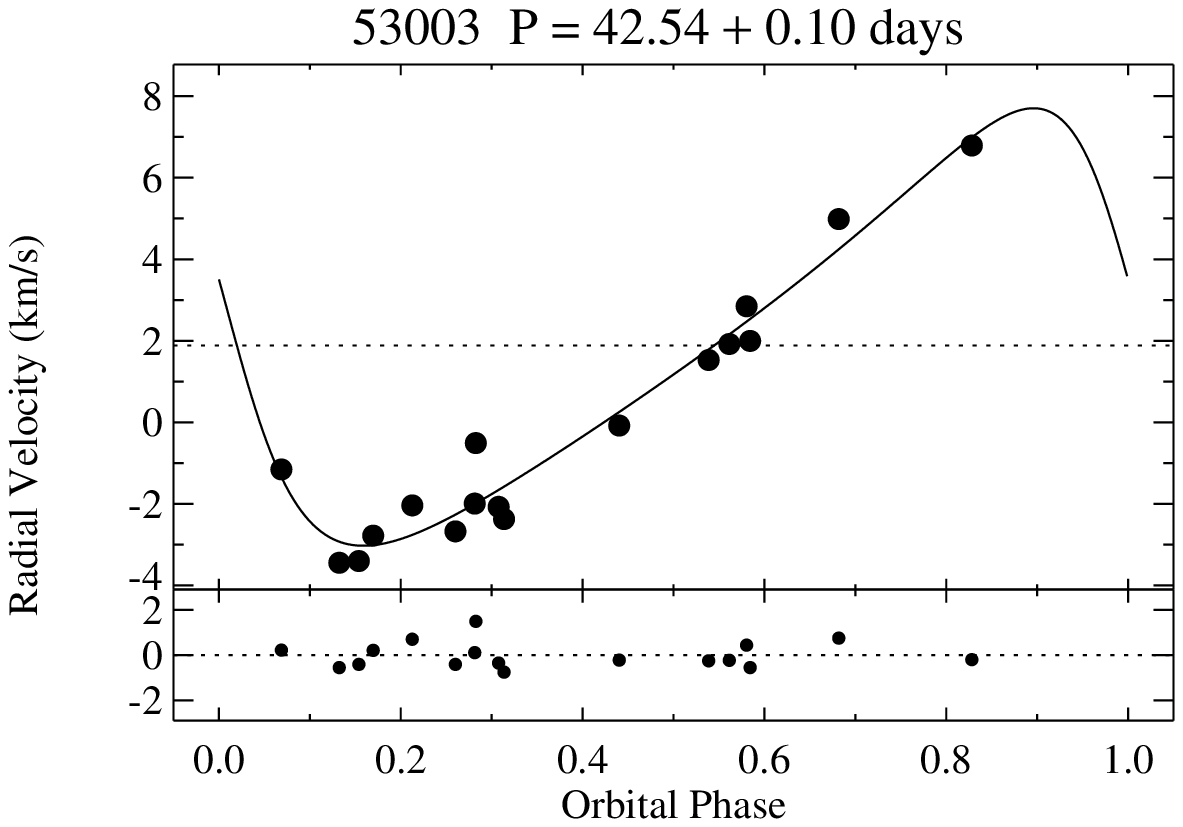}
\includegraphics[width=0.3\linewidth]{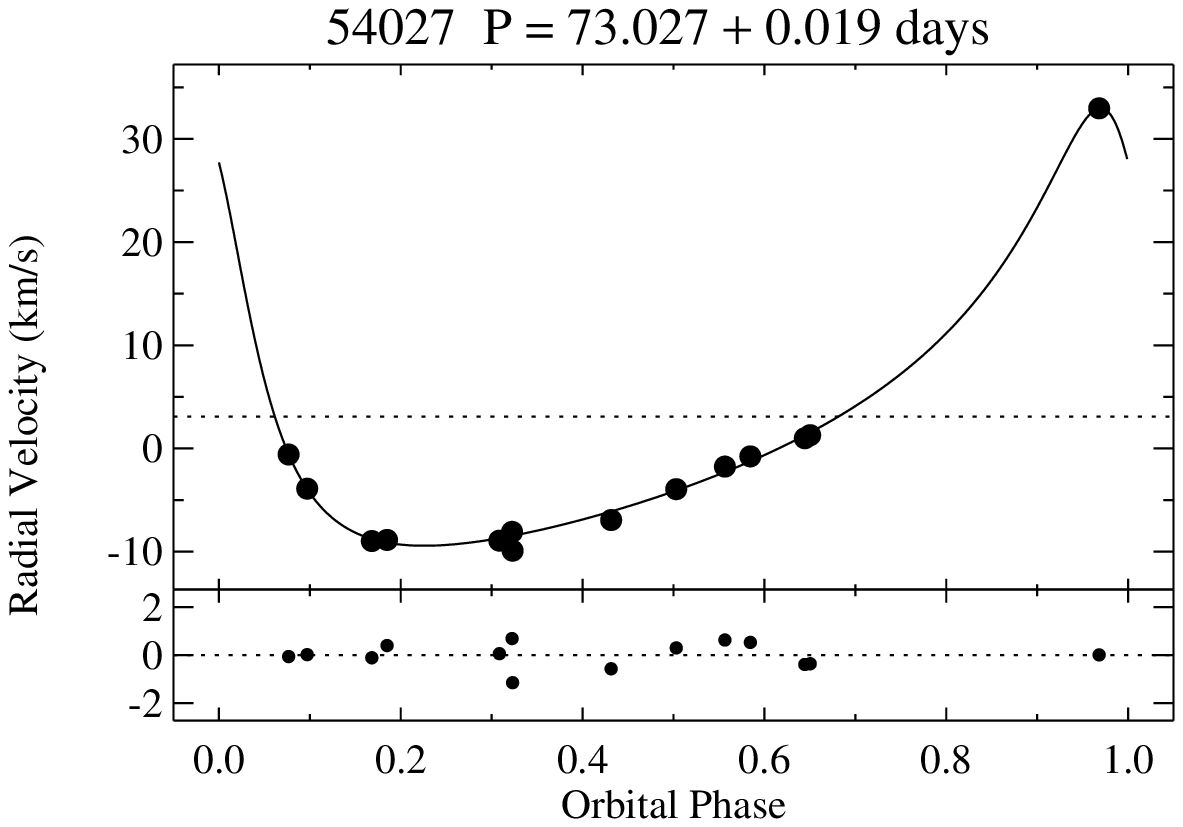}
\includegraphics[width=0.3\linewidth]{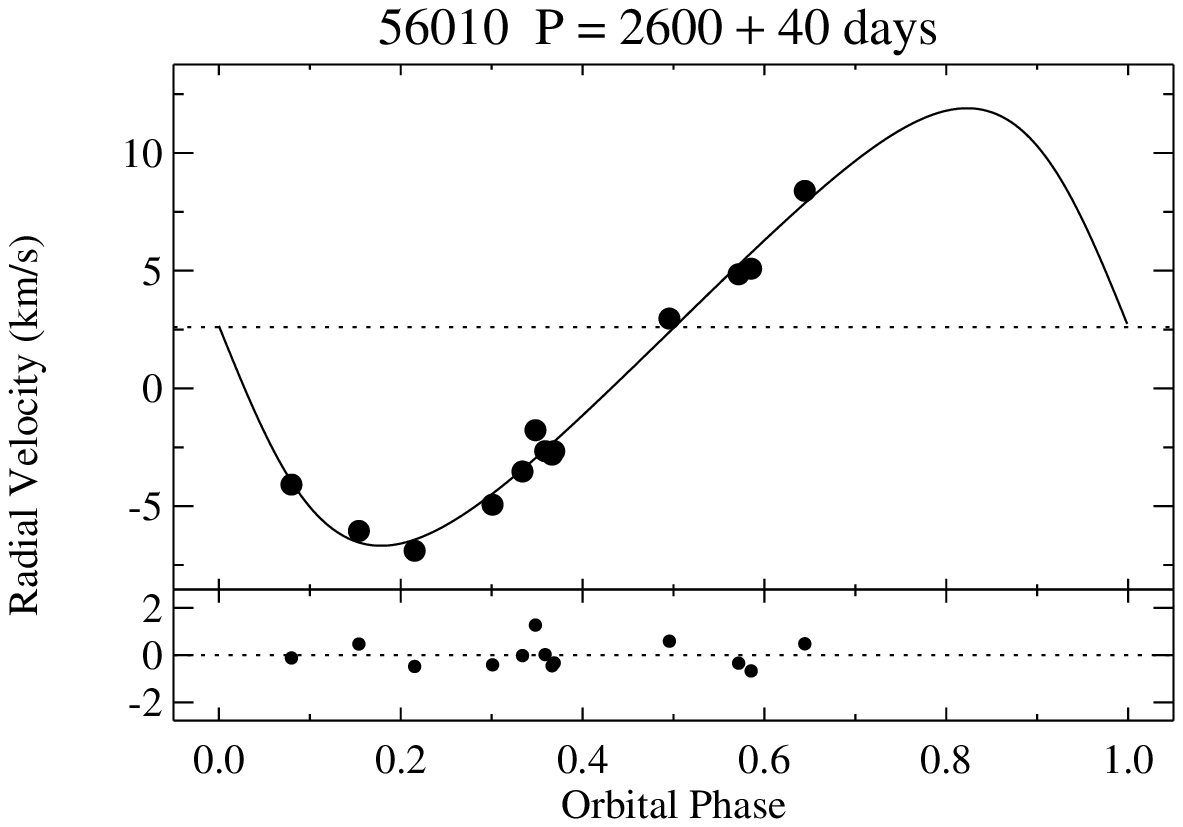}
\includegraphics[width=0.3\linewidth]{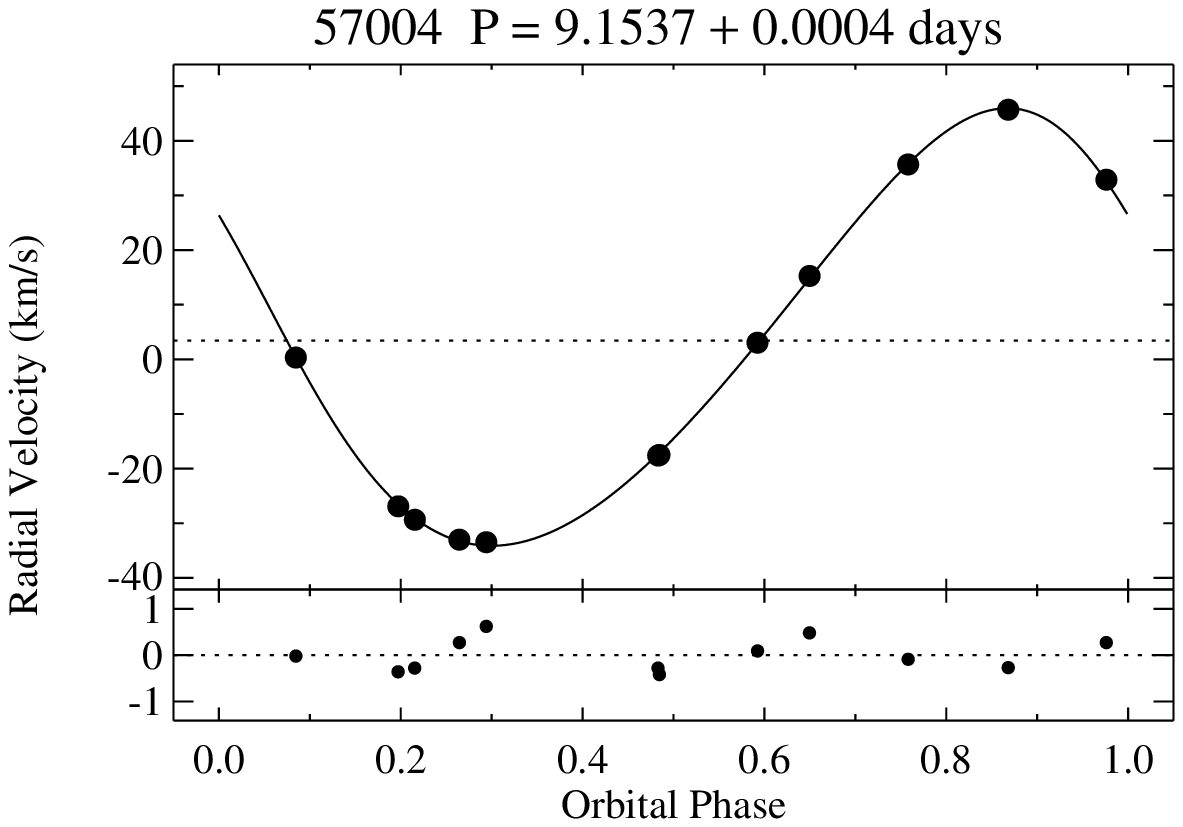}
\includegraphics[width=0.3\linewidth]{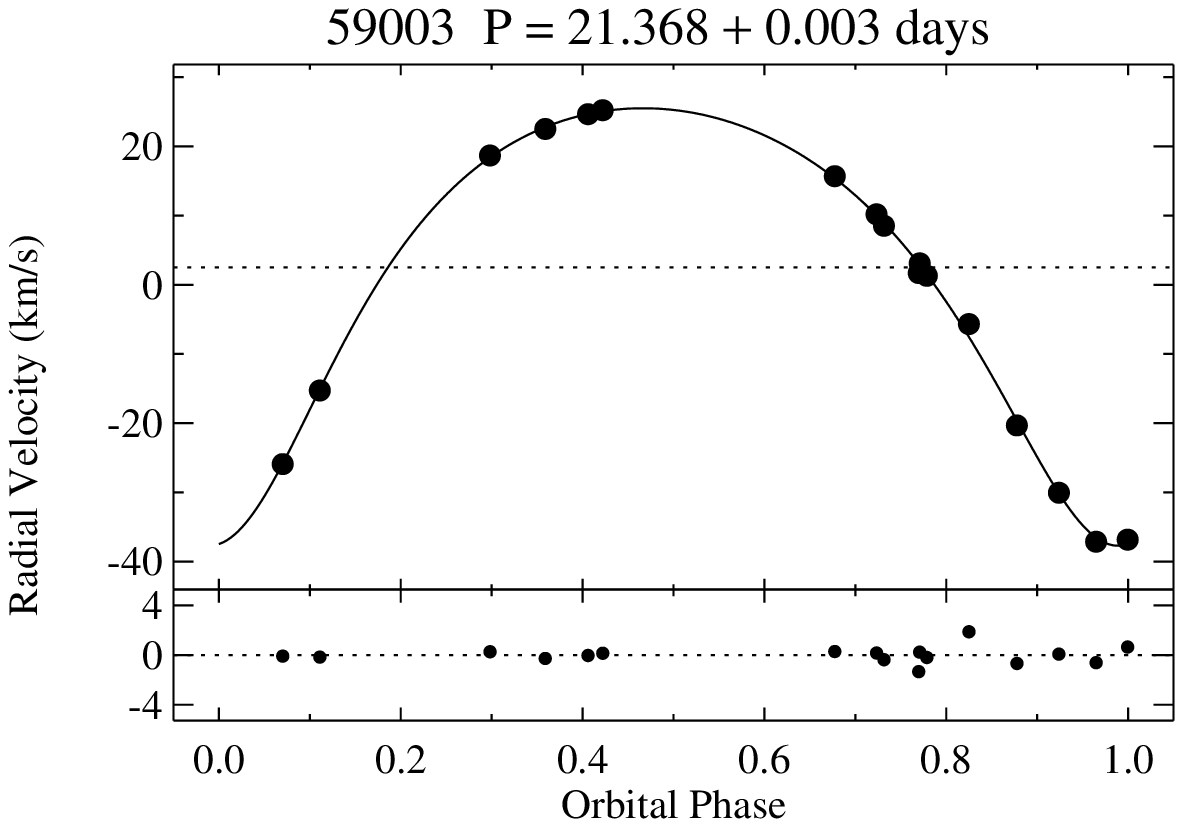}

{\bf{Figure 7.} (Continued)}
\end{center}
\end{figure*}
\begin{figure*}
\begin{center}
\includegraphics[width=0.3\linewidth]{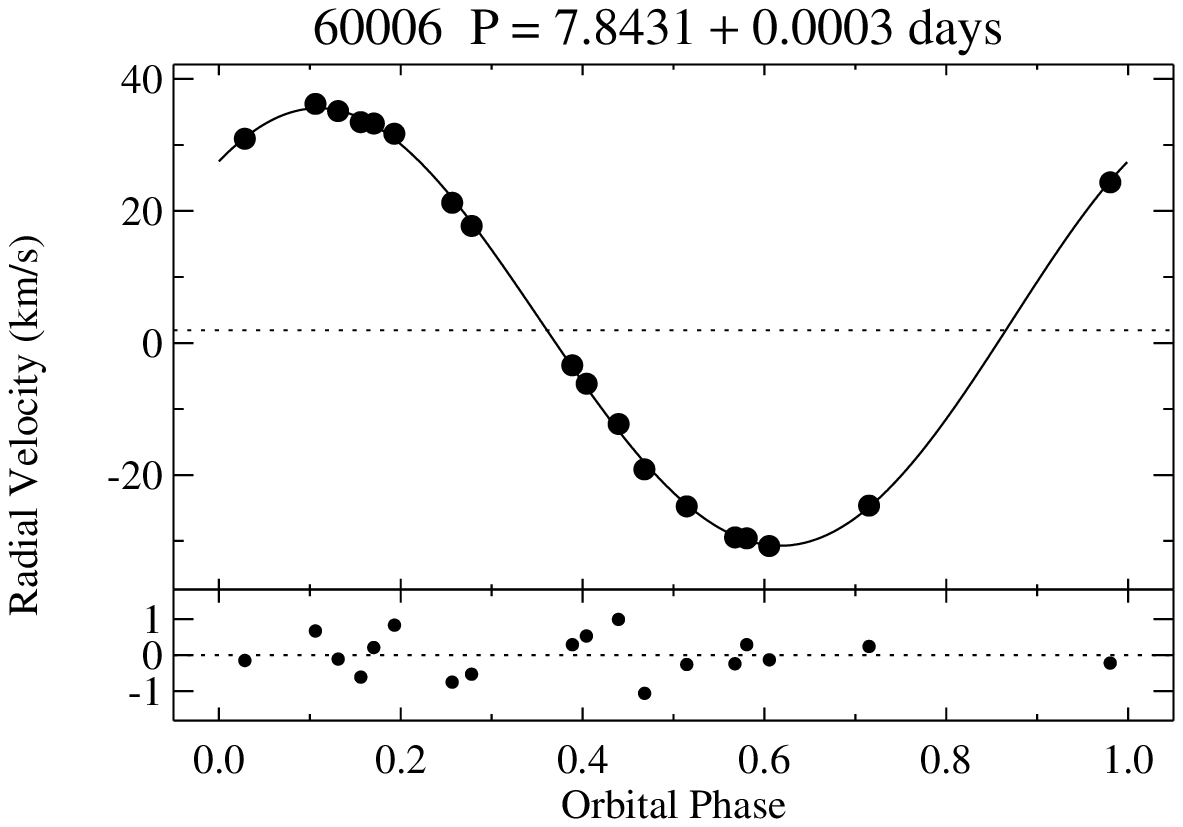}
\includegraphics[width=0.3\linewidth]{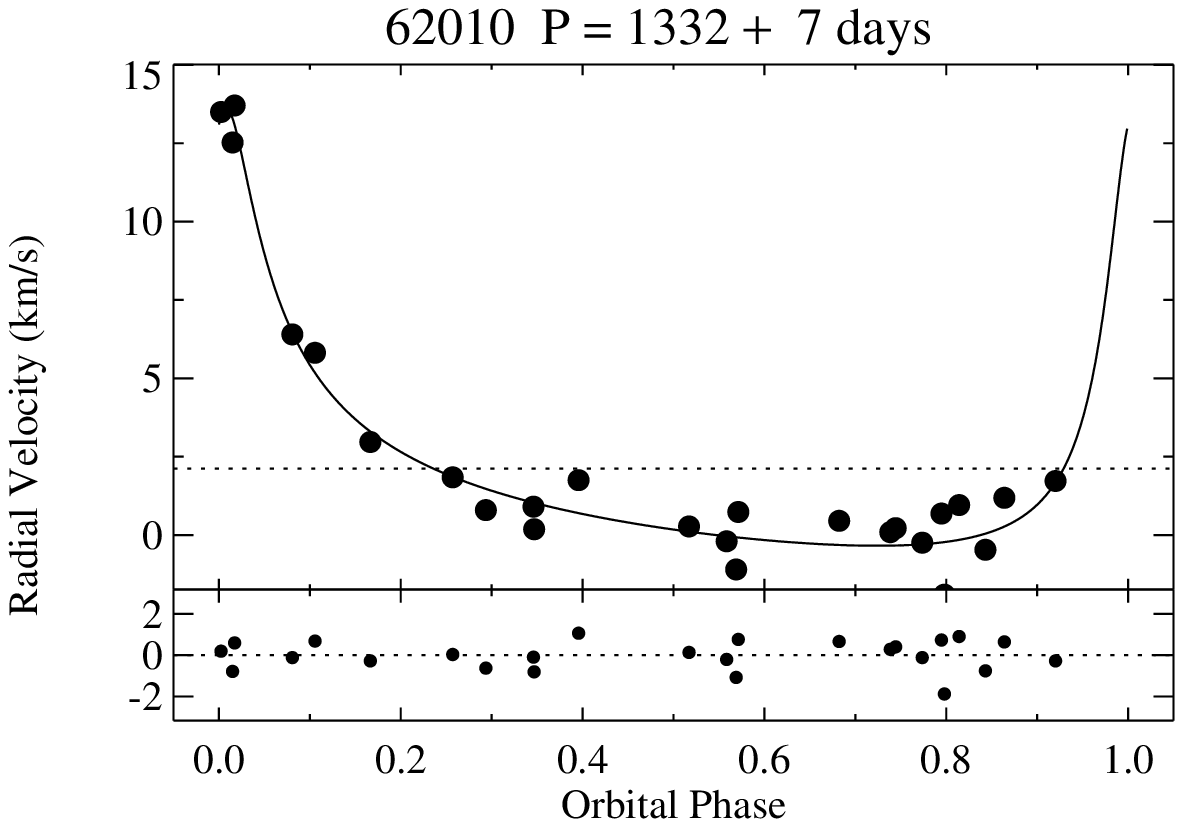}
\includegraphics[width=0.3\linewidth]{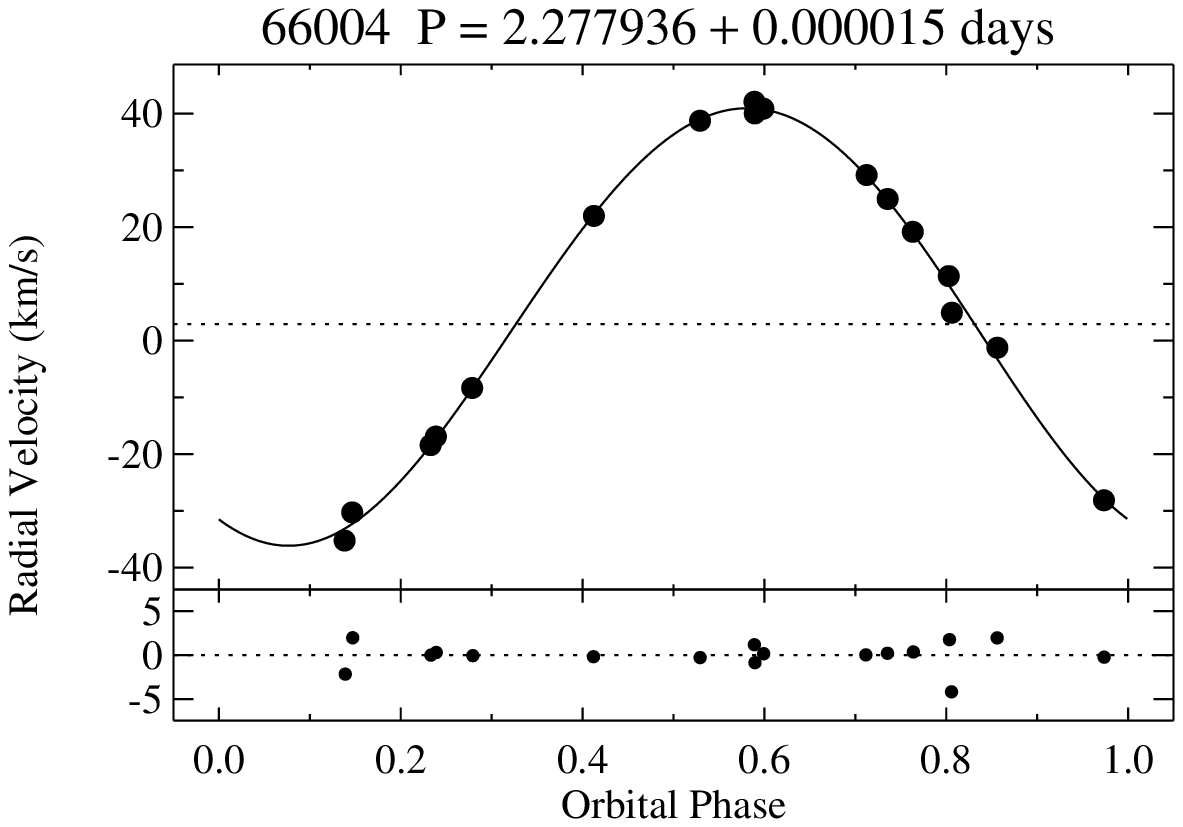}
\includegraphics[width=0.3\linewidth]{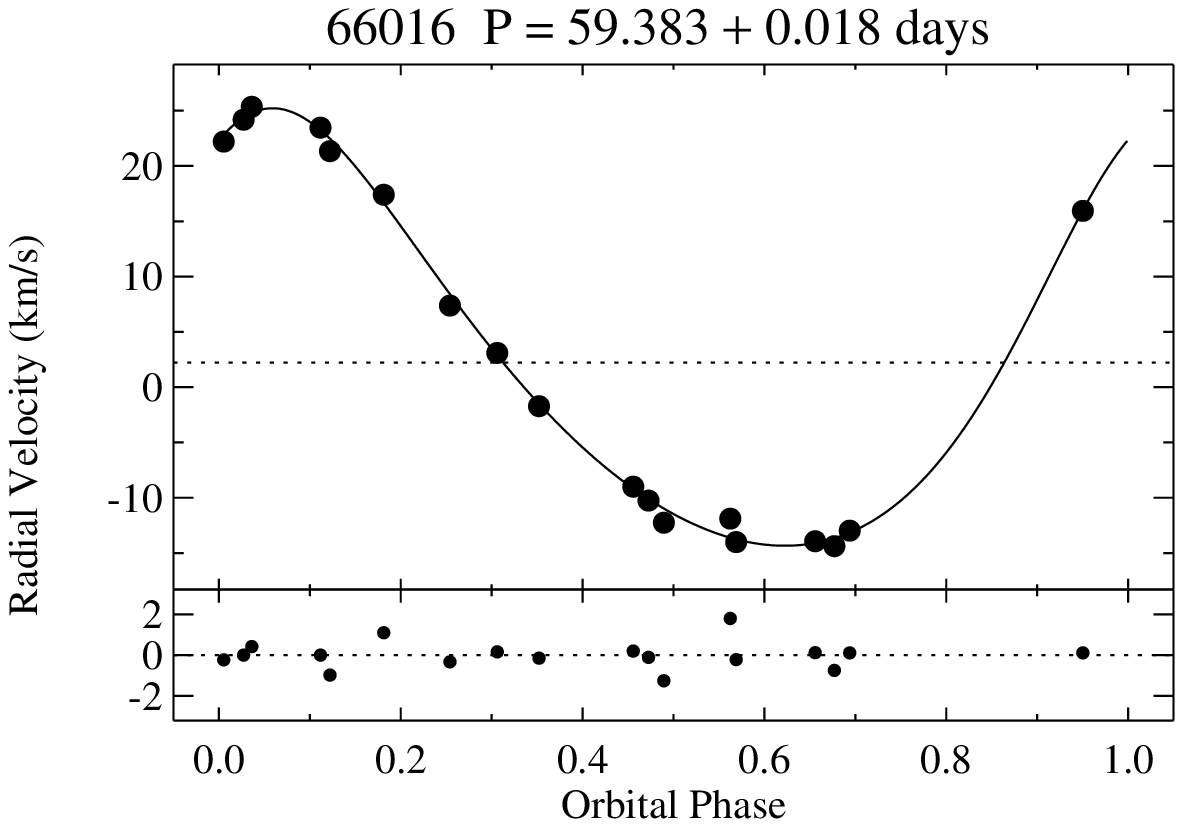}
\includegraphics[width=0.3\linewidth]{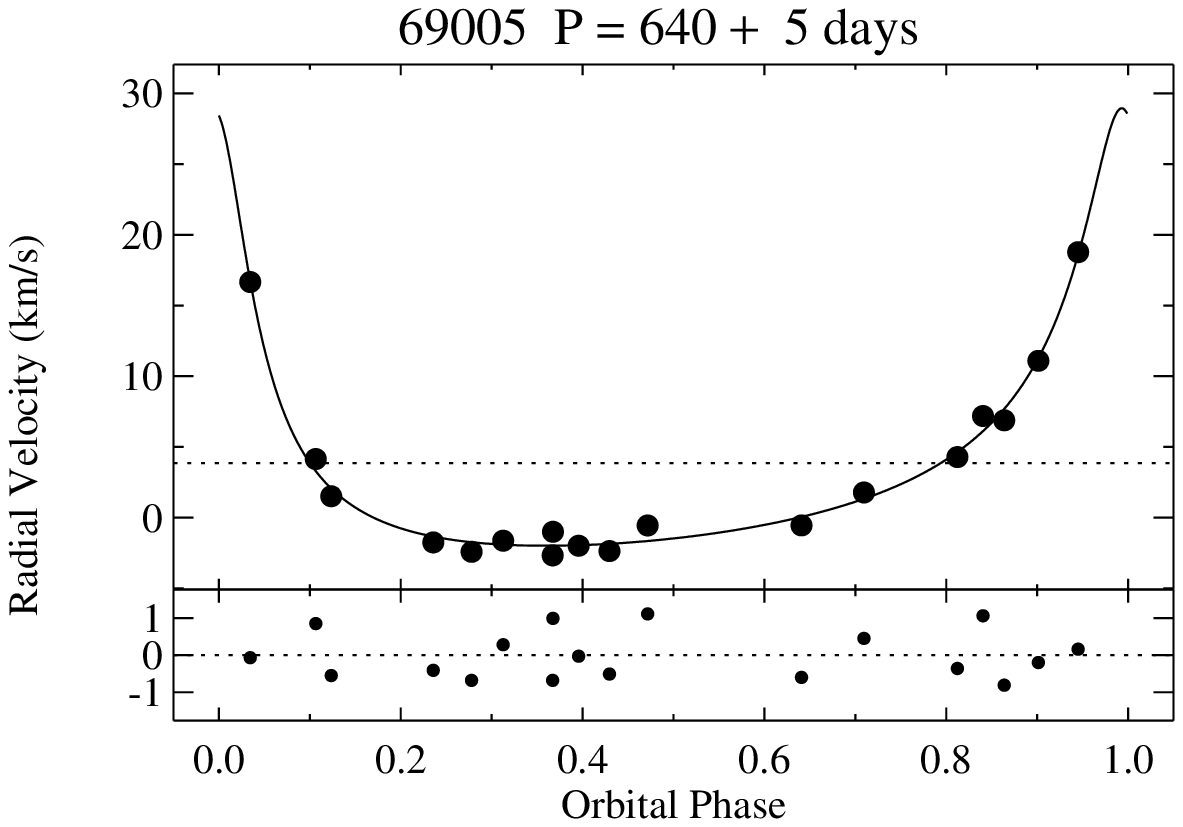}
\includegraphics[width=0.3\linewidth]{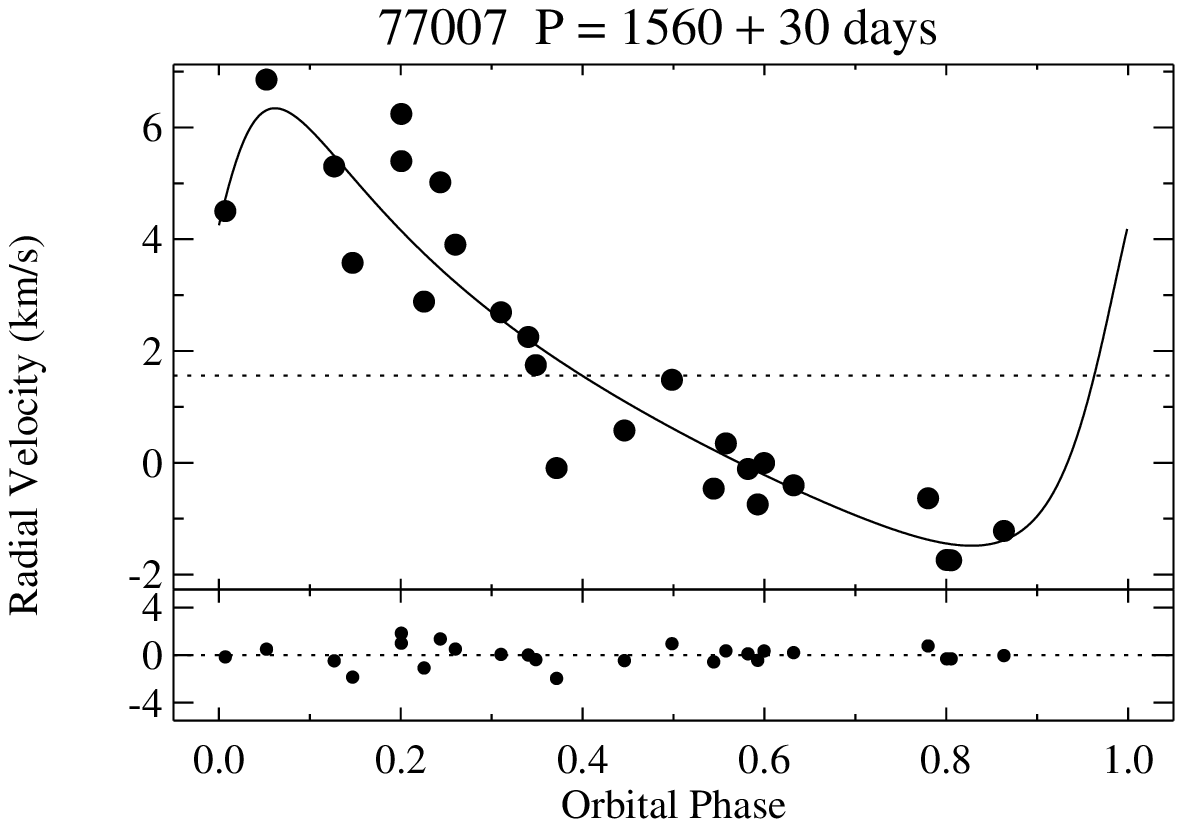}

{\bf{Figure 7.} (Continued)}
\end{center}
\end{figure*}
\begin{figure*}
\begin{center}
\subfigure{\includegraphics[width=0.3\linewidth]{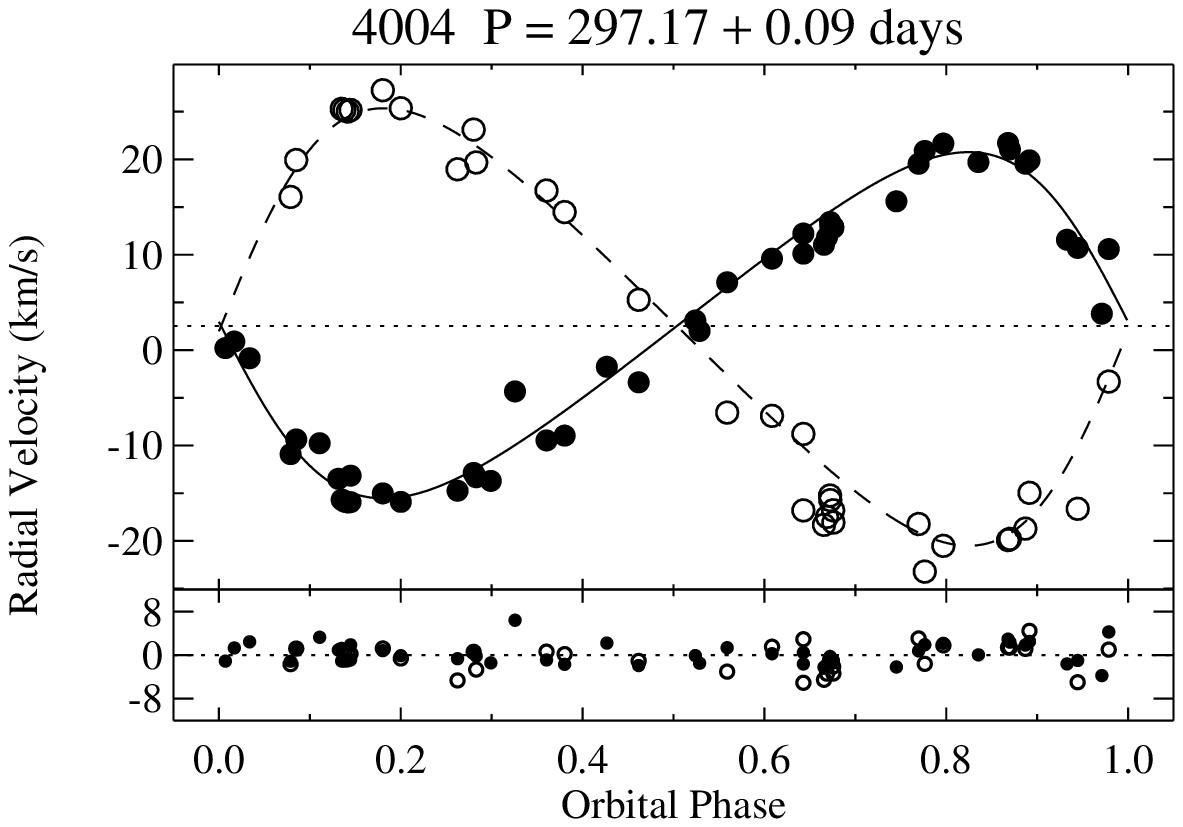}}
\subfigure{\includegraphics[width=0.3\linewidth]{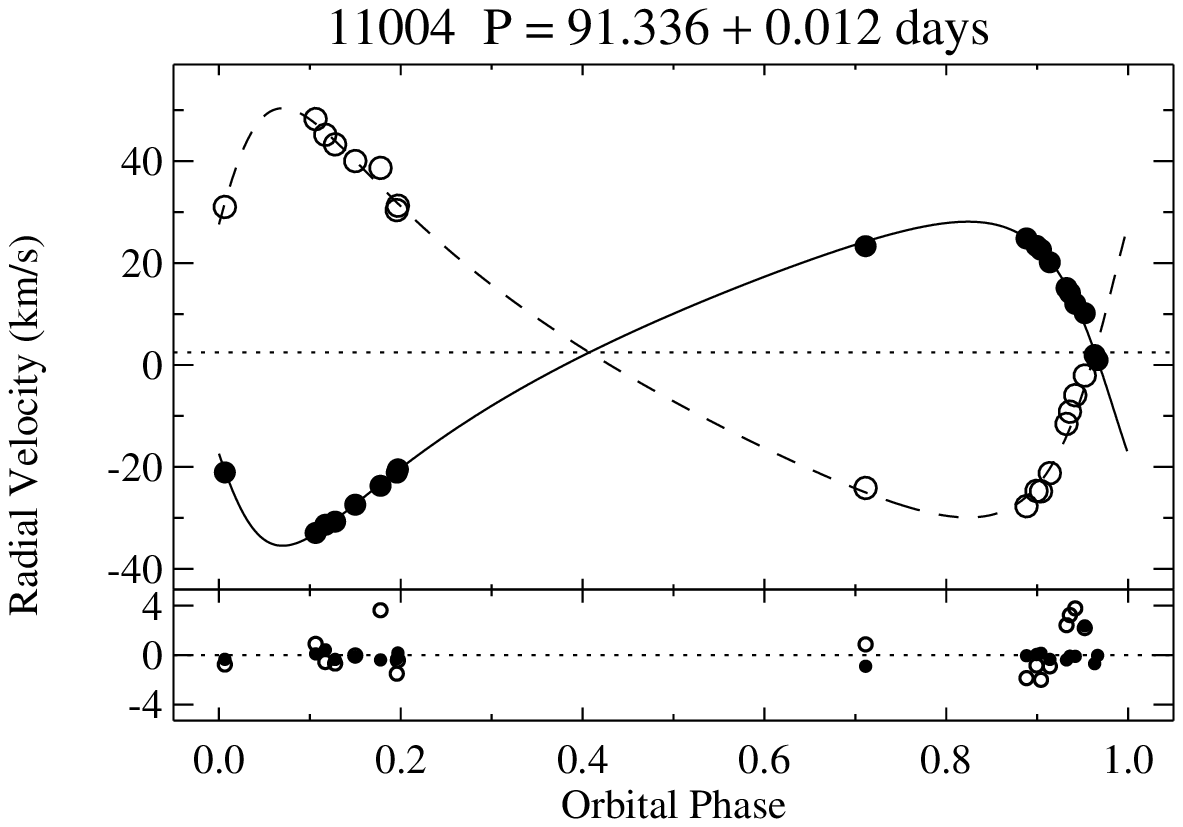}}
\subfigure{\includegraphics[width=0.3\linewidth]{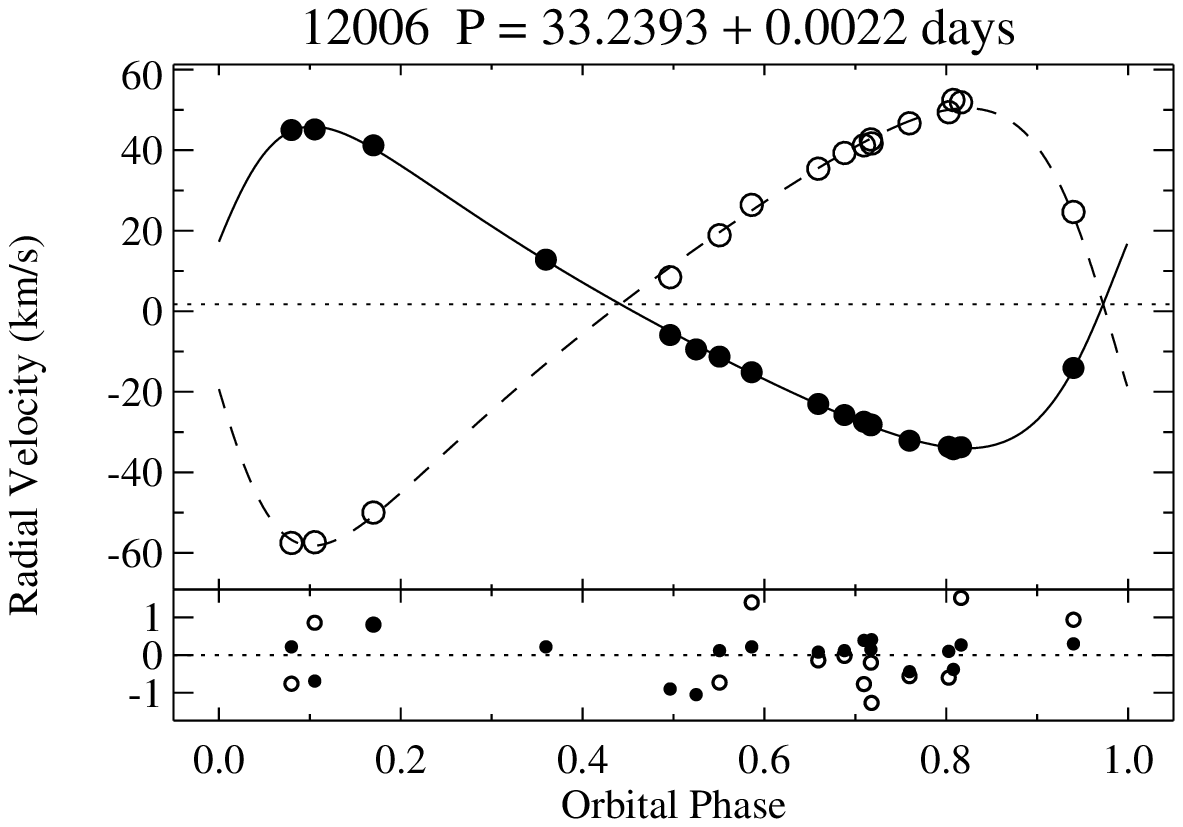}}
\subfigure{\includegraphics[width=0.3\linewidth]{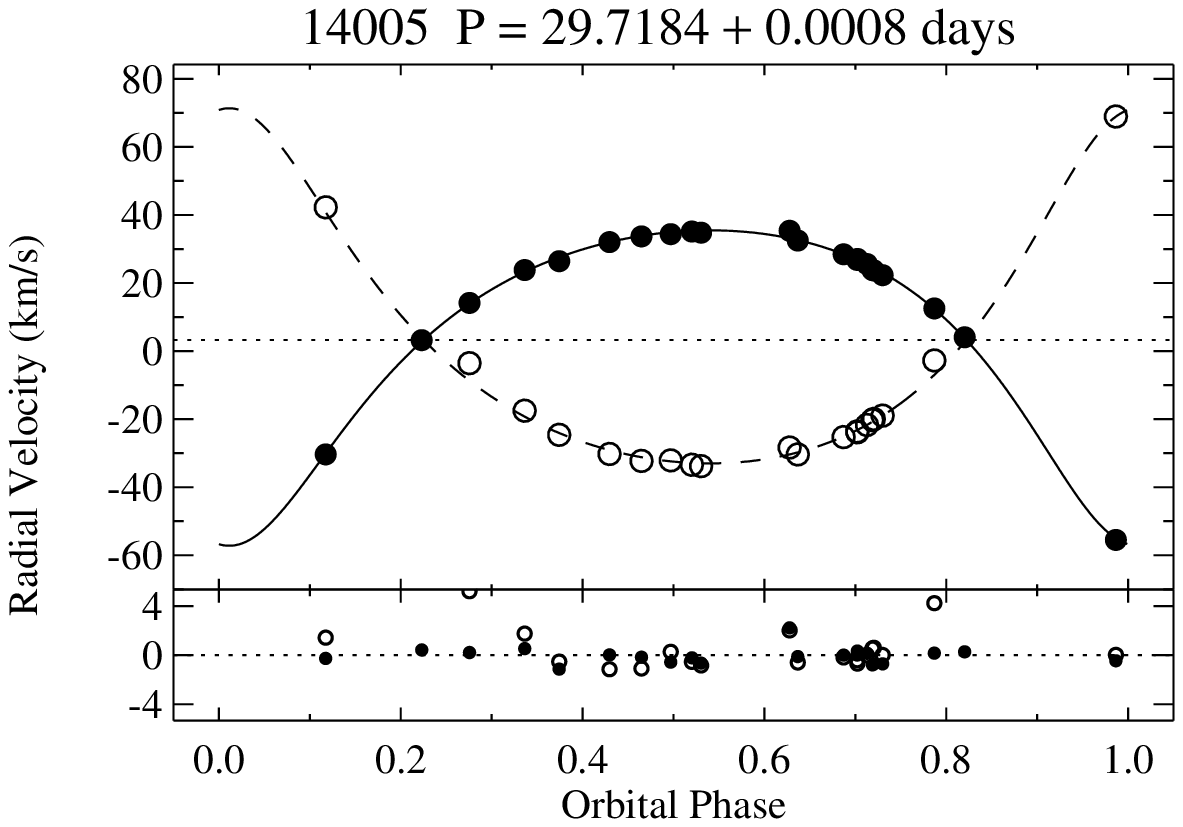}}
\subfigure{\includegraphics[width=0.3\linewidth]{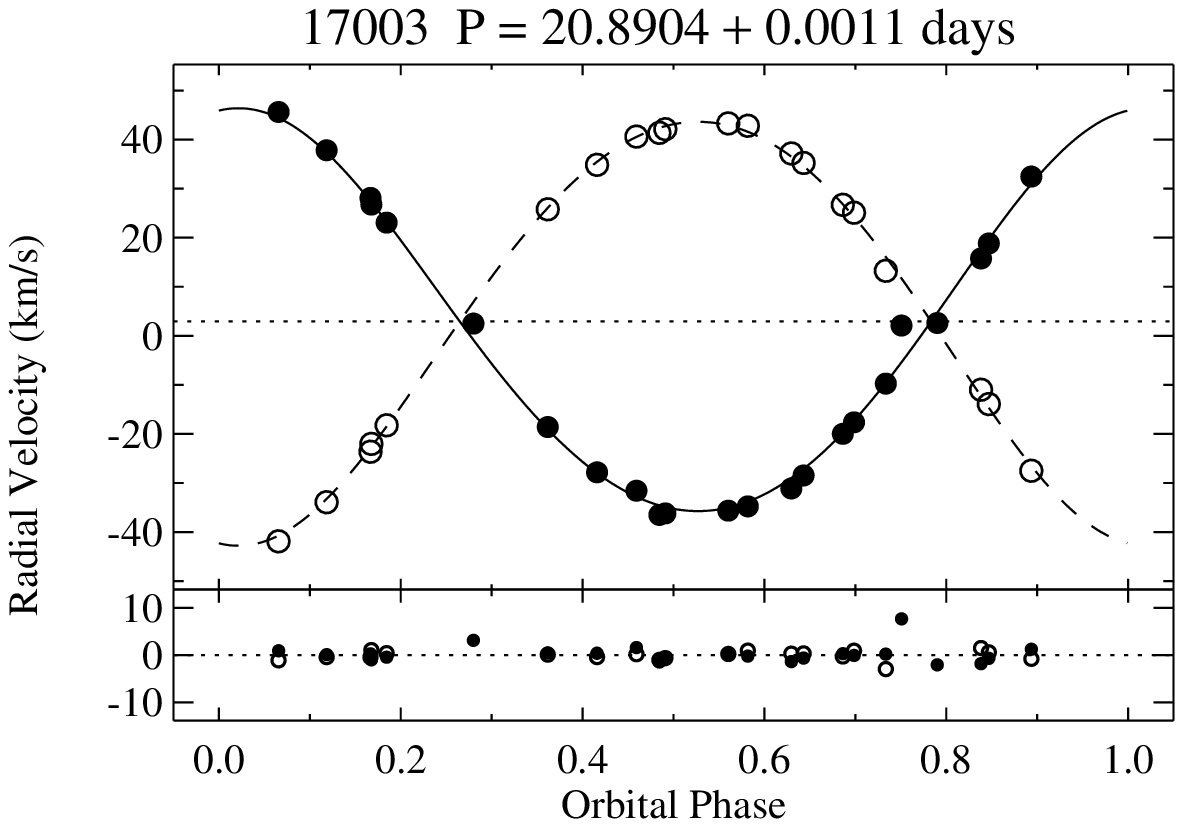}}
\subfigure{\includegraphics[width=0.3\linewidth]{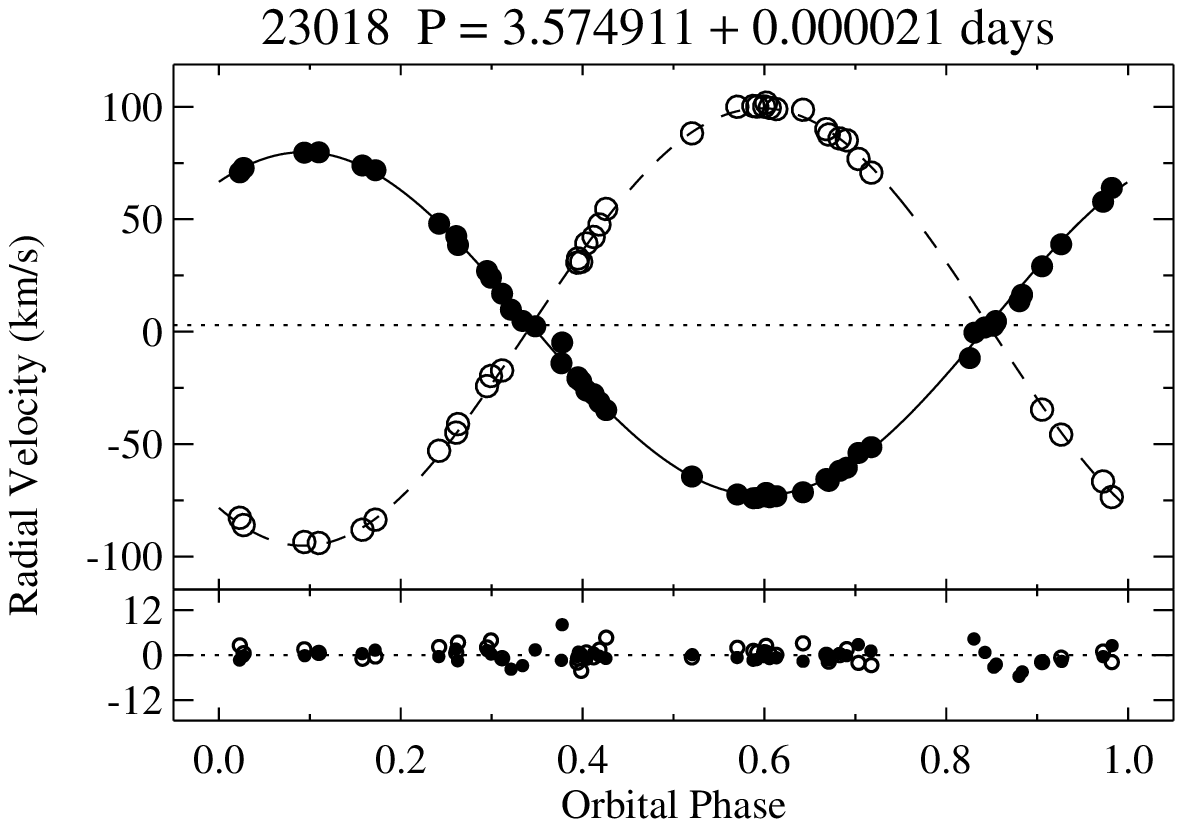}}
\subfigure{\includegraphics[width=0.3\linewidth]{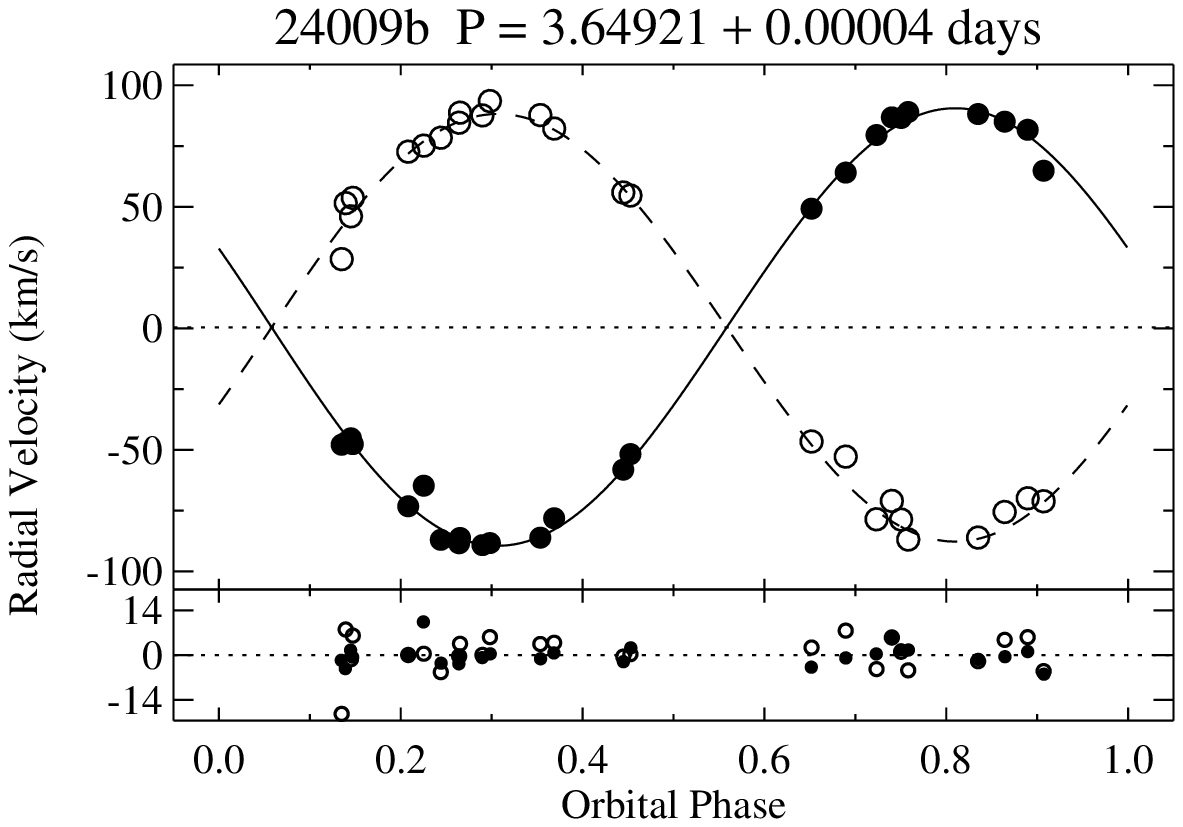}}
\subfigure{\includegraphics[width=0.3\linewidth]{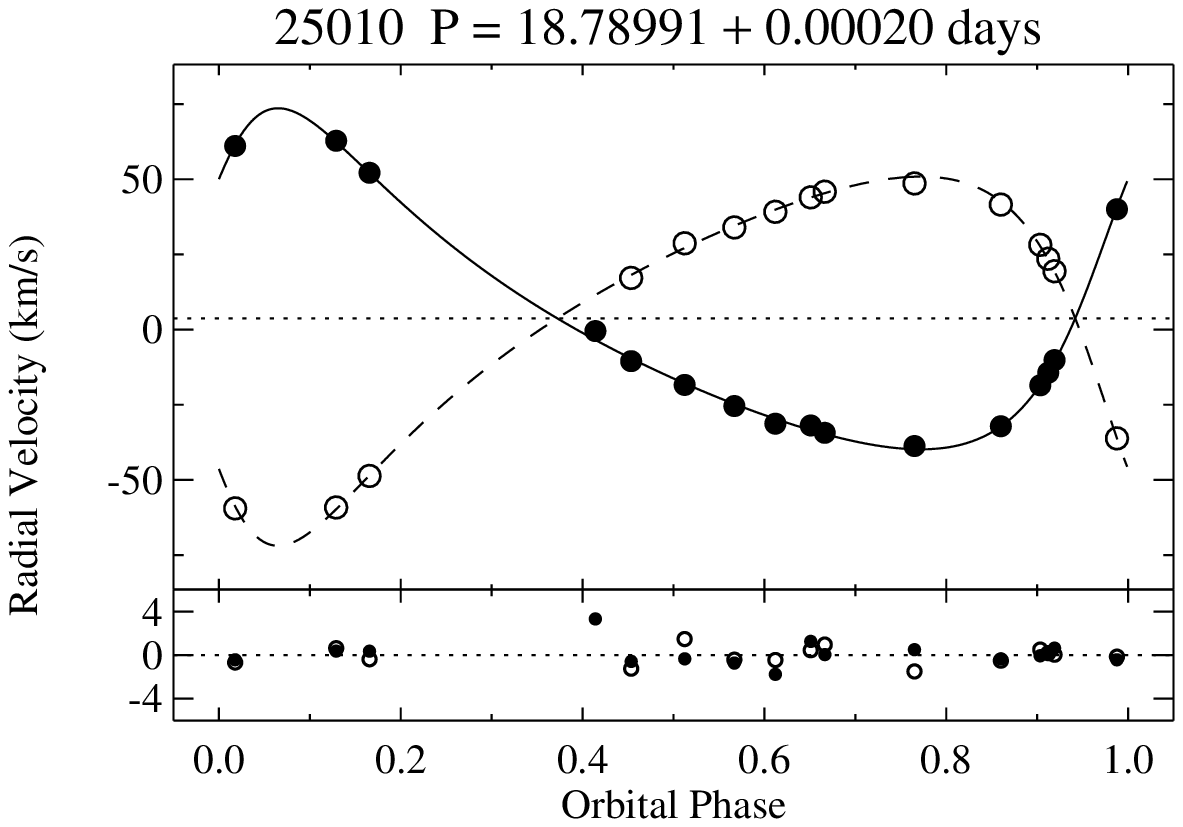}}
\subfigure{\includegraphics[width=0.3\linewidth]{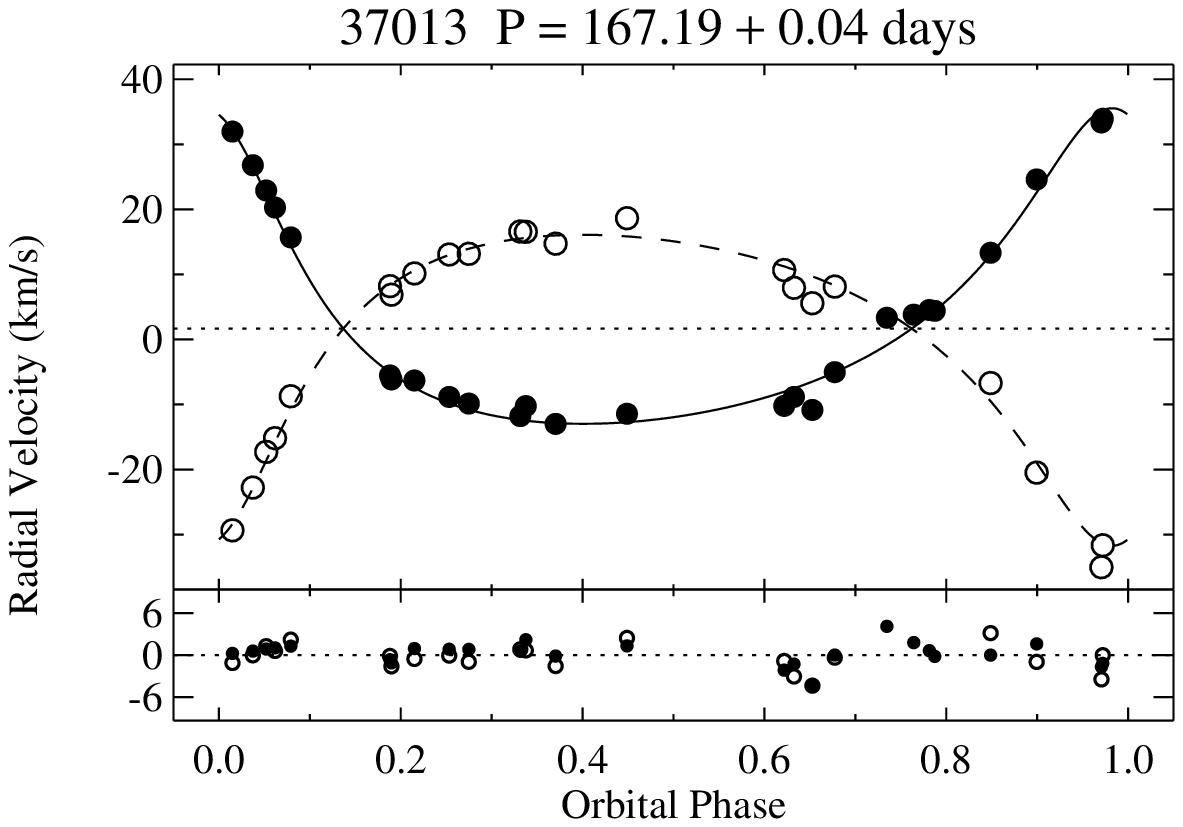}}
\subfigure{\includegraphics[width=0.3\linewidth]{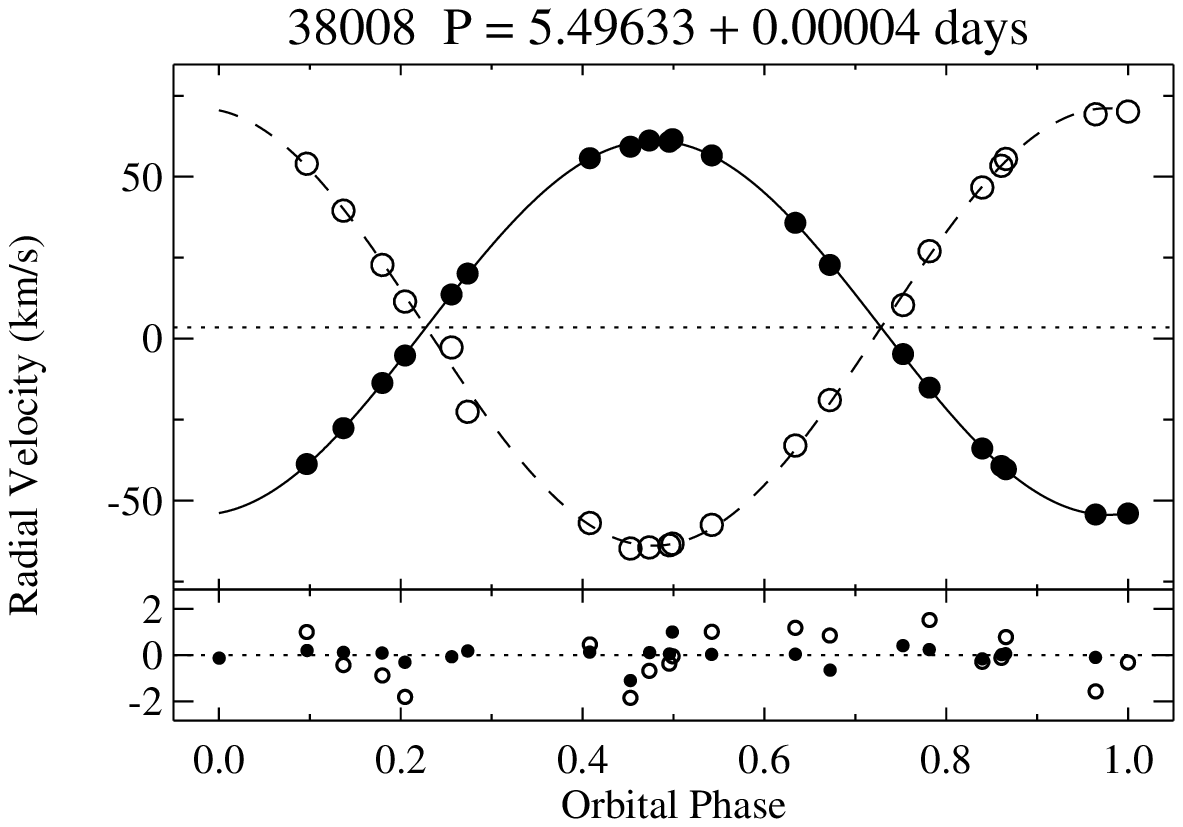}}
\subfigure{\includegraphics[width=0.3\linewidth]{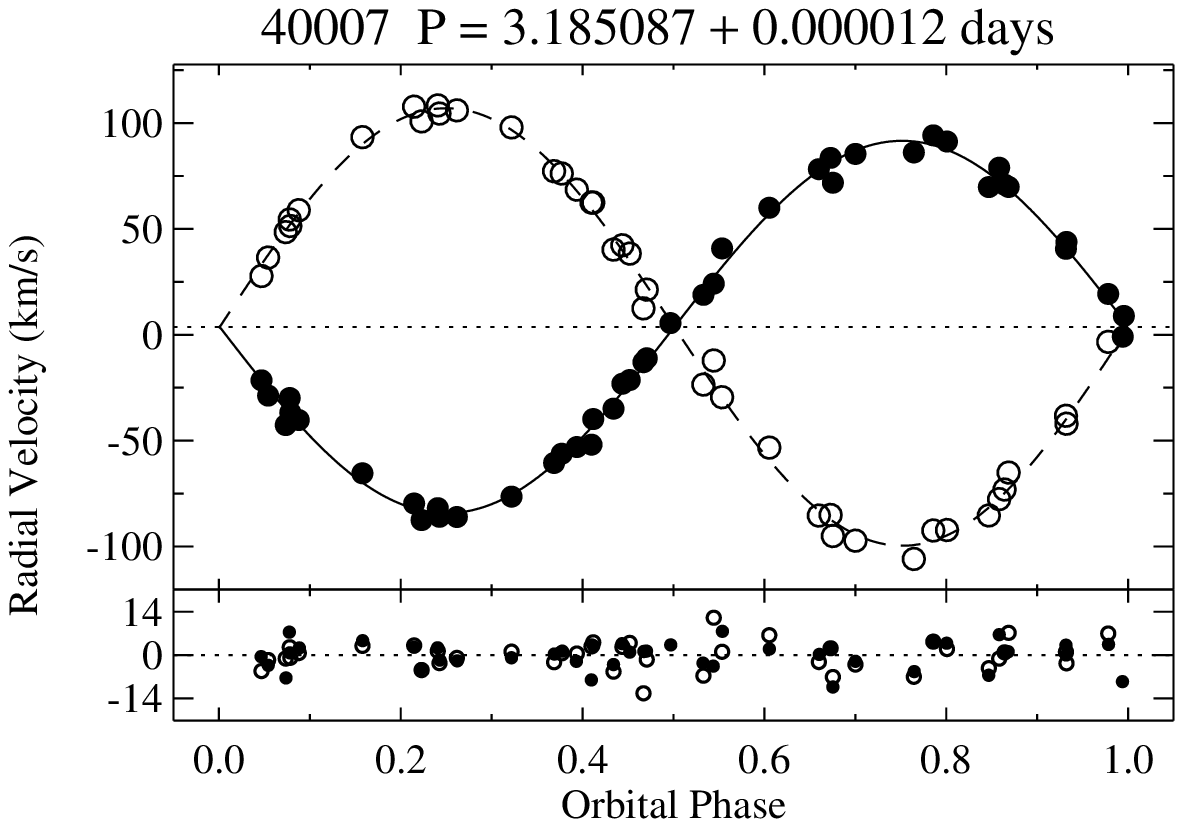}}
\subfigure{\includegraphics[width=0.3\linewidth]{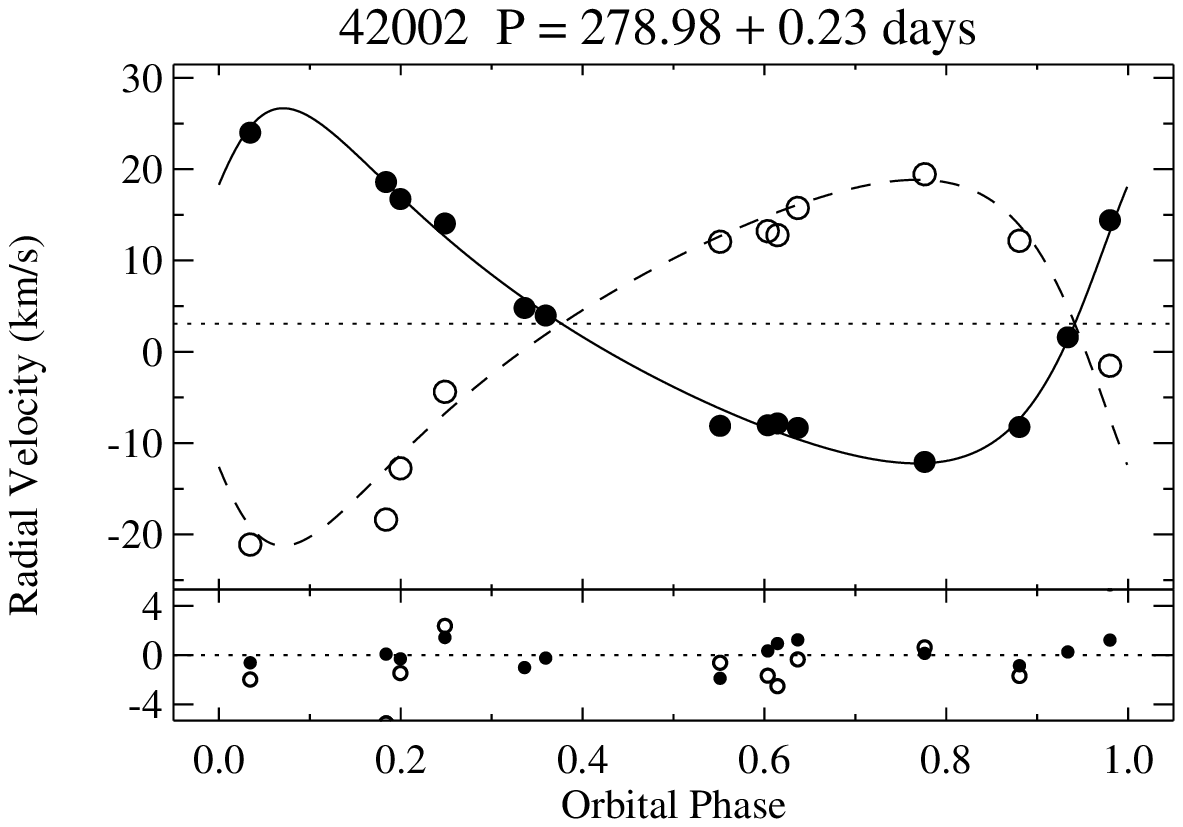}}
\subfigure{\includegraphics[width=0.3\linewidth]{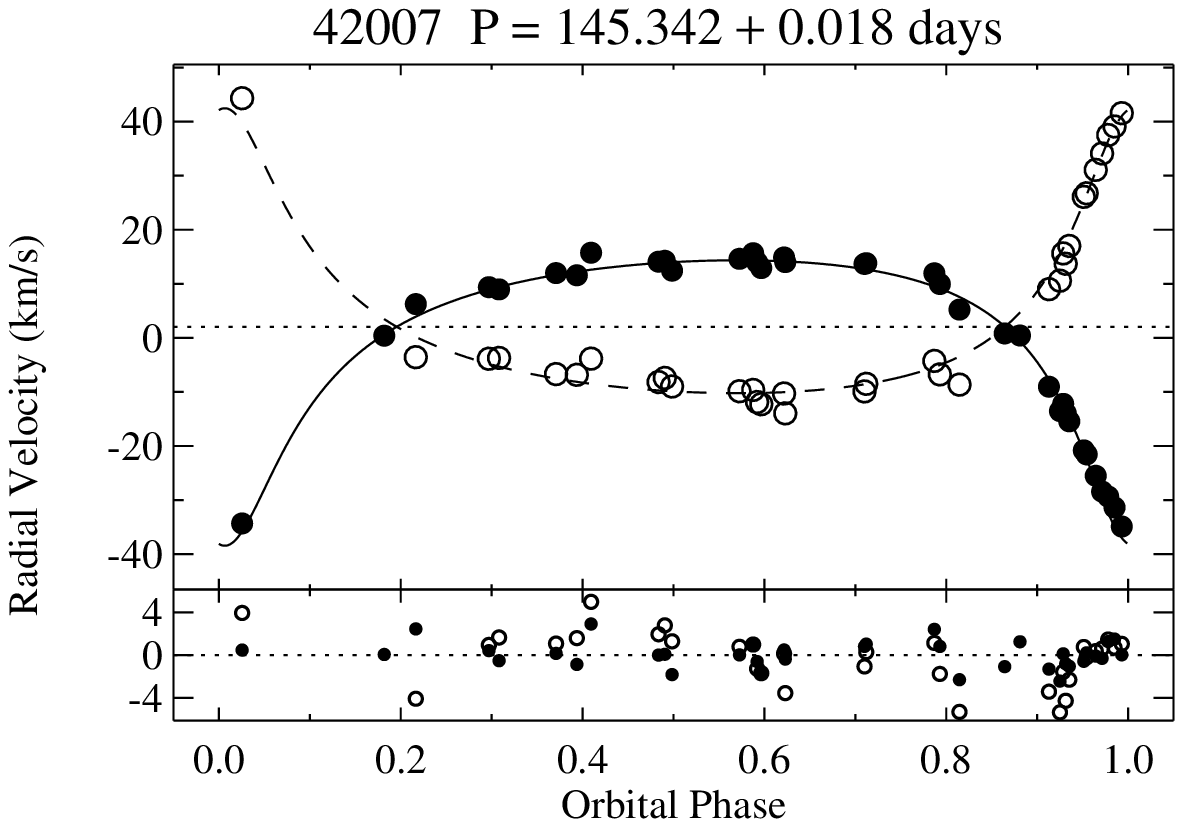}}
\subfigure{\includegraphics[width=0.3\linewidth]{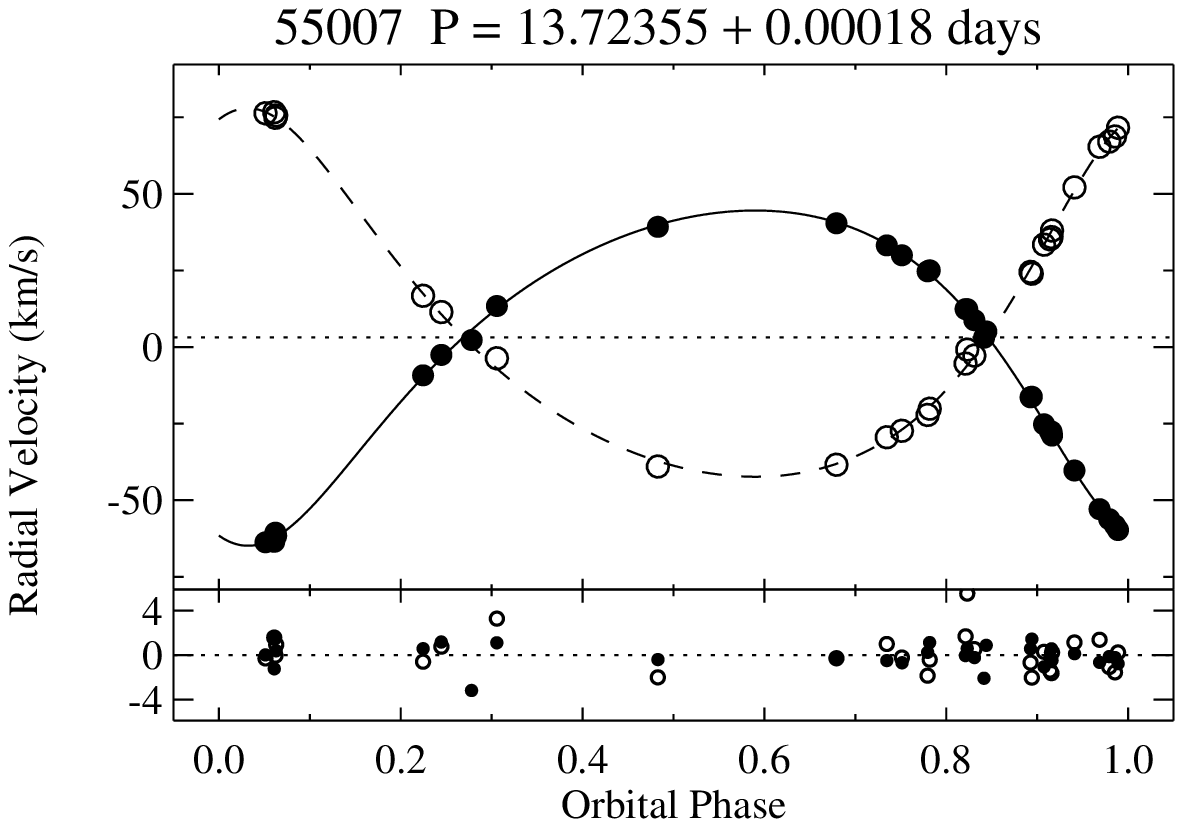}}
\subfigure{\includegraphics[width=0.3\linewidth]{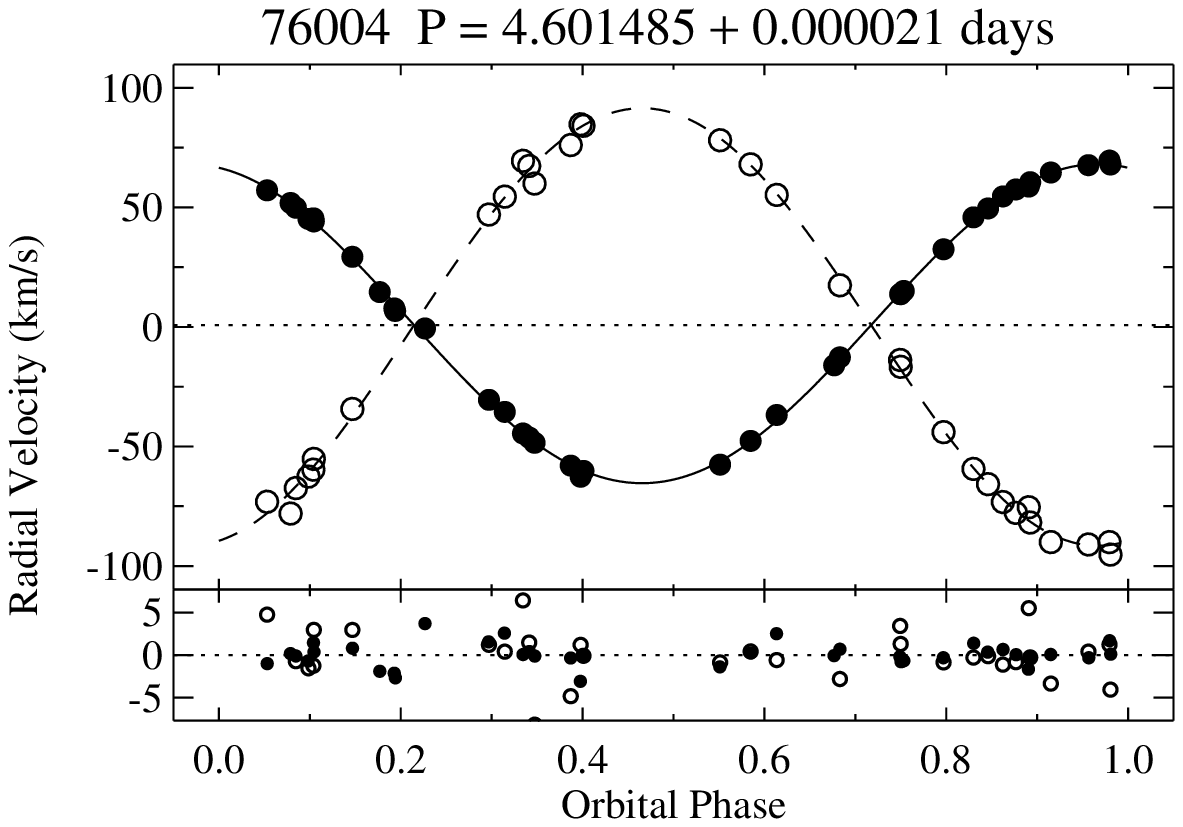}}
\end{center}
\caption{\footnotesize NGC 6819 SB2 orbit plots.  For each binary, we plot RV against orbital phase, showing the primary-star data points with filled circles and the secondary-star data points with open circles. The orbital fits to the data are plotted in the solid and dashed lines for the primary and secondary stars, respectively. The dotted line marks the $\gamma$-velocity.  Beneath each orbital plot, we show the residuals from the fit.  Above each plot, we give the binary WOCS ID and orbital period.\normalsize}
\label{fig:Sb2plots}
\end{figure*}

\clearpage
\begin{deluxetable*}{l r c r r r r r r r r c c}
\tabletypesize{\footnotesize}
\centering
\tablecaption{Orbital Parameters For NGC 6819 Double-lined Binaries\label{SB2tab}}
\tablehead{\colhead{ID} & \colhead{$P$} & \colhead{Orbital} & \colhead{$\gamma$} & \colhead{$K$} & \colhead{$e$} & \colhead{$\omega$} & \colhead{$T_\circ$} & \colhead{$a \sin i$} & \colhead{$m \sin^3 i$} & \colhead{$q$} & \colhead{$\sigma$} & \colhead{$N$} \\
\colhead{} & \colhead{(days)} & \colhead{Cycles} & \colhead{(\kms)} & \colhead{(\kms)} & \colhead{} & \colhead{(deg)} & \colhead{(HJD-2400000 d)} & \colhead{(10$^6$ km)} & \colhead{(\Msolar)} & \colhead{} & \colhead{(\kms)} & \colhead{}}
\startdata
    4004 &          297.17 &   27.4 &             2.5 &            18.1 &           0.227 &              89 &           51907 &            72.2 &            1.10 &            0.79 &  2.10 &   38 \\
         &      $\pm$ 0.09 &       &       $\pm$ 0.3 &       $\pm$ 0.5 &     $\pm$ 0.017 &         $\pm$ 5 &         $\pm$ 4 &       $\pm$ 2.0 &      $\pm$ 0.06 &      $\pm$ 0.03 &       &      \\
         &                 &       &                 &            22.9 &                 &                 &                 &            91.3 &            0.87 &                 &  1.56 &   21 \\
         &                 &       &                 &       $\pm$ 0.4 &                 &                 &                 &       $\pm$ 1.9 &      $\pm$ 0.05 &                 &       &      \\
   11004 &          91.336 &   20.2 &            2.48 &            31.8 &           0.447 &           115.6 &        51635.69 &            35.7 &            1.41 &           0.792 &  0.72 &   19 \\
         &     $\pm$ 0.012 &       &      $\pm$ 0.22 &       $\pm$ 0.3 &     $\pm$ 0.007 &       $\pm$ 1.5 &      $\pm$ 0.20 &       $\pm$ 0.4 &      $\pm$ 0.06 &     $\pm$ 0.017 &       &      \\
         &                 &       &                 &            40.1 &                 &                 &                 &            45.1 &            1.11 &                 &  2.14 &   17 \\
         &                 &       &                 &       $\pm$ 0.8 &                 &                 &                 &       $\pm$ 0.9 &      $\pm$ 0.04 &                 &       &      \\
   12006 &         33.2393 &  111.3 &            1.75 &           39.97 &           0.366 &           286.5 &        52256.78 &           17.00 &            1.34 &           0.735 &  0.54 &   18 \\
         &    $\pm$ 0.0022 &       &      $\pm$ 0.14 &      $\pm$ 0.21 &     $\pm$ 0.006 &       $\pm$ 0.8 &      $\pm$ 0.08 &      $\pm$ 0.09 &      $\pm$ 0.03 &     $\pm$ 0.008 &       &      \\
         &                 &       &                 &            54.3 &                 &                 &                 &           23.11 &           0.986 &                 &  1.31 &   16 \\
         &                 &       &                 &       $\pm$ 0.5 &                 &                 &                 &      $\pm$ 0.21 &     $\pm$ 0.014 &                 &       &      \\
   14005 &         29.7184 &  145.0 &            3.27 &            46.3 &           0.308 &           171.9 &        53425.03 &           18.02 &            1.34 &           0.888 &  0.73 &   22 \\
         &    $\pm$ 0.0008 &       &      $\pm$ 0.20 &       $\pm$ 0.4 &     $\pm$ 0.007 &       $\pm$ 1.4 &      $\pm$ 0.13 &      $\pm$ 0.15 &      $\pm$ 0.05 &     $\pm$ 0.016 &       &      \\
         &                 &       &                 &            52.2 &                 &                 &                 &            20.3 &            1.19 &                 &  1.89 &   20 \\
         &                 &       &                 &       $\pm$ 0.8 &                 &                 &                 &       $\pm$ 0.3 &      $\pm$ 0.03 &                 &       &      \\
   17003 &         20.8904 &  106.5 &             2.9 &            41.1 &           0.059 &             351 &         52489.2 &            11.8 &            0.66 &            0.95 &  2.70 &   15 \\
         &    $\pm$ 0.0011 &       &       $\pm$ 0.3 &       $\pm$ 1.3 &     $\pm$ 0.015 &        $\pm$ 10 &       $\pm$ 0.6 &       $\pm$ 0.4 &      $\pm$ 0.03 &      $\pm$ 0.04 &       &      \\
         &                 &       &                 &            43.2 &                 &                 &                 &           12.39 &            0.63 &                 &  1.29 &   12 \\
         &                 &       &                 &       $\pm$ 0.6 &                 &                 &                 &      $\pm$ 0.21 &      $\pm$ 0.05 &                 &       &      \\
   23018 &        3.574911 &  609.4 &             2.9 &            76.3 &           0.010 &             326 &        55064.78 &            3.75 &           1.086 &           0.785 &  2.39 &   49 \\
         &  $\pm$ 0.000021 &       &       $\pm$ 0.3 &       $\pm$ 0.5 &     $\pm$ 0.005 &        $\pm$ 25 &      $\pm$ 0.24 &      $\pm$ 0.03 &     $\pm$ 0.014 &     $\pm$ 0.007 &       &      \\
         &                 &       &                 &            97.3 &                 &                 &                 &           4.783 &           0.852 &                 &  2.04 &   37 \\
         &                 &       &                 &       $\pm$ 0.5 &                 &                 &                 &     $\pm$ 0.024 &     $\pm$ 0.013 &                 &       &      \\
   24009b &         3.64921 &  983.5 &             0.4 &            90.0 &           0.005 &              69 &         54836.1 &            4.52 &            1.05 &           1.023 &  4.86 &   26 \\
         &   $\pm$ 0.00004 &       &       $\pm$ 0.7 &       $\pm$ 1.3 &     $\pm$ 0.011 &       $\pm$ 140 &       $\pm$ 1.4 &      $\pm$ 0.07 &      $\pm$ 0.04 &     $\pm$ 0.023 &       &      \\
         &                 &       &                 &            88.0 &                 &                 &                 &            4.42 &            1.08 &                 &  5.00 &   22 \\
         &                 &       &                 &       $\pm$ 1.3 &                 &                 &                 &      $\pm$ 0.07 &      $\pm$ 0.04 &                 &       &      \\
   25010 &        18.78991 &  136.0 &            3.72 &            56.7 &           0.397 &           305.8 &        53192.52 &           13.45 &            1.29 &           0.924 &  1.16 &   12 \\
         &   $\pm$ 0.00020 &       &      $\pm$ 0.21 &       $\pm$ 0.6 &     $\pm$ 0.006 &       $\pm$ 1.4 &      $\pm$ 0.05 &      $\pm$ 0.16 &      $\pm$ 0.03 &     $\pm$ 0.013 &       &      \\
         &                 &       &                 &            61.4 &                 &                 &                 &           14.56 &            1.19 &                 &  0.88 &   11 \\
         &                 &       &                 &       $\pm$ 0.4 &                 &                 &                 &      $\pm$ 0.13 &      $\pm$ 0.03 &                 &       &      \\
   37013 &          167.19 &   24.0 &             1.7 &            24.3 &           0.414 &              17 &         54072.7 &            50.8 &            0.73 &            1.02 &  1.76 &   23 \\
         &      $\pm$ 0.04 &       &       $\pm$ 0.3 &       $\pm$ 0.7 &     $\pm$ 0.019 &         $\pm$ 3 &       $\pm$ 0.9 &       $\pm$ 1.6 &      $\pm$ 0.06 &      $\pm$ 0.04 &       &      \\
         &                 &       &                 &            23.9 &                 &                 &                 &            50.0 &            0.74 &                 &  2.08 &   19 \\
         &                 &       &                 &       $\pm$ 0.8 &                 &                 &                 &       $\pm$ 1.8 &      $\pm$ 0.06 &                 &       &      \\
   38008 &         5.49633 &  269.3 &            3.47 &           57.75 &           0.002 &             190 &         51994.5 &           4.365 &           0.604 &           0.855 &  0.45 &   19 \\
         &   $\pm$ 0.00004 &       &      $\pm$ 0.11 &      $\pm$ 0.15 &     $\pm$ 0.002 &        $\pm$ 70 &       $\pm$ 1.1 &     $\pm$ 0.013 &     $\pm$ 0.017 &     $\pm$ 0.012 &       &      \\
         &                 &       &                 &            67.5 &                 &                 &                 &            5.10 &           0.516 &                 &  2.62 &   19 \\
         &                 &       &                 &       $\pm$ 0.8 &                 &                 &                 &      $\pm$ 0.07 &     $\pm$ 0.008 &                 &       &      \\
   40007 &        3.185087 & 1023.6 &             3.7 &            87.9 &           0.002 &              90 &         54329.3 &            3.85 &            1.25 &           0.851 &  4.25 &   40 \\
         &  $\pm$ 0.000012 &       &       $\pm$ 0.5 &       $\pm$ 1.0 &     $\pm$ 0.008 &       $\pm$ 300 &       $\pm$ 2.4 &      $\pm$ 0.05 &      $\pm$ 0.03 &     $\pm$ 0.014 &       &      \\
         &                 &       &                 &           103.3 &                 &                 &                 &            4.53 &            1.06 &                 &  4.87 &   37 \\
         &                 &       &                 &       $\pm$ 1.1 &                 &                 &                 &      $\pm$ 0.05 &      $\pm$ 0.03 &                 &       &      \\
   42002 &          278.98 &   14.5 &             3.1 &            19.4 &           0.375 &             305 &           53522 &            69.2 &            0.72 &            0.97 &  1.09 &   14 \\
         &      $\pm$ 0.23 &       &       $\pm$ 0.3 &       $\pm$ 0.5 &     $\pm$ 0.018 &         $\pm$ 6 &         $\pm$ 4 &       $\pm$ 2.1 &      $\pm$ 0.13 &      $\pm$ 0.09 &       &      \\
         &                 &       &                 &            20.0 &                 &                 &                 &              71 &            0.70 &                 &  3.36 &   11 \\
         &                 &       &                 &       $\pm$ 1.5 &                 &                 &                 &         $\pm$ 6 &      $\pm$ 0.08 &                 &       &      \\
   42007 &         145.342 &   30.7 &            2.05 &            26.4 &           0.541 &           170.8 &         52910.5 &            44.4 &            0.66 &            1.00 &  1.37 &   29 \\
         &     $\pm$ 0.018 &       &      $\pm$ 0.25 &       $\pm$ 0.8 &     $\pm$ 0.019 &       $\pm$ 1.7 &       $\pm$ 0.5 &       $\pm$ 1.0 &      $\pm$ 0.06 &      $\pm$ 0.05 &       &      \\
         &                 &       &                 &            26.4 &                 &                 &                 &            44.3 &            0.66 &                 &  2.80 &   26 \\
         &                 &       &                 &       $\pm$ 1.2 &                 &                 &                 &       $\pm$ 1.9 &      $\pm$ 0.04 &                 &       &      \\
   55007 &        13.72355 &  232.8 &            3.17 &            54.7 &           0.259 &           160.0 &        54360.34 &            9.97 &           1.017 &           0.910 &  1.18 &   23 \\
         &   $\pm$ 0.00018 &       &      $\pm$ 0.25 &       $\pm$ 0.4 &     $\pm$ 0.006 &       $\pm$ 1.2 &      $\pm$ 0.04 &      $\pm$ 0.09 &     $\pm$ 0.023 &     $\pm$ 0.011 &       &      \\
         &                 &       &                 &            60.1 &                 &                 &                 &           10.96 &           0.925 &                 &  1.68 &   20 \\
         &                 &       &                 &       $\pm$ 0.5 &                 &                 &                 &      $\pm$ 0.11 &     $\pm$ 0.020 &                 &       &      \\
   76004 &        4.601485 &  724.1 &            0.79 &            66.7 &           0.009 &              12 &         54704.7 &           4.222 &           1.092 &           0.729 &  1.44 &   38 \\
         &  $\pm$ 0.000021 &       &      $\pm$ 0.23 &       $\pm$ 0.4 &     $\pm$ 0.005 &        $\pm$ 25 &       $\pm$ 0.3 &     $\pm$ 0.024 &     $\pm$ 0.023 &     $\pm$ 0.008 &       &      \\
         &                 &       &                 &            91.5 &                 &                 &                 &            5.79 &           0.796 &                 &  3.48 &   32 \\
         &                 &       &                 &       $\pm$ 0.8 &                 &                 &                 &      $\pm$ 0.06 &     $\pm$ 0.012 &                 &       &      
\enddata
\end{deluxetable*}

\section{Binaries of Note}
\label{sec:BoN}
\subsection{WOCS 3002}
WOCS 3002 is an SB1 binary located near the red clump, with $V$ = 12.73 and $(V-I)$ = 1.29. This system has a circular orbit and a short period of 17.7 days and is most likely a cluster member with \PPM\ = 99\% and \PRV\ = 94\%. As mentioned in \cite{Hole2009}, the current orbital separation is too small for the primary to have evolved to the tip of the RGB without overfilling its Roche lobe. \cite{Gosnell2007} explores the likelihood that the current state of this system is a result of a dynamical exchange between a horizontal branch star and a binary system. They found that this scenario is possible but that only 0.1\% of the parameter space could reproduce this system. Based on optical, UV, and X-ray information \cite{Gosnell2012} classify this system as currently an RS Canum Venaticorum type object, a tightly orbiting binary system with two chromospherically active late-type stars.

\subsection{WOCS 13001}
WOCS 13001 has $V$ = 15.87 and $(V-I)$ = 0.85 with a \PPM\ = 99\% and is a likely triple system. The light contribution from two stars is clearly seen in the spectra. The RVs for the more luminous of the two stars have an average velocity of 1.870 \kms\ and standard deviation of 1.386 \kms, resulting in an SM classification with a \PRV\ of 93\%. The velocity values of the second cross-correlation peak have much greater variability. Fitting an orbital solution to just the velocities of the second peak, as shown in Figure \ref{fig:Sb1plots}, gives a 4.24 day orbit with an eccentricity of 0.048 and \PRV\ = 94\% based on the center-of-mass velocity of the solution. It is unlikely to have a chance alignment between a binary system and an SM of this cluster within our 3$''$ fiber diameter. We suggest that the 4.24 day binary orbit is the inner binary of a long period hierarchical triple system, where the more luminous star is the tertiary companion. 

\subsection{WOCS 1006, WOCS 4008, and WOCS 6002}
WOCS 1006, WOCS 4008, and WOCS 6002 are all SB1 binaries. They are identified by \cite{Corsaro2012} as seismic outliers from the period spacing of mixed modes, lying away from the 17 other NGC 6819 stars in a $\Delta P_\mathrm{obs}-\Delta\nu$ diagram. 

WOCS 4008 has a \PPM\ of 98\%, a \PRV\ of 88\%, and is located slightly fainter and to the blue of the red clump. \cite{Corsaro2012} conclude that WOCS 4008 is a binary composed of a RGB member and a fainter less evolved companion, with the anomalous $\Delta P_\mathrm{obs}$ potentially originating from a binary interaction that altered the core structure of the primary star. WOCS 4008 has a very low eccentricity ($e$ = 0.06 $\pm$ 0.03 ) at a period of almost four years, potentially indicating a history of mass transfer in this system. Taking the primary mass of 1.75 \Msolar\ from \cite{Corsaro2012}, we find the lower limit for the secondary mass to be 0.46 \Msolar. 

WOCS 6002 has a \PPM\ of 99\% and \PRV\ of 94\% and WOCS 1006 has a \PPM\ of 99\% and a \PRV\ of 93\%. Both are located near the red clump and have $\Delta P_\mathrm{obs}$ that place them in the sequence of He-core burning stars, but with much higher masses (WOCS 6002: 2.3 \Msolar\ and WOCS 1006: 2.45 \Msolar; \citealt{Corsaro2012}) than the other red clump stars in NGC 6819 ($\sim$1.65 \Msolar; \citealt{Miglio2011}). This supports the suggestion by \cite{RV1998} that stars in that region of the CMD, brighter and bluer than the zero-age horizontal branch, may be the more evolved descendants of past blue stragglers.
We find WOCS 6002 and WOCS 1006 to have periods of 3360 days and 1524 days, respectively. This is consistent with the trend found for the blue straggler population in NGC 188 (\citealt{Geller2012}) where 13 of the 16 blue stragglers in binaries have periods of order 1000 days. From the primary mass estimates by \cite{Corsaro2012}, we put the lower limit for the secondary mass for WOCS 6002 at 0.45 \Msolar\ and for WOCS 1006 at 0.59 \Msolar. Again this is consistent with the companion-mass distribution for the NGC 188 blue straggler binaries with periods of order 1000 days, which is narrow and peaked near 0.5 \Msolar~(\citealt{GellerNature2011}). \cite{GellerNature2011} suggest this distribution indicates white dwarf companions and indeed white dwarf companions were found for three NGC 188 blue stragglers by \cite{Gosnell2014}.

\subsection{WOCS 7009 and WOCS 8007}
\cite{Basu2011} used $Kepler$ observations taken from 2009 May to 2009 December to study the oscillations of 21 red giant stars in NGC 6819 and derive estimates of mass, radius, and log($g$). Two of the giants listed in \cite{Basu2011}, WOCS 7009 and WOCS 8007, have completed SB1 orbital solutions (Figure~\ref{fig:Sb1plots}). 7009 has a \PRV\ of 58\% placing it near our member cutoff threshold, but it has \PPM\ of 99\% and \cite{Stello2011} identify it as a member from seismic data. \cite{Basu2011} estimate the primary mass to be 1.74 \Msolar. This mass combined with the mass function produced by the orbital solution gives a lower limit on the secondary mass of 0.88 \Msolar. 
WOCS 8007 has a \PPM\ of 94\%, \PRV\ of 88\%, and is also determined by \cite{Stello2011} to be a seismic member of NGC 6819. \cite{Basu2011} mass estimate for 1.62 \Msolar, putting the lower limit on the secondary mass at 0.108 \Msolar.

\subsection{WOCS 40007}
\begin{figure}[htbp]
\begin{center} 
\includegraphics[width=\linewidth]{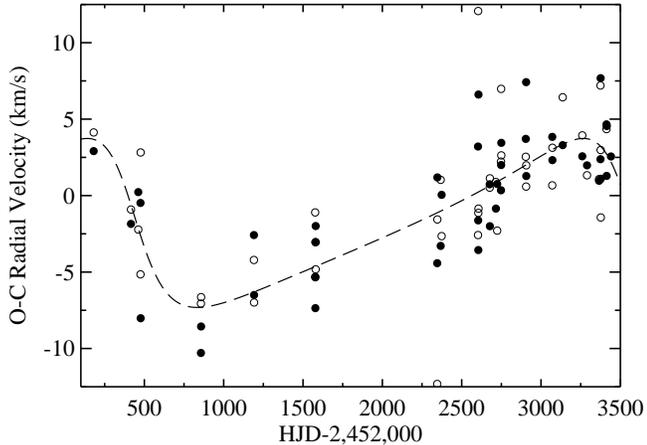}
\caption{Observed RVs minus computed RVs for the primary (filled circles) and secondary (open circles) stars in WOCS 40007. These residuals reveal the signature of a tertiary companion fit with an orbital period of $\sim$ 3100 days (dashed line). Evidence for a tertiary is also present in the photometry and eclipse timing of this system (\citealt{Jeffries2013}).}
\label{fig:40007.triple}
\end{center}
\end{figure}

WOCS 40007 is a detached eclipsing SB2 located on the MS with $V$ = 15.65, $(V-I)$ = 0.84, and an orbital period of 3.185 days. It has a \PPM\ = 97\% and a \PRV\ = 89\%. Examining the observed minus expected RVs of the primary and secondary components reveals the influence of a tertiary companion in this system with an orbital period greater than $\sim$ 3000 days (Figure~\ref{fig:40007.triple}). 

\cite{Jeffries2013} also find evidence for a triple companion in the $Kepler$ eclipse timing of primary and secondary stars. \cite{Jeffries2013} used extensive $B$$V$$R_{c}$ $I_{c}$ data (obtained on the 1 m telescope on at the Mount Laguna Observatory), WOCS RV data, and a simultaneous three-body fit to derive properties for the primary (p) and secondary (s) components. They find $M_{p}$ = 1.236 $\pm$ 0.020 \Msolar, $R_{p}$ = 1.399 $\pm$ 0.007 \Rsolar, $M_{s}$ of 1.086 $\pm$ 0.018 \Msolar, and $R_{s}$ = 1.098 $\pm$ 0.004 \Rsolar.

\subsection{WOCS 23009}
WOCS 23009 is a long-period ($P$ = 773 days) SB1 with a \PPM\ of 99\% and \PRV\ of 94\%, located near the MS turnoff with $(V-I)$ = 0.77 and $V$ = 15.11. A total eclipse was detected in quarter 4 of the $Kepler$ satellite. \cite{Sandquist2013} do a complete analysis of this system using the WOCS RV and $Kepler$ information. They find a secondary mass, $M_{s}$, of $\sim$ 0.45 \Msolar\ and use this system to estimate an age of 2.62 $\pm$ 0.25 Gyr for NGC 6819. 

\subsection{WOCS 24009}
\begin{figure}[htbp]
\begin{center} 
\includegraphics[width=\linewidth]{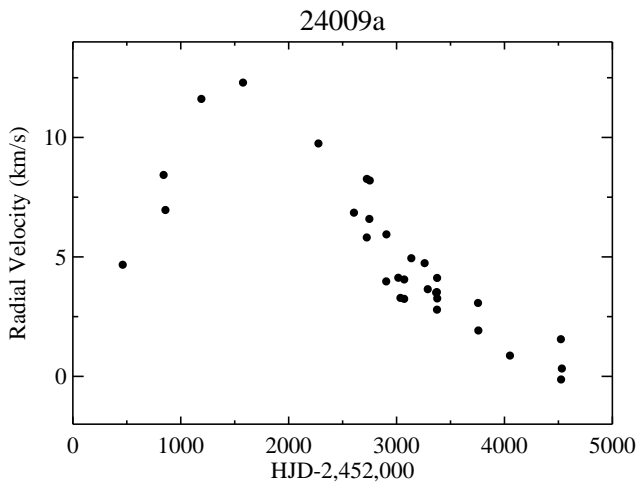}
\caption{WOCS 24009 spectra show the contributions from three sources, all with variable RVs. Radial velocities of the brightest source in the spectra, 24009a, vs. Heliocentric Julian Date (filled circles) show a binary star with a period greater than 4500 days.}
\label{fig:24009.sb1.hjd}
\end{center}
\end{figure}

WOCS 24009 is near the MS turnoff with a $V$ = 15.18 and $(V-I)$ = 0.80 and a \PPM\ of 99\%. The spectra show contributions from three sources. The brightest source, WOCS 24009a, is a long-period velocity-variable star with a $e/i$ = 7.60 and $\overline\mathrm{RV}$ = 4.911 \kms. Examining the RVs of WOCS 24009a versus HJD (Figure~\ref{fig:24009.sb1.hjd}) we find a slow change in RV over the course of more than 4000 days. The RVs peak at 12.3 \kms~and have an observed minimum of $\sim$ 0 \kms, but it appears we have not observed this binary for a full orbit and any orbital solution fit to the RVs has very large errors on all the parameters. With the current RV information this star is classified as a binary with unknown membership (BU).

The two dimmer sources in the spectra, WOCS 24009b, are very similar. The CCF consistently shows almost identical peak heights and FWHM for these sources. We are able to obtain velocities for these two sources using TODCOR when the CCFs are not too blended with the brightest source mentioned above. Using the date of periastron and period obtained from $Kepler$ observation of the eclipses in this inner binary (L. Brewer, private communication) we are able to obtain an orbital solution for this binary (Figure~\ref{fig:Sb1plots}). The $\gamma$-velocity for this short-period binary is 0.4 $\pm$ 0.7 \kms~and results in a \PRV\ of 66\%. This velocity is lower than the cluster average of \ClusAvgRV, but the \PRV\ meets our threshold for cluster membership. 

With a fiber diameter of 3$''$ it is unlikely that we have a chance alignment of a field and cluster binary or even two binary cluster members. This system has a high potential of being a hierarchical triple system in NGC 6819.

\subsection{WOCS 1010, WOCS 4004, and WOCS 14012}
In NGC 6819 we identify 17 potential blue stragglers as being brighter or bluer than the MS turnoff and not on the blue hook. Seven of these blue stragglers are velocity variable and for WOCS 1010, WOCS 4004, and WOCS 14012 we have obtained orbital solutions. 

WOCS 1010 has a $V$ = 12.82 and $(V-I)$ = 0.29, a \PRV\ 86\% and \PPM\ of 87\%. It is the bluest and one of the most luminous blue stragglers we have identified in NGC 6819. Its orbital period of 1,144 days and eccentricity of 0.55 are similar to those found for most of the blue stragglers in NGC 188 (\citealt{Geller2012}). WOCS 1010 falls on the evolutionary track for a 2.4 \Msolar~star, which is less than twice the $\sim$1.4 \Msolar~turnoff mass of NGC 6819 (\citealt{PARSEC2012}). However, we expect 2.4 \Msolar~to be a lower mass limit for WOCS 1010 based on evidence that blue stragglers are less luminous than normal stars of the same mass (\citealt{Sandquist2003}, \citealt{Geller2012}). 

WOCS 4004 is a double-lined spectroscopic binary with a \PRV\ of 94\%, \PPM\ of 99\%, $V$ = 13.98, and $(V-I)$ = 0.61. It has a period of 297.17 days, eccentricity of 0.23, and mass ratio, q = 0.79. In an effort to determine more information about the companion star, we cross-correlate the spectra of WOCS 4004 against a grid of synthetic spectra\footnote{The library is based on ATLAS9 (http://kurucz.harvard.edu) and the companion program SYNTHE used to compute the synthetic spectrum from the model atmosphere and line list.} (\citealt{Meibom2009}) with solar-metallicity, log($g$) = 4.0, and spanning appropriate ranges in temperature and projected rotation velocity. From the best matches (the highest cross-correlation peak height) we derive an effective temperature of 6900 K $\pm$ 125 K for the primary and 6650 K $\pm$ 160 K for the secondary. Based on a 2.5 Gyr solar-metallicity PARSEC isochrone (\citealt{PARSEC2012}) the hottest MS star in NGC 6819 has a T$_{eff}$ of 6382 K. Unfortunately since this temperature is within 2$\sigma$ of the temperature derived for the secondary star we are unable to conclude if the secondary star is another blue straggler or an MS star. But clearly the secondary is not a white dwarf. 

WOCS 14012 has a \PRV\ of 90\%, \PPM\ of 99\%, $V$ = 14.85, and $(V-I)$ = 0.58. It has the broadest absorption lines of any of the blue straggler candidates in NGC 6819. Using the same grid of synthetic spectra described above, we derive an effective temperature of 6465 K $\pm$ 130 K and a projected rotation velocity, $v$sin$i$, of 59 \kms~$\pm$ 6 \kms\ for WOCS 14012. This projected rotation velocity is much faster than the 10$-$20 \kms~of typical MS stars in NGC 6819, but is similar to the $v$sin$i$ values \cite{MathieuNature2009} found for some blue stragglers in NGC 188. We obtained more precise RVs, resulting in a more precise orbital solution, by cross-correlating the spectra of WOCS 14012 against a synthetic spectrum with $T_\mathrm{eff}$ of 6500 K and $v$sin$i$ of 60 \kms, instead of the typical solar template. The RVs from using the synthetic spectrum are presented in Table 1 and used for Figure~\ref{fig:Sb1plots}. The resulting orbital solution has a period of 762 days and an eccentricity of 0.13. Again, these are similar to the properties found for most of the blue straggler binaries in NGC 188 (\citealt{Geller2012}).

\section{Tidal Circularization in NGC 6819}
\label{sec:circ}
\begin{figure}[htbp]
\begin{center} 
\includegraphics[width=\linewidth]{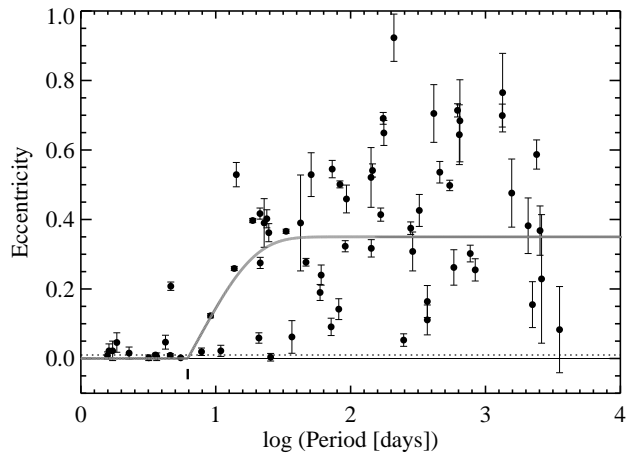}
\caption{Distribution of orbital eccentricity, $e$, as a function of the log of orbital period in days for 62 MS binary members of NGC 6819. The dashed line marks an eccentricity of 0.01, the thin solid line marks an eccentricity of zero, and the thick solid line represents the best-fit circularization function. The circularization period of NGC 6819, \Pcirc~days, is marked by a thick dash near the log($P$) axis.}
\label{fig:elogP.MS}
\end{center}
\end{figure}

We use the binary orbits of \numMSbin~MS stars to construct the distribution of orbital eccentricity versus orbital period ($e - logP$). The distribution in Figure~\ref{fig:elogP.MS} shows the usual transition from a wide range in eccentricities at longer periods to circular orbits at shorter periods, thought to result from tidal circularization over the age of the cluster. In particular, we find a tidal CP of \Pcirc~days for NGC 6819. We find the CP and the associated uncertainty using the information and fitting method of \cite{Meibom2005}. Specifically, we fit 
\begin{equation} 
	e(P) = \left\{
	\begin{array}{llr}
	0.0 & & $if$ ~P \leq P',\\
	\alpha~(1-e^{\beta(P'-P)})^{\Gamma} & & $if$ ~P > P',
	\end{array}
	\right.
\label{eq:pcirc}
\end{equation}  
to the observed period$-$eccentricity distribution. Based on Monte Carlo experiments and in order to minimize the sensitivity to the choice of eccentricity threshold, $\Gamma$ is set to 1.0 and $\beta$ is set to 0.14. The value of $\alpha$ is set to 0.35, the mean eccentricity of all observed binary orbits with periods longer than 50 days in the Pleiades, M35, Hyades, M67, and NGC 188. The location of the circularization function and $P'$ are determined by minimizing the total absolute deviation between the observed eccentricity and the model. The tidal CP is defined as the period at which the best-fit circularization function equals 0.01, $e$(CP) = 0.01. 

\begin{figure}[htbp]
\begin{center} 
\includegraphics[width=\linewidth]{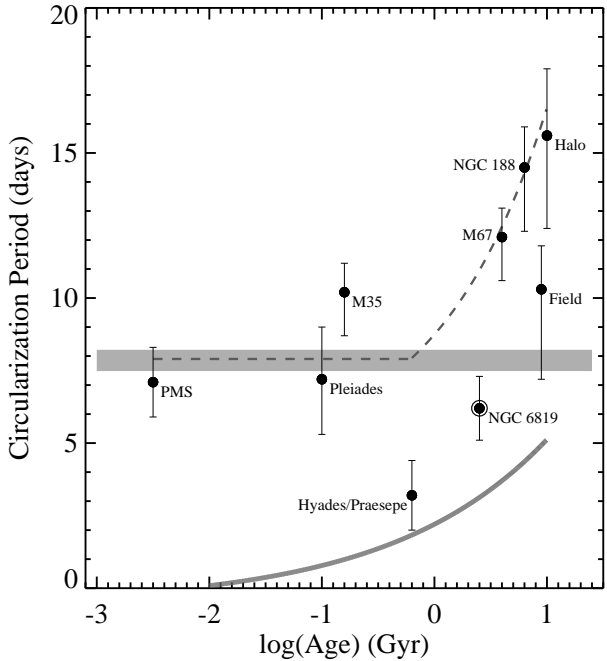}
\caption{Tidal circularization periods determined by \cite{Meibom2005} for eight different stellar populations (filled circles). The CP of NGC 6819 is plotted for comparison (large open circle around filled circle). The theoretical prediction of \cite{ZahnBouchet1989} in which tidal circularization is only significant during the PMS phase is indicated by the large gray line and the prediction of the Binary-Star Evolution algorithm of \cite{Hurley2002} is shown as the dark gray curve. The dashed line shows the ad hoc relationship introduced by \cite{Geller2013} that adds the pre-main-sequence circularization from \cite{Kroupa1995} to the Binary-Star Evolution algorithm with convective damping increased by a factor of 100. This relationship is ad hoc, but fits the increase in tidal circularization observed in the older populations.}
\label{fig:cluster.cp}
\end{center}
\end{figure}

NGC 6819 is an intermediate-age open cluster, log(Age(Gyr))= 0.4, and falls roughly in the middle of the age range of the eight MS binary populations studied by \cite{Meibom2005}. Figure~\ref{fig:cluster.cp} shows tidal CP as a function of age for all of these systems. Also plotted are three theoretical predictions. The flat grey line is from \cite{ZahnBouchet1989}, in which tidal circularization is only significant during the pre-main-sequence (PMS) phase. The other predictions plotted are from the Binary-Star Evolution algorithm used in \cite{Hurley2002} and the ad hoc tidal energy dissipation rate used by \cite{Geller2013}. The ad hoc rate from \cite{Geller2013} is a combination of the PMS circularization algorithm from \cite{Kroupa1995} and the Binary-Star evolution algorithm of \citealt{Hurley2002} with the convective damping term increased $ad$ $hoc$ by a factor of 100 over the default value in order to match the distribution of tidal CP with age (e.g., Figure~\ref{fig:cluster.cp}). 

A $\chi^{2}$-analysis shows that the CPs for the five youngest binary populations $-$ PMS binaries, Pleiades, M35, Hyades/Praesepe, and NGC 6819 $-$ are consistent with the theoretical prediction of \cite{ZahnBouchet1989} that MS tidal circularization will be negligible and the CP will remain constant as the system ages. On the other hand, Figure~\ref{fig:cluster.cp} indicates that for binary populations older than NGC 6819 tidal circularization acts to increase their tidal CPs. As yet no self-consistent theory is able to explain the distribution of cutoff periods in Figure 12 (\citealt{Meibom2005}).

\section{Binary Frequency}
\label{sec:bin.detect}
\subsection{Completeness in Binary Detection}
In order to characterize the completeness of binary detections we use a Monte Carlo approach to produce simulated observations of a set of artificial binaries, following the procedure outlined in \cite{Geller2012}. Specifically, we generated artificial binaries with period and eccentricity distributions of the Galactic field binary population found by \cite{R2010}. We used 1 \Msolar\ for the primary mass and selected the mass ratio, q, from a flat distribution. Orbital inclination, longitude of periastron, and phases were chosen randomly. Appropriate to NGC 6819, we set the eccentricity to zero for periods under 6.2 days to reflect the CP of NGC 6819 discussed in Section~\ref{sec:circ}, we used the actual observation dates in the Monte Carlo, and the precision values of 0.7 \kms\ and 0.4 \kms\ for observations made at the CfA and WIYN, respectively, were incorporated. Given these assumptions, RV measurements were produced for a large number of synthetic binaries and their detectability assessed. 

The resulting binary detection success rate varies most notably with distance from the cluster center due to using actual observation dates in the simulation and the less frequent observation near the outskirts of the cluster, and with increasing orbital period as long period binaries generally have lower RV variations. Our overall binary detection percentages are 89\% for binaries with periods under 1,000 days, 81\% for periods under 3,000 days, and 67\% for periods under 10,000 days. These numbers are in good agreement to those for NGC 188 found by \cite{Geller2012}.

\subsection{Binary Frequency}
We have identified 679 members and likely members of NGC 6819; 113 of these are binaries (BMs and BLMs). On the MS, we find 460 single and 85 binary member and likely member stars. To calculate our MS binary frequency we first remove four binary stars from the sample that, by deconvolving the combined light with possible companions, contain a primary star fainter than our magnitude cutoff of $V$ = 16.5. Next, we incorporate our 67\% binary-detection percentage for binaries with periods less than 10$^{4}$ days. This results in an incompleteness-corrected MS binary frequency of 22\% $\pm$ 3\% for binaries with periods less than 10$^{4}$ days. This is consistent with the MS binary frequency of 24\% $\pm$ 3\% found in M35 (150 Myr; \citealt{Geller2010}) and 29\% $\pm$ 3\% found for NGC 188 (7 Gyr; \citealt{Geller2012}) for binaries with periods under 10$^{4}$ days. This is also consistent with the observable MS binary frequency predicted for the age of NGC 6819 by $N$-body modeling of the older cluster NGC 188 \cite{Geller2013}. 

Using a fiducial fitting method \cite{K2001} quantified the location of the equal-mass binaries in NGC 6819 at $\sim$0.75 mag above the MS track and found an equal-mass binary percentage of $\sim$11\% $\pm$ 2\%. Using MS SB2 binaries as a proxy for equal-mass binary systems, we find the incompleteness-corrected equal-mass binary frequency for $P$ $<$ 10$^{4}$ days to be much lower at 3\% $\pm$ 1\% (11/545). Investigating the MS equal-mass binaries in NGC 188 we find a similar incompleteness-corrected frequency of 5\% $\pm$ 2\% for binaries with period under 10$^{4}$ days. 

\section{Summary}
In this paper we present the current state of the WOCS RV survey of the rich open cluster NGC 6819 (2.5 Gyr) including the orbital solutions for \numcatBM~cluster binaries. This paper builds on the work of \cite{Hole2009}, and presents the entire current WOCS database of RV measurements. We use our RV results, the proper-motion information of~\cite{Platais2013}, and the photometry of~\cite{Yang2013} to produce a cleaned CMD in Figure~\ref{fig:cmd.member} that includes BMs, BLMs, and SM stars. The CMD shows a clear MS turnoff, blue hook, red clump and potential blue stragglers, a large fraction of which are RV variable stars. 
We call special attention to a handful of interesting binaries in NGC 6819 including eclipsing binaries, $Kepler$ asteroseismology targets, potential higher order multiple star systems and possible descendants of past blue stragglers. We find the incompleteness-corrected binary fraction for all MS binaries with periods less than 10$^{4}$ days to be 22\% $\pm$ 3\%. This binary fraction is similar to the results for the young open cluster M35 and the old open cluster NGC 188 (Section~\ref{sec:bin.detect}). With the MS binary orbits and the techniques of \cite{Meibom2005} we determine a CP of \Pcirc~days for NGC 6819. The addition of this CP indicates that tidal circularization of MS binary orbits does not present itself until after 2-3 Gyr. As part of WOCS we plan to continue to observe NGC 6819 and in the future we hope to address the mass segregation, blue straggler population, and tidal circularization of the cluster in more detail.

The authors express their gratitude to the WIYN Observatory staff for the many long nights and excellent spectra. Also, we would like to thank the many undergraduate and graduate students who helped obtain and reduce the many hours WIYN data that underlie this paper, including E. Braden, T. Hole, E. Leiner and B. Tofflemire. This work was funded by the National Science Foundation grant AST-0908082 to the University of Wisconsin-Madison. A.M.G. is funded by a National Science Foundation Astronomy and Astrophysics Postdoctoral Fellowship under Award No. AST-1302765. K.E.M. is supported in part by the National Space Grant College and Fellowship Program and the Wisconsin Space Grant Consortium. 

\bibliographystyle{mn2e}
\bibliography{wocs.lx}

\begin{thebibliography}{43}
\expandafter\ifx\csname natexlab\endcsname\relax\def\natexlab#1{#1}\fi

\bibitem[{{Anthony-Twarog} {et~al}\mbox{.}(2013){Anthony-Twarog}, {Deliyannis},
  {Rich}, \& {Twarog}}]{AnthonyTwarog2013}
{Anthony-Twarog} B.~J., {Deliyannis} C.~P., {Rich} E., {Twarog} B.~A., 2013,
  \apjl, 767, L19

\bibitem[{{Barden} {et~al}\mbox{.}(1994){Barden}, {Armandroff}, {Muller},
  {et~al.}}]{Barden1994}
{Barden} S.~C., {Armandroff} T., {Muller} G., {et~al.}, 1994, in Proc. SPIE,
  Vol. 2198, Instrumentation in Astronomy VIII, {Crawford} D.~L., {Craine}
  E.~R., eds., pp. 87--97

\bibitem[{{Basu} {et~al}\mbox{.}(2011){Basu}, {Grundahl}, {Stello},
  {et~al.}}]{Basu2011}
{Basu} S., {Grundahl} F., {Stello} D., {et~al.}, 2011, \apjl, 729, L10

\bibitem[{{Bragaglia} {et~al}\mbox{.}(2001){Bragaglia}, {Carretta}, {Gratton},
  {et~al.}}]{Brag2001}
{Bragaglia} A., {Carretta} E., {Gratton} R.~G., {et~al.}, 2001, \aj, 121, 327

\bibitem[{{Bressan} {et~al}\mbox{.}(2012){Bressan}, {Marigo}, {Girardi},
  {et~al.}}]{PARSEC2012}
{Bressan} A., {Marigo} P., {Girardi} L., {et~al.}, 2012, \mnras, 427, 127

\bibitem[{{Clem} \& {Landolt}(2013)}]{Clem2013}
{Clem} J.~L., {Landolt} A.~U., 2013, \aj, 146, 88

\bibitem[{{Corsaro} {et~al}\mbox{.}(2012){Corsaro}, {Stello}, {Huber},
  {et~al.}}]{Corsaro2012}
{Corsaro} E., {Stello} D., {Huber} D., {et~al.}, 2012, \apj, 757, 190

\bibitem[{{Everett}, {Howell} \& {Kinemuchi}(2012){Everett}, {Howell}, \&
  {Kinemuchi}}]{UBV2012}
{Everett} M.~E., {Howell} S.~B., {Kinemuchi} K., 2012, \pasp, 124, 316

\bibitem[{{Geller}, {Hurley} \& {Mathieu}(2013){Geller}, {Hurley}, \&
  {Mathieu}}]{Geller2013}
{Geller} A.~M., {Hurley} J.~R., {Mathieu} R.~D., 2013, \aj, 145, 8

\bibitem[{{Geller} \& {Mathieu}(2011)}]{GellerNature2011}
{Geller} A.~M., {Mathieu} R.~D., 2011, \nat, 478, 356

\bibitem[{{Geller} \& {Mathieu}(2012)}]{Geller2012}
{Geller} A.~M., {Mathieu} R.~D., 2012, \aj, 144, 54

\bibitem[{{Geller} {et~al}\mbox{.}(2010){Geller}, {Mathieu}, {Braden},
  {et~al.}}]{Geller2010}
{Geller} A.~M., {Mathieu} R.~D., {Braden} E.~K., {et~al.}, 2010, \aj, 139, 1383

\bibitem[{{Geller} {et~al}\mbox{.}(2008){Geller}, {Mathieu}, {Harris}, \&
  {McClure}}]{Geller2008}
{Geller} A.~M., {Mathieu} R.~D., {Harris} H.~C., {McClure} R.~D., 2008, \aj,
  135, 2264

\bibitem[{{Gosnell} {et~al}\mbox{.}(2007){Gosnell}, {DiPompeo}, {Braden},
  {et~al.}}]{Gosnell2007}
{Gosnell} N., {DiPompeo} M.~A., {Braden} E.~K., {et~al.}, 2007, in BAAS,
  Vol.~39, American Astronomical Society Meeting Abstracts, p. 839

\bibitem[{{Gosnell} {et~al}\mbox{.}(2014){Gosnell}, {Mathieu}, {Geller},
  {et~al.}}]{Gosnell2014}
{Gosnell} N.~M., {Mathieu} R.~D., {Geller} A.~M., {et~al.}, 2014, \apjl, 783,
  L8

\bibitem[{{Gosnell} {et~al}\mbox{.}(2012){Gosnell}, {Pooley}, {Geller},
  {et~al.}}]{Gosnell2012}
{Gosnell} N.~M., {Pooley} D., {Geller} A.~M., {et~al.}, 2012, \apj, 745, 57

\bibitem[{{Hole} {et~al}\mbox{.}(2009){Hole}, {Geller}, {Mathieu},
  {et~al.}}]{Hole2009}
{Hole} K.~T., {Geller} A.~M., {Mathieu} R.~D., {et~al.}, 2009, \aj, 138, 159

\bibitem[{{Hurley}, {Tout} \& {Pols}(2002){Hurley}, {Tout}, \&
  {Pols}}]{Hurley2002}
{Hurley} J.~R., {Tout} C.~A., {Pols} O.~R., 2002, \mnras, 329, 897

\bibitem[{{Jeffries} {et~al}\mbox{.}(2013){Jeffries}, {Sandquist}, {Mathieu},
  {et~al.}}]{Jeffries2013}
{Jeffries}, Jr. M.~W., {Sandquist} E.~L., {Mathieu} R.~D., {et~al.}, 2013, \aj,
  146, 58

\bibitem[{{Kalirai} {et~al}\mbox{.}(2001){Kalirai}, {Richer}, {Fahlman},
  {et~al.}}]{K2001}
{Kalirai} J.~S., {Richer} H.~B., {Fahlman} G.~G., {et~al.}, 2001, \aj, 122, 266

\bibitem[{{Kozhurina-Platais} {et~al}\mbox{.}(1995){Kozhurina-Platais},
  {Girard}, {Platais}, {et~al.}}]{Koz1995}
{Kozhurina-Platais} V., {Girard} T.~M., {Platais} I., {et~al.}, 1995, \aj, 109,
  672

\bibitem[{{Kroupa}(1995)}]{Kroupa1995}
{Kroupa} P., 1995, \mnras, 277, 1507

\bibitem[{{Mathieu}(1983)}]{Mathieu1983}
{Mathieu} R.~D., 1983, \apjl, 267, L97

\bibitem[{{Mathieu}(2000)}]{Mathieu2000}
{Mathieu} R.~D., 2000, in ASP Conf. Ser., Vol. 198, Stellar Clusters and
  Associations: Convection, Rotation, and Dynamos, {Pallavicini} R., {Micela}
  G., {Sciortino} S., eds., p. 517

\bibitem[{{Mathieu} \& {Geller}(2009)}]{MathieuNature2009}
{Mathieu} R.~D., {Geller} A.~M., 2009, \nat, 462, 1032

\bibitem[{{Mathieu} {et~al}\mbox{.}(2003){Mathieu}, {van den Berg}, {Torres},
  {et~al.}}]{Mathieu2003}
{Mathieu} R.~D., {van den Berg} M., {Torres} G., {et~al.}, 2003, \aj, 125, 246

\bibitem[{{Meibom} {et~al}\mbox{.}(2009){Meibom}, {Grundahl}, {Clausen},
  {et~al.}}]{Meibom2009}
{Meibom} S., {Grundahl} F., {Clausen} J.~V., {et~al.}, 2009, \aj, 137, 5086

\bibitem[{{Meibom} \& {Mathieu}(2005)}]{Meibom2005}
{Meibom} S., {Mathieu} R.~D., 2005, \apj, 620, 970

\bibitem[{{Miglio} {et~al}\mbox{.}(2012){Miglio}, {Brogaard}, {Stello},
  {et~al.}}]{Miglio2011}
{Miglio} A., {Brogaard} K., {Stello} D., {et~al.}, 2012, \mnras, 419, 2077

\bibitem[{{Platais} {et~al}\mbox{.}(2013){Platais}, {Gosnell}, {Meibom},
  {et~al.}}]{Platais2013}
{Platais} I., {Gosnell} N.~M., {Meibom} S., {et~al.}, 2013, \aj, 146, 43

\bibitem[{{Raghavan} {et~al}\mbox{.}(2010){Raghavan}, {McAlister}, {Henry},
  {et~al.}}]{R2010}
{Raghavan} D., {McAlister} H.~A., {Henry} T.~J., {et~al.}, 2010, \apjs, 190, 1

\bibitem[{{Rosvick} \& {Vandenberg}(1998)}]{RV1998}
{Rosvick} J.~M., {Vandenberg} D.~A., 1998, \aj, 115, 1516

\bibitem[{{Sanders}(1972)}]{Sanders1972}
{Sanders} W.~L., 1972, \aap, 19, 155

\bibitem[{{Sandquist} {et~al}\mbox{.}(2003){Sandquist}, {Latham}, {Shetrone},
  \& {Milone}}]{Sandquist2003}
{Sandquist} E.~L., {Latham} D.~W., {Shetrone} M.~D., {Milone} A.~A.~E., 2003,
  \aj, 125, 810

\bibitem[{{Sandquist} {et~al}\mbox{.}(2013){Sandquist}, {Mathieu}, {Brogaard},
  {et~al.}}]{Sandquist2013}
{Sandquist} E.~L., {Mathieu} R.~D., {Brogaard} K., {et~al.}, 2013, \apj, 762,
  58

\bibitem[{{Sarrazine} {et~al}\mbox{.}(2003){Sarrazine}, {Deliyannis},
  {Sarajedini}, \& {Platais}}]{Sarrazine2003}
{Sarrazine} A.~R., {Deliyannis} C.~P., {Sarajedini} A., {Platais} I., 2003, in
  BASS, Vol.~35, American Astronomical Society Meeting Abstracts, p. 1230

\bibitem[{{Stello} {et~al}\mbox{.}(2010){Stello}, {Basu}, {Bruntt},
  {et~al.}}]{Stello2010}
{Stello} D., {Basu} S., {Bruntt} H., {et~al.}, 2010, \apjl, 713, L182

\bibitem[{{Stello} {et~al}\mbox{.}(2011){Stello}, {Meibom}, {Gilliland},
  {et~al.}}]{Stello2011}
{Stello} D., {Meibom} S., {Gilliland} R.~L., {et~al.}, 2011, \apj, 739, 13

\bibitem[{{Street} {et~al}\mbox{.}(2002){Street}, {Horne}, {Lister},
  {et~al.}}]{Street2002}
{Street} R.~A., {Horne} K., {Lister} T.~A., {et~al.}, 2002, \mnras, 330, 737

\bibitem[{{Street} {et~al}\mbox{.}(2005){Street}, {Horne}, {Lister},
  {et~al.}}]{Street2005}
{Street} R.~A., {Horne} K., {Lister} T.~A., {et~al.}, 2005, \mnras, 358, 795

\bibitem[{{Yang} {et~al}\mbox{.}(2013){Yang}, {Sarajedini}, {Deliyannis},
  {et~al.}}]{Yang2013}
{Yang} S.-C., {Sarajedini} A., {Deliyannis} C.~P., {et~al.}, 2013, \apj, 762, 3

\bibitem[{{Zahn} \& {Bouchet}(1989)}]{ZahnBouchet1989}
{Zahn} J.-P., {Bouchet} L., 1989, \aap, 223, 112

\bibitem[{{Zucker} \& {Mazeh}(1994)}]{Zucker1994}
{Zucker} S., {Mazeh} T., 1994, \apj, 420, 806

\end{thebibliography}
\end{document}